\documentclass[twocolumn,prd,floats,preprintnumbers,showpacs,nofootinbib]{revtex4-1}  

\usepackage[english]{babel}
\usepackage{graphicx}
\usepackage{amsmath,amssymb}
\usepackage{color}

\begin{document}

\newcommand{\etal}{et al.\ }

 \newcommand{\nc}{\newcommand}
 \nc{\8}{\"{a}}
 \nc{\bea}{\begin{eqnarray}} \nc{\eea}{\end{eqnarray}}
 \nc{\beq}{\begin{equation}} \nc{\eeq}{\end{equation}}
 \nc{\bi}{\begin{itemize}}\nc{\ei}{\end{itemize}}
 \nc{\ben}{\begin{enumerate}}\nc{\een}{\end{enumerate}}
 \nc{\nn}{\nonumber}
 \nc{\exercise}{{\bf (exercise)}}
 \nc{\const}{\mbox{const.}}
 
 \nc{\G}[3]{\Gamma^#1_{#2#3}}
 \nc{\bG}[3]{\bar{\Gamma}^#1_{#2#3}}
 \nc{\dG}[3]{\delta\Gamma^#1_{#2#3}}
 \nc{\Gd}[4]{\Gamma^#1_{#2#3,#4}}
 \nc{\bGd}[4]{\bar{\Gamma}^#1_{#2#3,#4}}
 \nc{\dGd}[4]{\delta\Gamma^#1_{#2#3,#4}}
 \nc{\piG}{\pi G}

 \nc{\zth}{$0^\mathrm{th}$\ }
 \nc{\fst}{$1^\mathrm{st}$\ }
 \nc{\snd}{$2^\mathrm{nd}$\ }

 \nc{\half}{{\textstyle \frac12}}
 \nc{\third}{{\textstyle \frac13}}
 \nc{\quarter}{{\textstyle \frac14}}
 \nc{\fifth}{{\textstyle \frac15}}
 \nc{\sixth}{{\textstyle \frac16}}
 \nc{\seventh}{{\textstyle \frac17}}
 \nc{\eighth}{{\textstyle \frac18}}
 \nc{\nineth}{{\textstyle \frac19}}
 \nc{\tenth}{{\textstyle \frac{1}{10}}}
 \nc{\threehalves}{{\textstyle \frac32}} 
 \nc{\ninehalves}{{\textstyle \frac92}}
 \nc{\twothirds}{{\textstyle \frac23}}
 \nc{\fourthirds}{{\textstyle \frac43}}
 \nc{\eightthirds}{{\textstyle \frac83}}
 \nc{\tenthirds}{{\textstyle \frac{10}{3}}}
 \nc{\threequarters}{{\textstyle \frac34}}
 \nc{\ninequarters}{{\textstyle \frac94}}
 \nc{\twofifths}{{\textstyle \frac25}}
 \nc{\threefifths}{{\textstyle \frac35}}
 \nc{\sixfifths}{{\textstyle \frac65}}
 
 \nc{\pardx}[3]{\biggl(\frac{\partial#1}{\partial#2}\biggr)_#3}
 \nc{\pard}[2]{\frac{\partial#1}{\partial#2}}
 \nc{\divg}{\nabla\cdot}
 \nc{\rot}{\nabla\times}
 \nc{\curl}{\nabla\times}

 \nc{\br}{{\bm r}}
 \nc{\bx}{{\bm x}}
 \nc{\bu}{{\bm u}}
 \nc{\bv}{{\bm v}}
 \nc{\bk}{{\bm k}}
 \nc{\ki}{{{\bm k} i}}

 \nc{\vk}{{\vec{k}}}
 \nc{\vx}{{\vec{x}}}
 \nc{\vv}{{\vec{v}}}

 \nc{\bgg}{\bar{g}_{\mu\nu}}
 \nc{\bgrho}{\bar{\rho}}
 \nc{\bgp}{\bar{p}}
 \nc{\bgphi}{\bar{\varphi}}
 \nc{\delphi}{{\delta\varphi}}
 \nc{\ddelphi}{\delta\dot{\varphi}}
 \nc{\dddelphi}{\delta\ddot{\varphi}}

 \nc{\keq}{k_\mathrm{eq}}
 \nc{\kdec}{k_\mathrm{dec}}
 \nc{\aeq}{a_\mathrm{eq}}
 \nc{\adec}{a_\mathrm{dec}}
 \nc{\aend}{a_\mathrm{end}} 
 \nc{\Heq}{H_\mathrm{eq}}
 \nc{\Hdec}{H_\mathrm{dec}}
 \nc{\teq}{t_\mathrm{eq}}
 \nc{\tdec}{t_\mathrm{dec}}
 \nc{\tend}{t_\mathrm{end}} 
 \nc{\etaeq}{\eta_\mathrm{eq}}
 \nc{\Hubeq}{{\cal H}_\mathrm{eq}}
 
 \nc{\Hub}{{\cal H}}
 \nc{\R}{{\cal R}}
 \nc{\Ord}{{\cal O}}
 \nc{\Pow}{{\cal P}}
 \nc{\Cov}{{\cal C}}
 \nc{\Se}{{\cal S}}
 \nc{\La}{{\cal L}}
 \nc{\Vol}{{\cal V}}
\newcommand{\h}{\mathcal{H}} 

 \nc{\koHub}{\left(\frac{k}{\Hub}\right)}
 \nc{\koHubeq}{\left(\frac{k}{\Hubeq}\right)}

 \nc{\M}{M_\mathrm{Pl}}
 \nc{\V}{V(\varphi)}
 \nc{\GeV}{\mbox{ GeV}}
 
 \nc{\ad}{{\hat{a}^\dagger}}
 \nc{\an}{{\hat{a}}}
 \nc{\vac}{{|0\rangle}}
 \nc{\vacd}{{\langle 0|}}
 \nc{\ophi}{{\hat{\varphi}}}
 \nc{\odelphi}{{\delta\hat{\varphi}}}
 
 \nc{\qand}{\quad\mbox{and}\quad}
 \nc{\qqand}{\qquad\mbox{and}\qquad}
 \nc{\qor}{\quad\mbox{or}\quad}
 \nc{\qqor}{\qquad\mbox{or}\qquad}
 \nc{\So}{\quad\Rightarrow\quad}
 
 \nc{\vfi}{\varphi}
 \nc{\bfi}{\bar{\varphi}}
 \nc{\delfi}{{\delta\varphi}}
 \nc{\veps}{\varepsilon}
 \nc{\dfi}{\dot{\varphi}}
 \nc{\ddfi}{\ddot{\varphi}}
 \nc{\dsigma}{\dot{\sigma}}
 \nc{\ddsigma}{\ddot{\sigma}} 
 \nc{\dtheta}{\dot{\theta}}
 \nc{\ddtheta}{\ddot{\theta}}
 \nc{\dchi}{\dot{\chi}}
 \nc{\ddchi}{\ddot{\chi}}

 \nc{\Vz}{V_{\sigma}}
 \nc{\Vzz}{V_{\sigma\sigma}}
 \nc{\Vzs}{V_{\sigma s}}
 \nc{\Vss}{V_{ss}} 
 \nc{\etzz}{\eta_{\sigma\sigma}}
 \nc{\etzs}{\eta_{\sigma s}}
 \nc{\etss}{\eta_{ss}}
 \nc{\xizzs}{\xi_{\sigma\sigma s}}
 \nc{\delsigma}{{\delta\sigma}}
 \nc{\dels}{\delta s}

 \nc{\snt}{\sin\theta}
 \nc{\cst}{\cos\theta}
 \nc{\tnt}{\tan\theta}
 
 \nc{\sgn}{\mbox{sign}}
 
\newcommand{\mr}[1]{\mathrm{#1}}
\newcommand{\mtc}[1]{\mathcal{#1}}
\newcommand{\mtr}[1]{\mathrm{#1}}

\newcommand{\mtb}[1]{\mathbf{#1}}
\newcommand{\nad}{n_{\mr{ad}}}
\newcommand{\nadI}{n_{\mr{ar}}}
\newcommand{\nadII}{n_{\mr{as}}}
\newcommand{\niso}{n_{\mr{iso}}}
\newcommand{\ncor}{n_{\mr{cor}}}
\newcommand{\acor}{\alpha_{\mr{cor}}}
\newcommand{\abs}[1]{\vert #1 \vert}
\newcommand{\comment}[1]{}
\nc{\Mpci}{{\mbox{Mpc}^{-1}}}

\title{Constraints on neutrino density and velocity isocurvature modes from WMAP-9 data}

\author{Matti Savelainen}
\email{matti.savelainen@helsinki.fi}

\author{Jussi V\"{a}liviita}

\author{Parampreet Walia}

\author{Stanislav Rusak}

\author{Hannu Kurki-Suonio}

\affiliation{Department of Physics and Helsinki Institute of Physics, University of Helsinki,
P.O. Box 64, FIN-00014 University of Helsinki, Finland}

\date{16th July 2013}

\begin{abstract}
We use WMAP 9-year and other CMB data to constrain cosmological models where the primordial perturbations have both an adiabatic and a (possibly correlated)
neutrino density (NDI), neutrino velocity (NVI), or cold dark matter density (CDI) isocurvature component.  
For NDI and CDI we use both a phenomenological approach, where  primordial perturbations are parametrized in terms of amplitudes at two scales, and a slow-roll two-field inflation approach, where slow-roll parameters are used as primary parameters.  For NVI we use only the phenomenological approach, since it is difficult to imagine a connection with inflation. We find that in the NDI and NVI cases larger isocurvature fractions are allowed than in the corresponding models with CDI.  For uncorrelated perturbations, the upper limit to the primordial NDI (NVI) fraction is 24\% (20\%) at $k = 0.002\, \mbox{Mpc}^{-1}$ and 28\% (16\%) at $k = 0.01\, \mbox{Mpc}^{-1}$.  For maximally correlated (anticorrelated) perturbations, the upper limit to the NDI fraction is 3.0\% (0.9\%).  The nonadiabatic contribution to the CMB temperature variance can be as large as 10\% (--13\%) for the NDI (NVI) modes. Bayesian model comparison favors pure adiabatic initial mode over the mixed primordial adiabatic and NDI, NVI, or CDI perturbations. At best, the betting odds for a mixed model (uncorrelated NDI) are 1:3.4 compared to the pure adiabatic model. For the phenomenological generally correlated mixed models the odds are about 1:100, whereas the slow-roll approach leads to 1:13 (NDI) and 1:51 (CDI). 
\end{abstract}

\preprint{HIP-2013-13/TH}

\maketitle

%
%

\section{Introduction \label{sec:intro}}

Cosmological models with cold dark matter isocurvature perturbations (CDI) were extensively studied in light of  Wilkinson Microwave Anisotropy Probe (WMAP) seven-year data in \cite{Valiviita:2012ub} both in a phenomenological and multi-field inflationary set-ups. 
Now we test within a similar framework what information the nine-year WMAP (WMAP-9) data \cite{Bennett:2012zja,Larson:2010gs} and other, smaller scale, cosmic microwave background (CMB) data \cite{Reichardt:2008ay, Brown:2009uy} give on possible deviations from adiabaticity, focusing on the less studied neutrino density (NDI)  and velocity (NVI) isocurvature modes. For completeness we also update the analysis of the CDI mode.
Recently, the CDI, NDI, and NVI modes were discussed in \cite{Ade:2013uln} in light of the Planck data, but only using a phenomenological approach and without simultaneously allowing for the possible tensor perturbation component.

Along with the adiabatic mode, the baryon density isocurvature (BDI), CDI, NDI and NVI modes are the only regular primordial
scalar perturbation modes \cite{Bucher:1999re}, others are either decaying modes or singular. The BDI and CDI modes are
indistinguishable at the linear level in the CMB  (see however \cite{Grin:2013uya,Grin:2011tf}) and the results for the CDI mode can easily be interpreted as constraints on the total
matter density isocurvature.

We define the different perturbation modes using the five perturbation quantities: 1) the curvature perturbation in the comoving gauge, $\mtc{R}$, 2) the cold dark matter entropy perturbation
 \beq
   S_{cr} \ \equiv \ \delta_c - \threequarters\delta_r \,,
 \label{eq:Scdef}  
 \eeq
3) the baryon entropy perturbation
 \beq
 	S_{br} \equiv \delta_b - \frac{3}{4}\delta_r \,,
 \label{eq:Sbdef}
 \eeq
4) the neutrino entropy perturbation
  \beq
   S_{\nu r} \ \equiv \ \threequarters(\delta_\nu - \delta_r) \,,
 \label{eq:Snudef}  
 \eeq
and 5) the relative neutrino heat flux \cite{cambnote}
 \beq
	q_{\nu r} \ \equiv \ \textstyle{\frac{4}{3}}  (v_\nu-v_r) \,,
 \eeq
where the $\delta_i$ are density contrasts of the different energy components, the $v_i$ are velocity perturbation potentials, and $r$ stands for radiation, i.e., photons ($\gamma$) and neutrinos ($\nu$).
The pure perturbation modes correspond to 4 of these 5 quantities vanishing initially (in the limit where conformal time $\tau\rightarrow 0$).

Thus in the adiabatic (or pure CDI, BDI, NDI, NVI) mode, initially only $\mtc{R}$ (or $S_{cr}$, $S_{br}$, $S_{\nu r}$, $q_{\nu r}$) $ \neq 0$.\footnote{%
Note that sometimes, for example in \cite{Langlois:2012tm}, these modes are defined in terms of 
$S_{c\gamma} \equiv \delta_c - \threequarters\delta_\gamma$, $S_{b\gamma}$, $S_{\nu\gamma}$, and $q_{\nu\gamma}$ instead,
which leads to a different definition of the NDI mode, since if $S_{\nu r} \neq 0$, then $S_{cr}$ and $S_{br}$ can vanish initially, but
$S_{c\gamma} = S_{b\gamma} \neq 0$ in the limit $\tau\rightarrow 0$. On the other hand,  if $S_{\nu \gamma} \neq 0$, then $S_{c\gamma}$ and $S_{b\gamma}$ can vanish initially, but $S_{cr} = S_{br} \neq 0$ in the limit $\tau\rightarrow 0$.
The neutrino perturbation quantities are related by $S_{\nu r} = (1-f_\nu)S_{\nu\gamma}$ and $q_{\nu r} = (1-f_\nu)q_{\nu\gamma}$, where $f_\nu \equiv \rho_\nu/(\rho_\gamma+\rho_\nu) \approx 0.4$ is the neutrino energy density fraction. CAMB uses the definitions with respect to the total radiation, i.e., $S_{\nu r}$ and $q_{\nu r} $ \cite{cambnote}.%
}

There is so far no evidence for a CDI, NDI, nor NVI perturbation component, and CMB temperature anisotropy observations require them to be subdominant compared to the adiabatic component  \cite{Enqvist:2000hp,Enqvist:2001fu,Beltran:2005xd,Beltran:2005gr,Keskitalo:2006qv,Trotta:2006ww,Seljak:2006bg,Lewis:2006ma,Bean:2006qz,Kawasaki:2007mb,Beltran:2008ei,Sollom:2009vd,Valiviita:2009bp,Castro:2009ej,Li:2010yb,Ade:2013uln}.

We assume a power-law power spectrum for the primordial curvature and isocurvature perturbations and for their correlation. We also allow for tensor perturbations, which are a natural prediction of inflationary models. However, we allow only one isocurvature mode at a time, since the more general cases, where we have a mixture of several isocurvature modes in addition to the adiabatic and tensor modes, are quite intractable with the present data. In addition to the generally correlated perturbations, we study special cases with no correlation or with $\pm$100\% correlation.
Assuming spatially flat geometry of the Universe, we perform full parameter scans of this \emph{mixed} (adiabatic and isocurvature) model, as well as the standard adiabatic $\Lambda$CDM model.
We present posterior probability densities of the standard cosmological parameters and the extra isocurvature parameters and report the Bayesian evidences for the models. These evidences give the betting odds for the pure adiabatic model against the various mixed (isocurvature) models.

We use two different approaches: 1) A phenomenological approach, where we make no reference to the origin of the primordial perturbations and just determine or constrain their amplitudes from the data, allowing the spectra of the adiabatic, isocurvature, and correlation components to be independent. 
2) For the CDI and NDI modes, we also use a slow-roll two-field inflation approach, where we assume the perturbations were generated by quantum fluctuations during two-field inflation, and the spectral indices are determined by the slow-roll parameters at the time the cosmological scales exited the horizon during inflation. This approach forces the spectra to be nearly scale invariant, since we assume the magnitude of the slow roll parameters to be small.  

The first approach is good for detecting non-adiabatic features in the data. 
If these were found, then a further investigation would be motivated. If not, then we can set an upper limit to the non-adiabaticity of the data. The second approach may give answers to questions directly related to inflation and inflationary potential.

We use the same notation as in \cite{Valiviita:2012ub}.
A summary of the symbols can be found in Table I of \cite{Valiviita:2012ub}. In Sec.~\ref{sec:model} we introduce our model and its phenomenological and inflationary parametrizations.
In Sec.~\ref{sec:mechanisms} we review two scenarios for generating the NDI mode. In Sec.~\ref{sec:data} we describe the data and sampling method. Secs.~\ref{sec:general_cmb} and \ref{sec:specialcases} are devoted to the results. We summarize the main findings in Sec.~\ref{sec:discussion} and discuss why WMAP-9 leads to tighter constraints than the recent Planck data.

\vspace{-2mm}
\section{The model \label{sec:model}}
\vspace{-1mm}

We assume that the primordial perturbation, here presented in Fourier space, is a superposition of the adiabatic mode (characterized by the comoving curvature perturbation $\mtc{R}(\mathbf{k})$) and an isocurvature mode characterized by $S(\mathbf{k})$, where $S$ is either $S_{cr}$, $S_{\nu r}$, or  $q_{\nu r}$.
The power spectrum
$\Pow =  \Pow_{\mtc{R}}  +  \mtc{C}_{\mtc{R}S}  +   \mtc{C}_{S\mtc{R}} + \Pow_{S}$ and its components are defined by the expectation value
\begin{multline}
\Big\langle [\mtc{R}(\mathbf{k}) + S(\mathbf{k})]^\ast [\mtc{R}(\mathbf{\tilde k}) + S(\mathbf{\tilde k})] \Big\rangle \equiv (2\pi)^3 \delta^{(3)}(\mathbf{k}-\mathbf{\tilde{k}}) \times \\
\frac{2\pi^2}{k^3}\left[ \Pow_{\mtc{R}}(k) + \mtc{C}_{\mtc{R}S}(k) +  \mtc{C}_{S\mtc{R}}(k) + \Pow_{S}(k)  \right]\,.
\end{multline}

Following \cite{Peiris:2003ff,Valiviita:2003ty,KurkiSuonio:2004mn,Keskitalo:2006qv,Kawasaki:2007mb,Valiviita:2009bp} we divide
 $\mtc{R}(\mathbf{k})$ into an uncorrelated part (``ar'') and a part fully correlated with $S$ (``as''),
and assume power-law forms for the power spectra:
 \beq
 	\Pow_\R(k) = \Pow_{\mr{ar}}(k) + \Pow_{\mr{as}}(k) \,,
 \eeq
where
 \bea
 	\Pow_{\mr{ar}}(k) & = & A_{r0}^2\left(\frac{k}{k_0}\right)^{\nadI-1} \,,\nn\\
	\Pow_{\mr{as}}(k) & = & A_{s0}^2\left(\frac{k}{k_0}\right)^{\nadII-1} \,.
 \eea
Since the spectral indices $\nadI$ and $\nadII$ are assumed constant (i.e. do not depend on $k$),
the effective single adiabatic spectral index  $n_{\mr{ad}}^{\mr{eff}}(k) \equiv
 \frac{d\ln{\cal P}_{\cal R}(k)}{d\ln k} + 1$ will depend on the scale (i.e., has running), in particular,
if  $\nadI$ and $\nadII$ differ a lot \cite{Valiviita:2003ty}. 

For the isocurvature and correlation we have
 \bea
 	\Pow_S(k) & = & B_0^2\left(\frac{k}{k_0}\right)^{\niso-1} \,\\
 	\Cov_{\R S}(k) & = & \Cov_{S \R}(k) = A_{s0}B_0\left(\frac{k}{k_0}\right)^{\ncor-1} \,,
 \eea
where
 \beq
 \ncor = \frac{\nadII+\niso}{2}\,.
 \eeq

We denote the values of power spectra at scale $k_i$ by $A_{ri}^2 \equiv \Pow_{\mr{ar}}(k_i)$, $A_{si}^2 \equiv \Pow_{\mr{as}}(k_i)$,
and $B_i^2 \equiv \Pow_S(k_i)$. In the following we choose three reference (i.e., pivot) scales
 \bea
 	k_1 & = & 0.002\, \mbox{Mpc}^{-1} \nn\\ 
	k_0 & = & 0.010\, \mbox{Mpc}^{-1} \nn\\ 
	k_2 & = & 0.050\, \mbox{Mpc}^{-1} \,.
 \eea
 
We further define the total primordial perturbation power
 \beq
 	A_i^2 \equiv A_{ri}^2+A_{si}^2+B_i^2\,,
 \eeq
 the primordial isocurvature fraction
\beq
      \alpha_i \equiv \frac{B_i^2}{A_i^2}\,,
\eeq
 and the ratio of the correlated adiabatic component to the total adiabatic power
 \beq
 	\gamma_i \equiv \sgn(A_{si}B_i)\frac{A_{si}^2}{A_{ri}^2+A_{si}^2} \,,
 \label{eq:a2frac}
 \eeq
so that
 \bea
	A_{ri}^2 & = & (1-|\gamma_i|)(A_{ri}^2+A_{si}^2) \ = \ (1-|\gamma_i|)(1-\alpha_i)A_i^2 \nn\\
 	A_{si}^2 & = & |\gamma_i|(A_{ri}^2+A_{si}^2) \ = \ |\gamma_i|(1-\alpha_i)A_i^2 \nn\\
	B_i^2 & = & \alpha_i A_i^2 \nn\\
        A_{si}B_i & = &  \alpha_{\mr{cor}i} A_i^2 =  {\mtc C}_{\mtc{R}S}(k_i) = {\mtc C}_{S\mtc{R}}(k_i)\,.
 \label{eq:frac2a}
 \eea
On the last line  we defined the relative amplitude of the primordial correlation between the adiabatic and isocurvature perturbations,
$\alpha_{\mr{cor}i} \equiv  \mathrm{sign} (\gamma_i) \sqrt {\alpha_i (1- \alpha_i) |\gamma_i|}$.

The total CMB temperature angular power spectrum can be written as
\begin{align}
  \label{eq:totCl}
  C_{\ell} &= A_0^{2} \bigl[ (1-\alpha_0)(1-\abs{\gamma_0})\hat{C}^{\mr{ar}}_{\ell} +
  (1-\alpha_0)\abs{\gamma_0}\hat{C}^{\mr{as}}_{\ell} \nonumber \\
  &\quad + \alpha_0 \hat{C}^{\mr{iso}}_{\ell} +
\alpha_{\mr{cor}0} \hat{C}^{\mr{cor}}_{\ell} 
+ (1-\alpha_0)r_0\hat{C}^{T}_{\ell} \bigr] \nonumber \\
  & \equiv
  C^{\mr{ar}}_{\ell} + C^{\mr{as}}_{\ell} + C^{\mr{iso}}_{\ell} + C^{\mr{cor}}_{\ell} + C^{T}_{\ell}
  \,,
\end{align}
where the $\hat{C}_{\ell}$ represent the different contributions to the 
angular power spectrum that would result from a corresponding primordial spectrum with unit amplitude at the pivot scale $k=k_0$ (see \cite{KurkiSuonio:2004mn}). $C^{T}_{\ell}$ comes from the primordial tensor perturbations.

The total non-adiabatic contribution to the CMB temperature variance,
\bea
\alpha_T &\equiv& \frac {\langle ( \delta T^{\mathrm{non-ad}})^2 \rangle} {\langle ( \delta T^{\mathrm{total\ from\ scalar\ perturbations}})^2 \rangle} \nonumber \\
&=& \frac {\sum_{\ell=2}^{2100} (2\ell + 1) (C_\ell^{\mathrm{iso}} + C_\ell^{\mathrm{cor}})     } {\sum_{\ell=2}^{2100} (2\ell + 1) ( C^{\mr{ar}}_{\ell} + C^{\mr{as}}_{\ell} + C^{\mr{iso}}_{\ell} + C^{\mr{cor}}_{\ell})}\,,
\label{eq:alphaT}
\eea
is our \emph{pivot-scale free measure of the non-adiabaticity}.

Our \emph{sign convention} for $\R$ and $S$ is such that for NDI and NVI a positive primordial correlation leads to a positive contribution to the final $C_\ell$ spectrum, i.e., a positive primordial $\gamma$ (or $\Cov_{\R S}(k) > 0$) gives $C^{\mr{cor}}_{\ell} >0$. In the case of CDI this is true in the Sachs-Wolfe region (at low multipoles), but at higher multipoles $C^{\mr{cor}}_{\ell}$ keeps changing its sign as a function of $\ell$, although the primordial correlation does not change its sign as a function of $k$ in our model.

\subsection{Phenomenological parametrization \label{sec:phenpar}}

The above model has six independent scalar perturbation parameters. In the \emph{amplitude parametrization} we
assign uniform priors to the following primary parameters:
\begin{gather}
 	\ln (10^{10}A_1^2),\, \ln (10^{10}A_2^2) \in (1,7), \quad \alpha_1,\, \alpha_2 \in (0,1),\nonumber\\
       \gamma_1 \in (-1,1), \quad\mbox{and}\quad |\gamma_2| \in (0,1) \,.
\label{eq:ampparams}
\end{gather}
As we assume power law spectra, the sign of correlation can not be a function of $k$. Therefore the sign of $\gamma_2$ has to be the same
as that of $\gamma_1$. The background is described by the usual $4$ $\Lambda$CDM background parameters, i.e.,
the physical baryon density, the physical CDM density, the sound horizon angle at last scattering $\theta$, and the optical depth:
\begin{gather}
\omega_b \equiv \Omega_b h^2 \in (0.01,0.05),\quad
\omega_c \equiv \Omega_c h^2 \in (0.02,0.30),\nonumber\\
100\theta \in (0.5,2.2),\quad
\tau \in (0.02,0.30)\,.\label{eq:bgparams}
\end{gather}

In addition, we include in the analysis the primordial tensor perturbations with a power law power spectrum
\beq
 	\Pow_T(k) \ = \ \Pow_T(k_0)\left(\frac{k}{k_0}\right)^{n_T} \,.
\eeq
Their amplitude has been traditionally parametrized by the tensor-to-scalar ratio
$r(k) \equiv {\Pow_T(k)}/{\Pow_\R(k)}$.
In principle, in the phenomenological treatment the tensor perturbations would add two extra parameters, $r_0$ and the tensor spectral index  $n_T$.
However, as no tensor nor isocurvature perturbations have been detected so far, allowing $n_T$ to be a free parameter would give us too many poorly constraint parameters to make this study feasible.
As the focus of this paper are the isocurvature perturbations,
we assume even in the phenomenological approach the first inflationary consistency relation which fixes $n_T$,
see, e.g., \cite{Lidsey:1995np,Bartolo:2001rt,Wands:2002bn,Cortes:2006ap,Kawasaki:2007mb}.
Therefore the tensor perturbations add only one extra free parameter while the tensor spectral index is a derived parameter
given by the consistency relation $n_{T0}  = -{r_0}/{[8(1-|\gamma_0|)]}$. We impose also the second consistency relation \cite{Cortes:2007ak,Cortes:2006ap}, which gives the running $q_{T0} \equiv d n_T / d \ln k|_{k=k_0} = n_{T0}[n_{T0} - (\nadI -1)]$.

Assigning a uniform prior for $r_0$ and $\gamma_{1,2}$ and using the
first inflationary consistency relation, as described above, would lead to an unphysical prior on $n_T$.
Namely,
the tensor spectral index would receive huge negative values whenever $|\gamma_0|$ was near to one. 
This is against the very motivation of using the consistency relation, which was to force the tensor spectrum to follow the typical inflationary prediction of near scale invariance, $|n_T| \ll 1$. In addition, the large scale (low multipole) CMB data will not allow for such a huge tensor contribution. Hence, the use of the first consistency relation would artificially exclude any models where $|\gamma|$ was near to one. (We will demonstrate this later in the end of Sec.~\ref{sec:general_cmb_ampl_par}.) In order to avoid these problems, 
we will not parametrize the tensor power by $r_0$, but instead by   
\beq 
     \tilde{r}_0 \equiv \frac{ {\Pow_T(k_0)} }{\Pow_{\mr{ar}}(k_0) } = \frac{r_0}{1-|\gamma_0|}
\eeq
for which we assign a uniform prior between $0$ and $1.35$. 
When connecting the model to inflation, $\tilde{r}$ is the ratio of tensor to curvature perturbations
generated at horizon exit, to the leading order in slow-roll parameters. So it reflects directly the inflationary physics, and hence a uniform prior on it is physically motivated. On the other hand, $r$ is a parameter directly related to observables.  With our new definition the first consistency relation reads
\beq
n_{T0}  = -\frac{\tilde{r}_0}{8}\,,
\eeq 
which leads to a uniform prior on the derived parameter $n_{T0}$ between $-0.17$ and $0$ without being affected by the correlation parameter $\gamma$. 

Furthermore, our new $\tilde r$ parametrization has an advantage when studying the fully (anti)correlated models, $\gamma=\pm1$. In these special cases there is no tensor contribution. So, we can turn off tensors in CAMB and set the tensor parameters to constant values, $r=0$ and $n_T=0$. However, in the ``old parametrization'' we cannot recover these special cases from the general case by taking the limit $|\gamma|\rightarrow 1$ and $r_0\rightarrow 0$, since this would correspond to $n_T = 0 / 0$. In the new parametrization the $|\gamma|\rightarrow 1$ tensorless limit gives naturally $r_0 = (1-|\gamma_0|)\tilde r_0 = 0\times0 = 0$ and $n_T = 0$. So, although we do the special cases as separate MultiNest runs, we can also see the behavior of the posterior in the limit  $|\gamma|\rightarrow 1$ from our generally correlated runs.

\subsection{Inflationary slow-roll parametrization \label{sec:SlowrollParametrization}}

In the inflationary slow-roll approach we assume that during inflation there exists at least two ``active'' fields.
The field space can be locally rotated so that the perturbations can be described by an adiabatic component, which is a perturbation in the direction of the background trajectory $\sigma$, and an ``isocurvature'' component which is a perturbation in the perpendicular direction $s$. Now we can define four slow-roll parameters that are calculated from the inflationary potential $V(\sigma, s)$ at the time the interesting scale exits the horizon as follows: 
\begin{gather}
\etzz=\textstyle \frac{1}{8\pi G}\frac{\partial_\sigma\partial_\sigma V}{V},  \quad
\etzs=\textstyle \frac{1}{8\pi G}\frac{\partial_\sigma\partial_s V}{V}, \quad
\etss=\textstyle \frac{1}{8\pi G}\frac{\partial_s\partial_s V}{V},\nn\\
\veps=\textstyle \frac{1}{16\pi G}(\frac{\partial_\sigma V}{V})^2.
\end{gather}

From the slow-roll parameters we can determine  the spectral indices and the tensor-to-scalar ratio at horizon exit, $\tilde{r}$:
\bea
 	\nadI & = & 1 -6\veps + 2\etzz \nn\\
	\nadII & = & 1 -2\veps + 2\etss - 4\etzs\tan\Delta \nn\\
	\niso & = & 1 -2\veps +2\etss \nn\\
	\tilde{r} & = & 16\veps\ \nn\\
	n_T & = & -2\veps \,,
 \label{eq:BWresults}
 \eea
where the primordial correlation angle $\Delta$ is defined by
 \beq
 	\cos\Delta \equiv 
\frac{{\mtc{C}_{\mtc{R}S}}}{\Pow_{\mtc{R}}^{1/2} \Pow_{S}^{1/2}}
= \sgn(\gamma) \sqrt{|\gamma|}\,,
 \eeq
with $0 \leq \Delta \leq \pi$.
The relations (\ref{eq:BWresults}) are valid to first order in slow-roll parameters, see \cite{Byrnes:2006fr,Bartolo:2001rt,Wands:2002bn,vanTent:2003mn,Peterson:2010np}
 and note that, e.g., in Byrnes \& Wands \cite{Byrnes:2006fr} $n=0$ stands for a
 scale-invariant spectrum whereas we have added the conventional 1 (except for tensors) and we use
 a different sign convention for $\cos\Delta$ and thus also for $\tan\Delta  =  \sgn(\gamma){\sqrt{1-|\gamma|}}/{\sqrt{|\gamma|}}$.

In the the slow-roll approach we have the same background parameters as in the phenomenological approach,  Eq. (\ref{eq:bgparams}),
but the perturbations are parametrized by the four slow-roll parameters (three $\eta_{ij}$ whose prior is uniform from $-0.075$ to $+0.075$ and $\veps$ with the prior range from $0$ to $0.075$), $\gamma_0 \in (-1,1)$ and $\ln (10^{10} A_0^2) \in (1,7)$.  Note that unlike in the fully numerical treatments, see e.g. \cite{Norena:2012rs}, we can only allow small magnitudes for the slow-roll parameters, so that Eq.~(\ref{eq:BWresults}) is accurate enough. Our choice of prior ranges should guarantee that the second order corrections to $n_{\mtr{ar,iso}}-1$ are less than $\mtc{O}(10\%)$.

\section{Mechanisms that may produce neutrino isocurvature \label{sec:mechanisms}}

There are various mechanisms that can produce (correlated) isocurvature and adiabatic perturbations, see, e.g., Refs.
\cite{Choi:2008et,Hamazaki:2007eq,DiMarco:2007pb,Lalak:2007vi,Choi:2007su,Byrnes:2006fr,Rigopoulos:2005us,Bassett:2005xm,Hattori:2005ac,DiMarco:2005nq,Parkinson:2004yx,Gruzinov:2004ad,Bartolo:2003ad,Vernizzi:2003vs,Lee:2003ed,vanTent:2003mn,Mazumdar:2003iy,Malik:2002jb,DiMarco:2002eb,Ashcroft:2002vj,Bernardeau:2002jy,Wands:2002bn,Tsujikawa:2002nf,Starobinsky:2001xq,Bartolo:2001rt,GrootNibbelink:2001qt,Bartolo:2001vw,Hwang:2001fb,Hwang:2000jh,Gordon:2000hv,Taylor:2000ze,Finelli:2000ya,Liddle:1999pr,Bassett:1999ta,Pierpaoli:1999zj,Langlois:1999dw,Felder:1999pv,Perrotta:1998vf,Chiba:1997ij,Nakamura:1996da,Polarski:1994rz}. 
The common ingredient of these models is that at least one extra degree of freedom is needed in addition to the one degree of freedom provided by the single-field slow-roll inflation, which can give rise to only the adiabatic mode. Multi-field inflationary models and curvaton/spectator field models are natural candidates for generating primordial CDI or NDI perturbations.  In \cite{Valiviita:2012ub} we reviewed several such scenarios focusing on the CDI case. Here we will provide examples of stimulating the NDI mode.

It has been suggested \cite{Lyth:2002my} (see also \cite{Gordon:2003hw, Kawasaki:2011rc, DiValentino:2011sv}) that NDI might be generated from inhomogeneous lepton asymmetry in the context of the curvaton scenario. In the curvaton
scenario the light curvaton field $\chi$ remains subdominant during inflation but may become important once the inflaton has decayed into radiation. This happens because once the Hubble parameter becomes smaller than the mass of the curvaton, the curvaton starts oscillating and behaves like dust and thus loses energy slower than the radiation fluid. This causes the perturbations in the curvaton to be transferred to the curvature perturbation,
\begin{equation}
\mathcal R \simeq \frac{2R}{3}\frac{\delta\chi_*}{\chi_*},
\end{equation}
where $R \equiv \left(\frac{3\rho_{\chi}}{4\rho_r + 3\rho_{\chi}}\right)_{\text{dec}}$ is evaluated at the time of curvaton decay. If the lepton number is generated before
curvaton decay the NDI perturbation is
\begin{equation}
S_{\nu r} \simeq -\lambda \mathcal R, \quad \text{with} \quad \lambda = \frac{135}{7}(1-f_{\nu})\left(\frac{\xi}{\pi}\right)^2,
\end{equation}
where $\xi \equiv \mu/T$ is the neutrino asymmetry parameter.\footnote{%
The non-zero chemical potential $\mu$ of neutrinos affects the effective number of neutrino species by
\beq 
\textstyle
N_\nu^{\mathrm{eff}} \rightarrow \tilde N_\nu^{\mathrm{eff}} \simeq \left[1 + \frac{30}{7}\left(\frac{\xi}{\pi}\right)^2 + \frac{15}{7}\left(\frac{\xi}{\pi}\right)^4 \right]N_\nu^{\mathrm{eff}}\,.
\eeq
 Assuming the standard $N_\nu^{\mathrm{eff}} = 3.046$ and the Big Bang nucleosynthesis (BBN) constraint $|\xi| < 0.07$, we find that the corrected number would be $\tilde N_\nu^{\mathrm{eff}} = 3.052$. Even as large $|\xi|$ as 0.2, would lead to quite a small correction, $\tilde N_\nu^{\mathrm{eff}} = 3.099$. Anyway, we checked that our NDI $\gamma=-1$ runs (see Sec.~\ref{sec:fullycorrelated}) led to virtually identical results with $N_\nu^{\mathrm{eff}} = 3.046$ and $3.100$. So, we can safely perform the analysis with the standard $N_\nu^{\mathrm{eff}}=3.046$.}
Perturbations are fully anticorrelated 
and the isocurvature fraction is
\begin{equation}
\alpha = \frac{\lambda^2}{1+\lambda^2}.
\end{equation}
BBN constrains $|\xi| < 0.07$ \cite{Orito:2002hf}, which implies $\alpha\lesssim10^{-4}$, too small to be observed.

If the lepton number is created directly from curvaton decay, however, the neutrino isocurvature perturbation is \cite{Lyth:2002my}
\begin{equation}
S_{\nu r} \simeq \lambda \left(\frac{1-R}{R}\right) \mathcal R,
\end{equation}
and the perturbations are fully correlated with isocurvature fraction
\begin{equation}
\alpha = \frac{\lambda^2(1-R)^2}{R^2+\lambda^2(1-R)^2}.
\label{eq:alphalambda}
\end{equation}
Now the isocurvature fraction can be significant if the curvaton decays sufficiently early. However, non-Gaussianity in the curvaton scenario is
\begin{equation}
f_{\mathrm{NL}}^{\mathrm{local}} = \frac{5}{4R} - \frac{5}{3} - \frac{5R}{6},
\label{eq:fnl}
\end{equation}
which implies $R>0.078$ from the latest Planck 2$\sigma$ constraint $f_{\mathrm{NL}}^{\mathrm{local}} < 14.3$ \cite{Ade:2013ydc}. 
Eq.~(\ref{eq:alphalambda}) gives largest $\alpha$ when $\lambda^2$ is the largest possible. Hence, saturating $\lambda$ with the BBN constraint,
the above result for $R$ leads to $\alpha<0.0045$.

The NVI mode is more difficult to motivate theoretically because it would have to be generated after neutrino decoupling,
and there are no proposed theoretical models to date. Thus we study the NVI mode only in the phenomenological set-up.

\section{Data and sampling method \label{sec:data}}

We employ the CMB temperature and polarization anisotropy data: WMAP-9 data \cite{Bennett:2012zja}, the Arcminute Cosmology Bolometer Array Receiver (ACBAR) data \cite{Reichardt:2008ay}, and QUEST at DASI (QUaD) data \cite{Brown:2009uy} (QUEST stands for Q and U Extragalactic Survey Telescope, and DASI for Degree Angular Scale Interferometer). 
The additional CMB dataset are the same as in \cite{Valiviita:2012ub} to make the comparison clear and in order to see whether there are significant differences between WMAP-7 and WMAP-9.

We sample the parameter space using the MultiNest nested sampling package \cite{Feroz:2008xx,MultiNest},  see also
\cite{Skilling:2004,Mukherjee:2005wg,Shaw:2007jj,Feroz:2007kg}. 
It is easy to interface with CAMB/CosmoMC codes \cite{CAMB,COSMOMC} that we have modified to handle
arbitrarily correlated mixtures of adiabatic and isocurvature perturbations.

\section{Results for the generally correlated mixed adiabatic and isocurvature models \label{sec:general_cmb}}

In this section we let $\gamma_{1,2}$ in the amplitude parametrization or $\gamma_0$ and $\etzs$ in the slow-roll parametrization to be free parameters, thus allowing for a general scale-dependent correlation amplitude between the adiabatic and one isocurvature mode (either CDI, NDI or NVI). These models have four extra parameters compared to the ``standard'' adiabatic $\Lambda$CDM model.

\subsection{Phenomenological approach --- Amplitude parametrization for CDI, NDI and NVI \label{sec:general_cmb_ampl_par}}

\begin{figure}[t]
  \centering
  \includegraphics[width=\columnwidth]{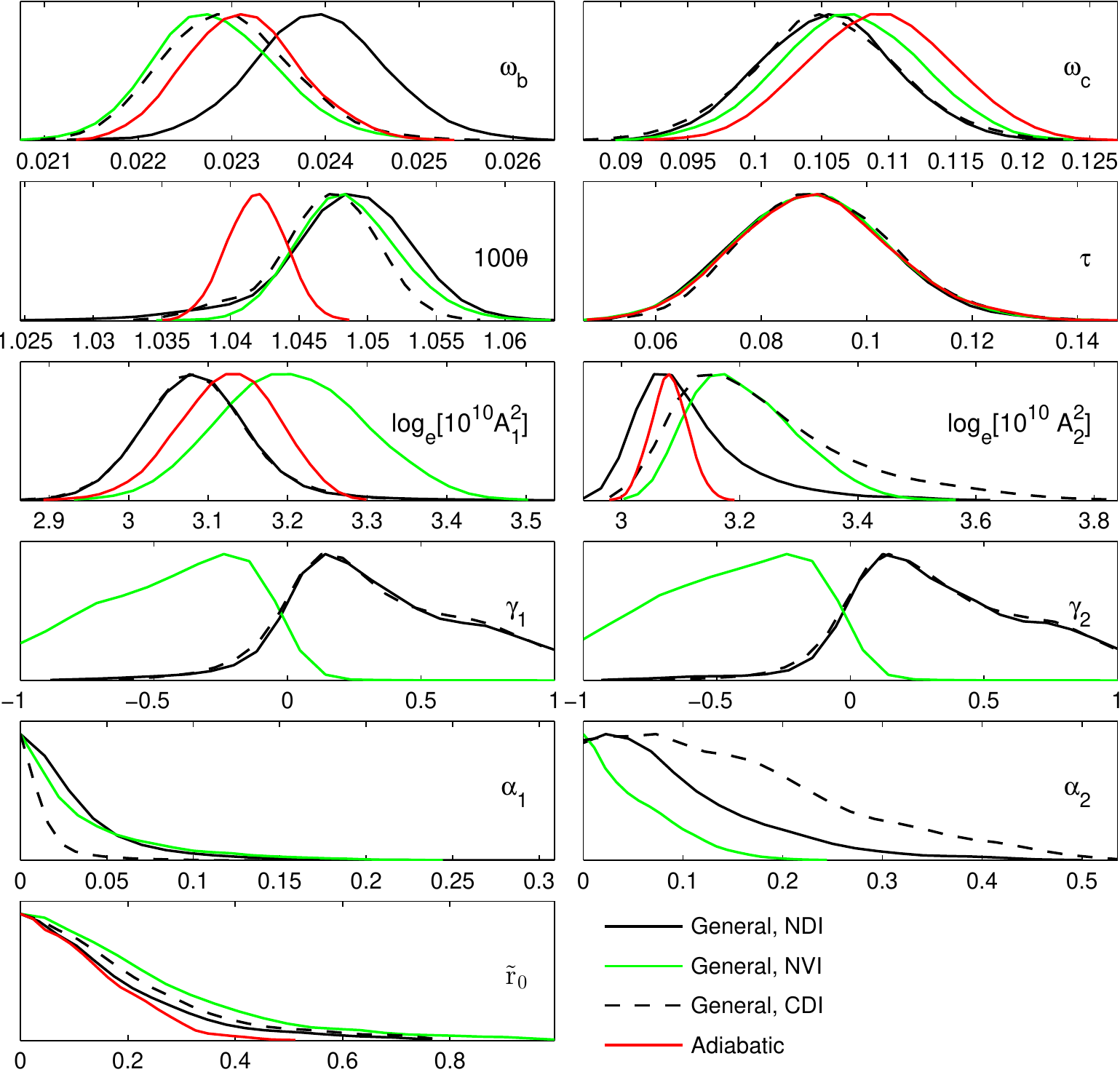}
  \caption{{\bf Amplitude parametrization, general correlation, primary parameters.} Marginalized 1-d posterior pdfs of the primary parameters of models with generally correlated mixture of primordial adiabatic and NDI (solid black), NVI (solid green), or CDI (dashed black) modes compared to the pure adiabatic model (solid red).
\label{fig:AmplPrimaryNeutDens}
}
\end{figure}

The marginalized 1-d posterior probability density functions (pdf) in the mixed adiabatic and isocurvature models (NDI, NVI, CDI) are compared to the pure adiabatic model in Fig.~\ref{fig:AmplPrimaryNeutDens}  (primary parameters) and Fig.~\ref{fig:AmplDerivedNeutDens} (selected derived parameters). In Appendix \ref{app:Tables1} we tabulate 68\% or 95\% confidence level (C.L.) intervals for selected parameters and the Bayesian evidence $-\log \mathcal{Z}$  (see the third column of tables \ref{tab:NDIamp}, \ref{tab:NVIamp} and \ref{tab:CDIamp} for the generally correlated mixed models).

\begin{figure*}[t]
\includegraphics[scale=0.255]{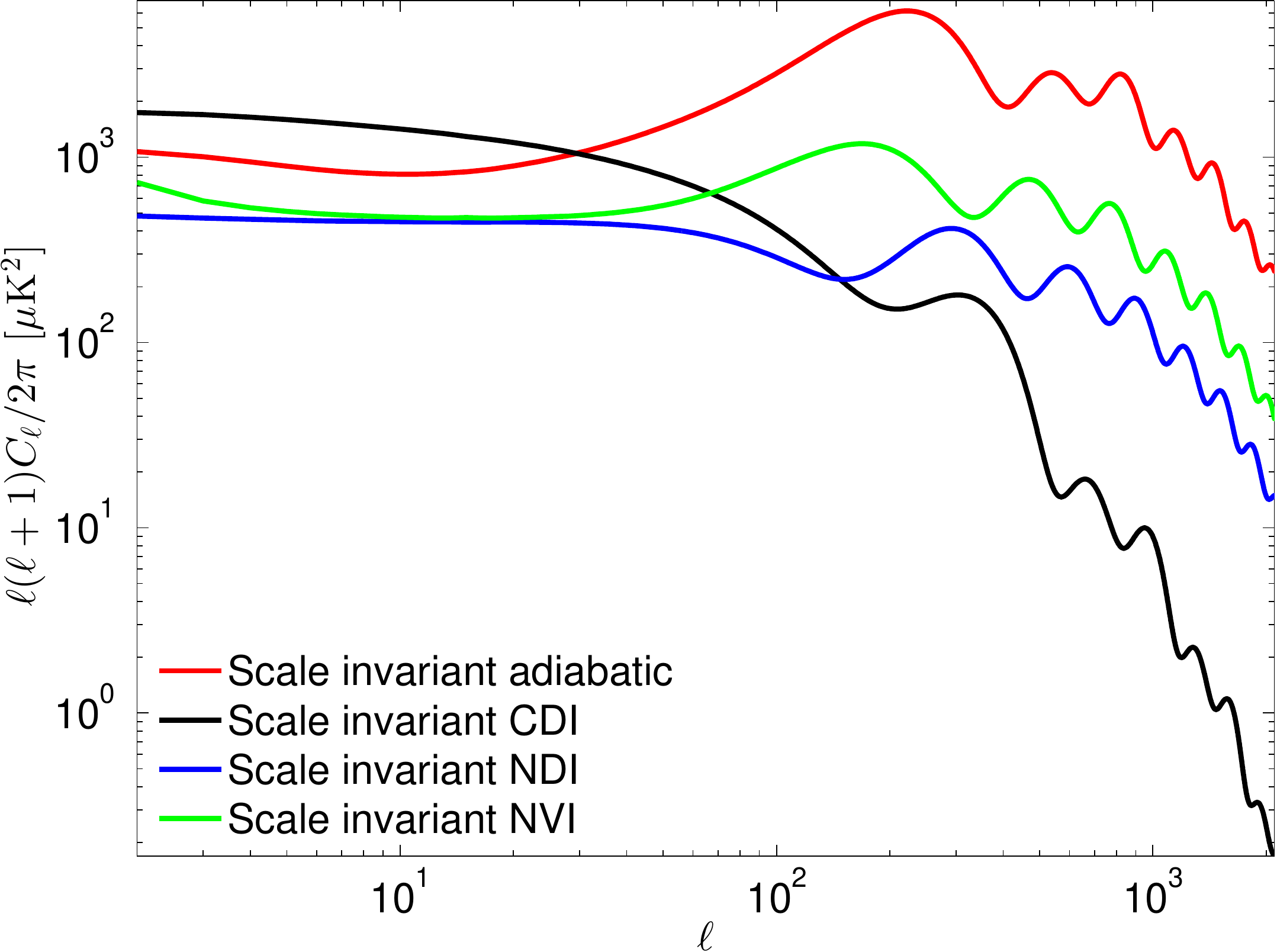}
\includegraphics[scale=0.255]{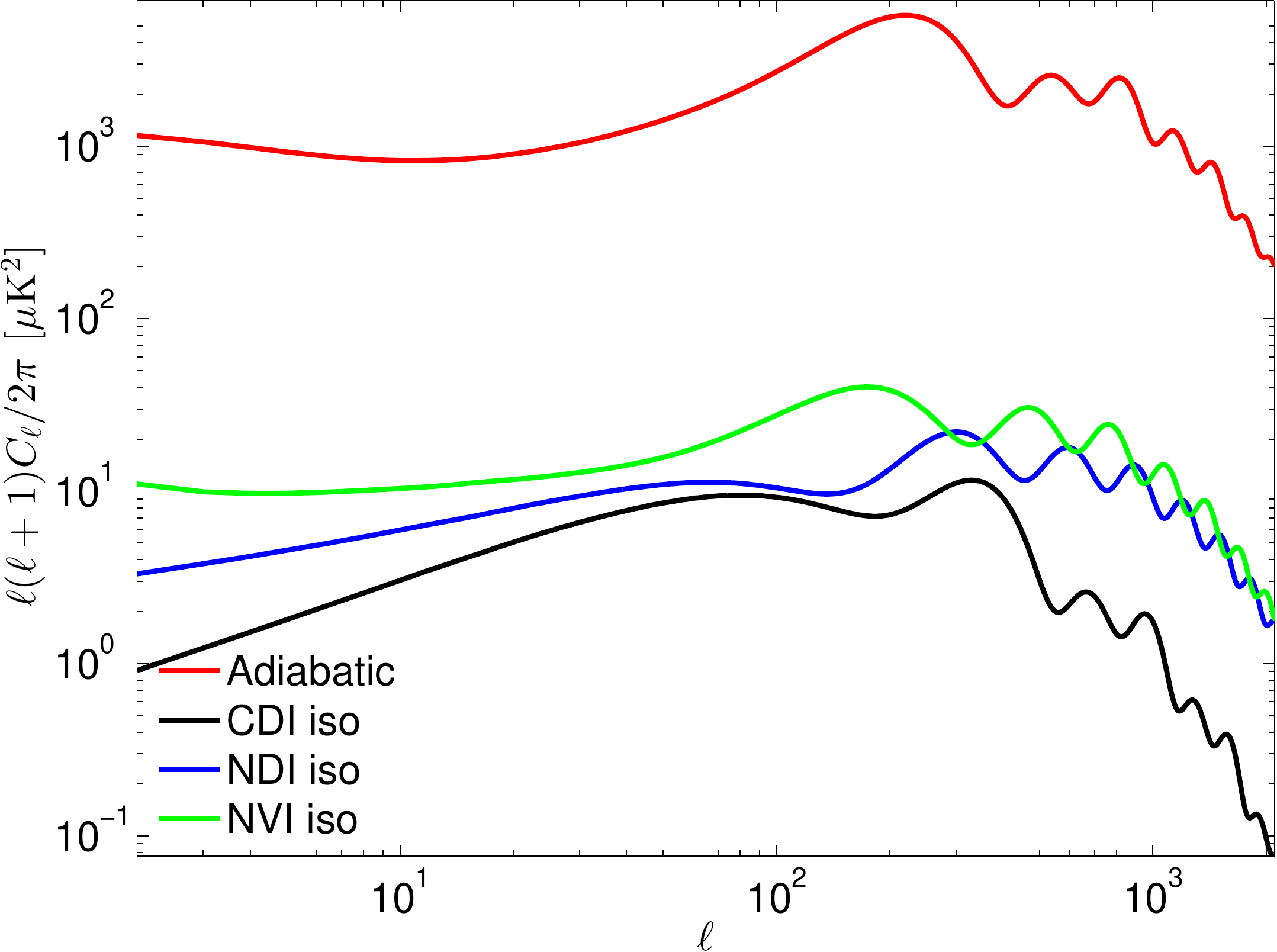}
\includegraphics[scale=0.255]{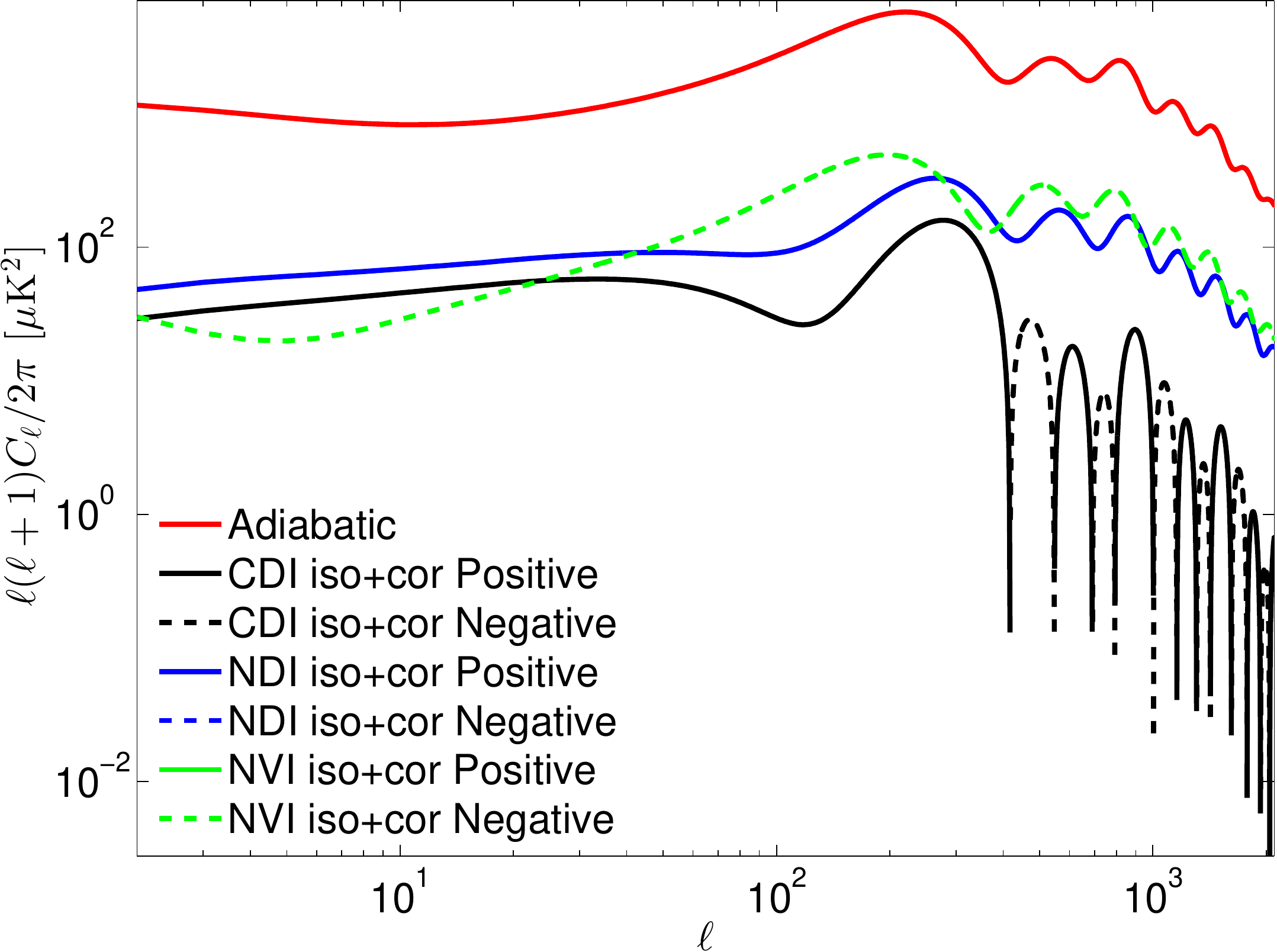}
   \caption{{\bf CMB temperature angular power spectra.} The left panel shows the angular power spectrum resulting from scale-invariant primordial \emph{pure} CDI (black), NDI (blue), or NVI (green) perturbations with $\mtc{P}_S = 2.4\times 10^{-9}$, and the \emph{pure} adiabatic spectrum (red) with $\mtc{P}_\mtc{R} = 2.4\times 10^{-9}$. The middle panel shows a typical well-fitting adiabatic model, and the isocurvature contributions in the typical well-fitting \emph{mixed models} (correlated adiabatic and CDI, NDI, or NVI primordial perturbations), found in the amplitude parametrization. The right panel is the same as the middle panel, except instead of the isocurvature contribution showing the total non-adiabatic contribution, i.e., the sum of isocurvature and correlation components.
\label{fig:Pure_Models}
}
\end{figure*}

For the NDI and CDI modes a positive correlation\footnote{%
Recall our sign convention presented after Eq.~(\ref{eq:alphaT}).
}
 with the curvature perturbations is preferred by the data, whereas in case of the NVI mode, a negative correlation is preferred. This can be explained by the temperature angular power spectra of Fig.~\ref{fig:Pure_Models}. In the left panel we plot the angular power spectra resulting from pure isocurvature or pure adiabatic scale-invariant primordial perturbations with the same background parameters. The CDI and NDI modes produce an acoustic peak that is to the right of the first acoustic peak of the adiabatic case, whereas the NVI mode leads to a peak that is slightly to the left. For a long time it has been known that in the CDI case the WMAP data prefer minimizing the CDI contribution everywhere \cite{KurkiSuonio:2004mn,Keskitalo:2006qv,Valiviita:2009bp,Valiviita:2012ub}. Since there is $\ell^{-2}$ damping of the CDI mode compared to the adiabatic mode (see again the left panel of Fig.~\ref{fig:Pure_Models}), the overall minimization of the isocurvature contribution is achieved by a relatively large $\niso$. With WMAP-9 and other CMB data used in this paper, the median of the posterior pdf for CDI is $\niso = 2.05$ (Fig.~\ref{fig:AmplDerivedNeutDens} and Table  \ref{tab:CDIamp}).

In the middle (right) panel of Fig.~\ref{fig:Pure_Models} we show a typical well-fitting adiabatic model and the isocurvature (non-adiabatic, i.e., isocurvature $+$ correlation) contributions of the well-fitting mixed adiabatic and isocurvature models. For these plots we used the median values of the 1-d pdf of each parameter. So, these $C_\ell$ curves are ``representative'' of the curves in the good-fit region. Compared to the correlation contribution, the actual isocurvature contribution is negligible. Therefore we focus on the right panel.  For the CDI mode, as well as for the NDI mode, the correlation component would push the first acoustic peak toward the right compared to the pure adiabatic case, which fits the data very well. To push the peak back toward the left (in order to fit the data) we need in these models a larger sound horizon angle, see the solid black (NDI) and dashed black (CDI) pdfs for $100\theta$ in Fig.~\ref{fig:AmplPrimaryNeutDens}. As explained, e.g., in \cite{KurkiSuonio:2004mn,Keskitalo:2006qv,Valiviita:2009bp,Valiviita:2012ub} this leads to a larger Hubble parameter $H_0$, a larger $\Omega_\Lambda$, and a smaller $\omega_c$ than in the pure adiabatic model, since we are studying models with a flat spatial geometry ($\Omega_{\rm tot} = 1$).

\begin{figure}[t]
  \centering
  \includegraphics[width=\columnwidth]{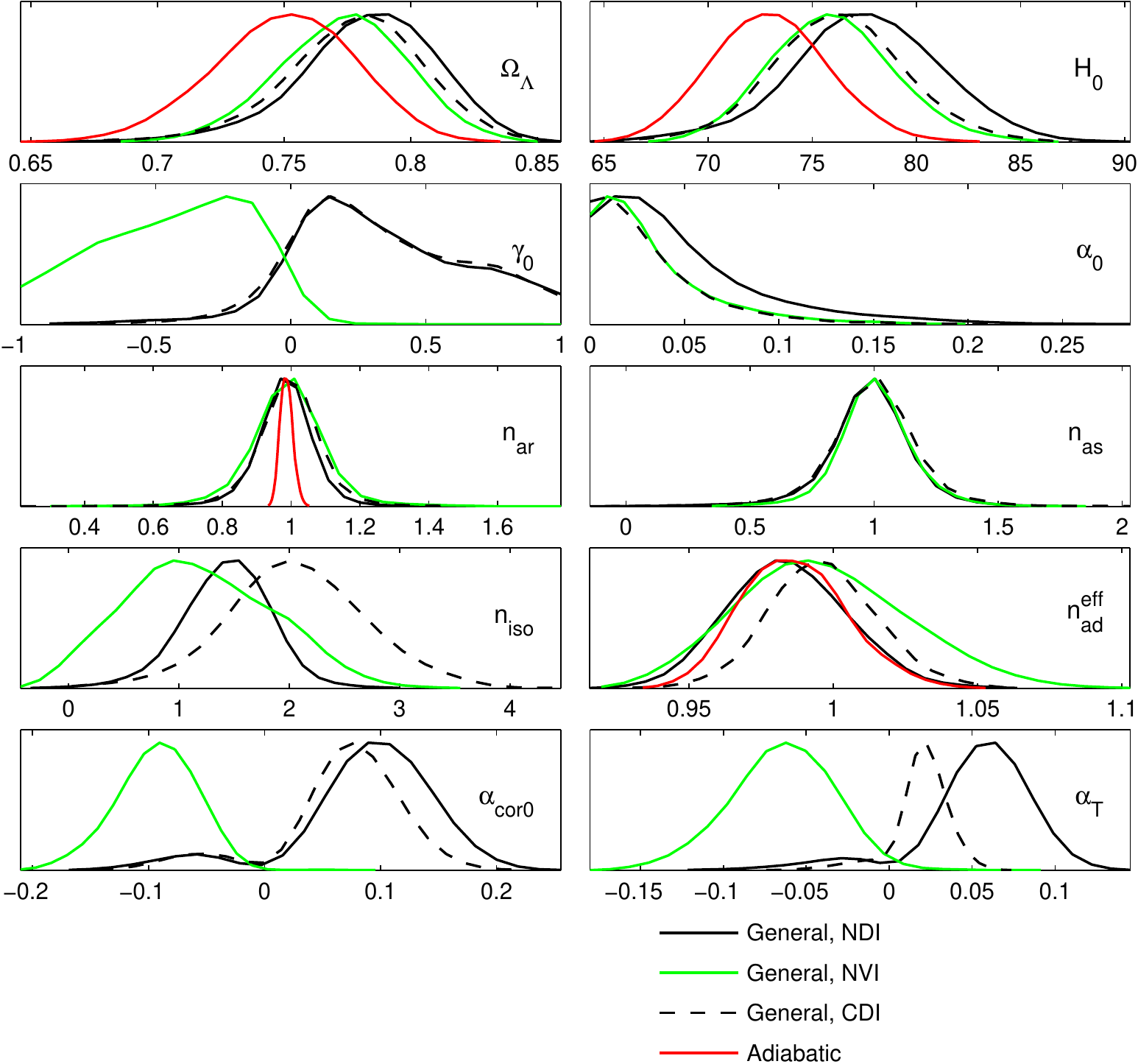}
  \caption{{\bf Amplitude parametrization, general correlation, derived parameters.} Marginalized 1-d posterior pdfs as in Fig.~\ref{fig:AmplPrimaryNeutDens}, but for selected derived parameters.
\label{fig:AmplDerivedNeutDens}
}
\end{figure}
\begin{figure}[t]
  \centering
 \includegraphics[width=\columnwidth]{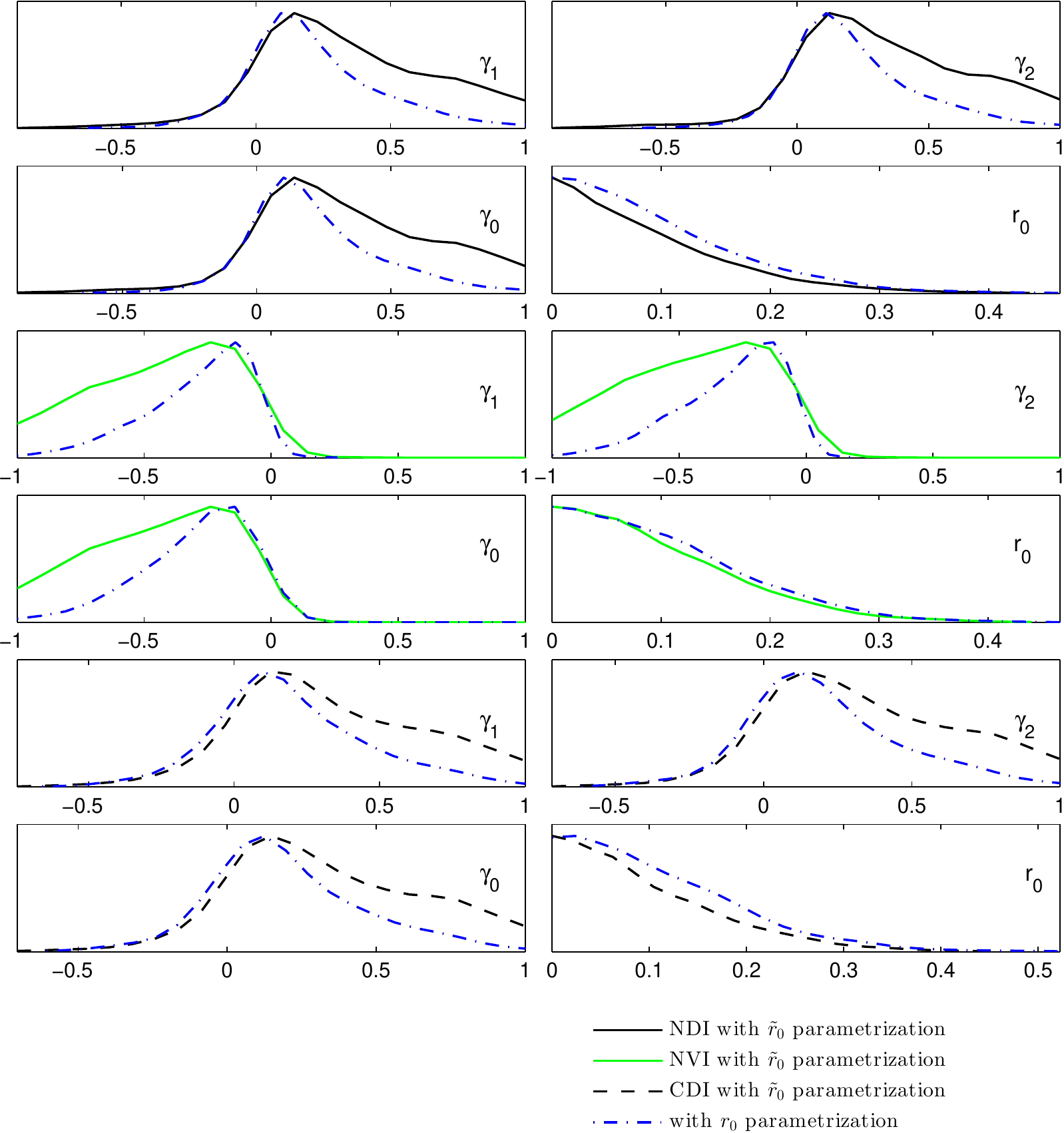}
    \caption{{\bf Amplitude parametrization, comparison of $\tilde r$ and $r$ tensor-to-scalar ratio parametrization.}  Marginalized 1-d posterior pdfs of $\gamma_1$, $\gamma_2$, $\gamma_0$, and the tensor-to-scalar ratio $r_0$ (the ratio of the tensor perturbation power to total curvature perturbation power at primordial time) in the mixed NDI (top four panels, solid black), NVI (middle four panels, solid green) and CDI (bottom four panels, dashed black) models with $\tilde r_0$  (the ratio of the tensor perturbation power to curvature perturbation power at horizon exit during inflation)  as a primary parameter, with a uniform prior on it.
The dot-dashed blue lines are with $r_0$ as a primary parameter.
\label{fig:rvstilder}
}
\end{figure}
\begin{figure}[t]
  \centering
  \includegraphics[width=\columnwidth]{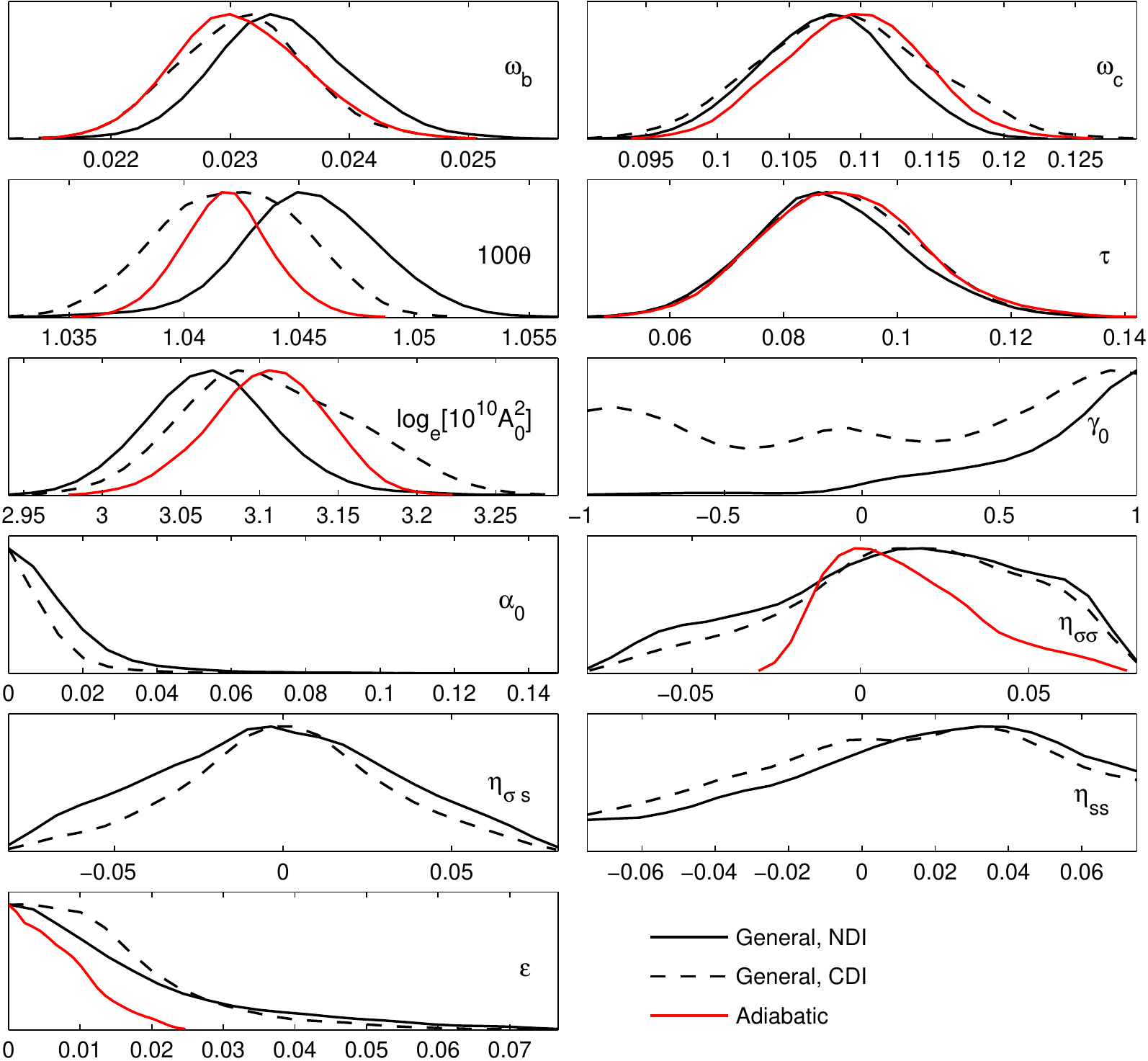}
\caption{{\bf Slow-roll parametrization, general correlation, primary parameters.} Marginalized 1-d posterior pdfs of the primary parameters of models with generally correlated mixture of primordial adiabatic and NDI (solid black) or CDI (dashed black) modes compared to the pure adiabatic model (solid red).
  \label{fig:InflPrimaryNeutDens}}
\end{figure}
\begin{figure}[t]
  \centering
  \includegraphics[width=\columnwidth]{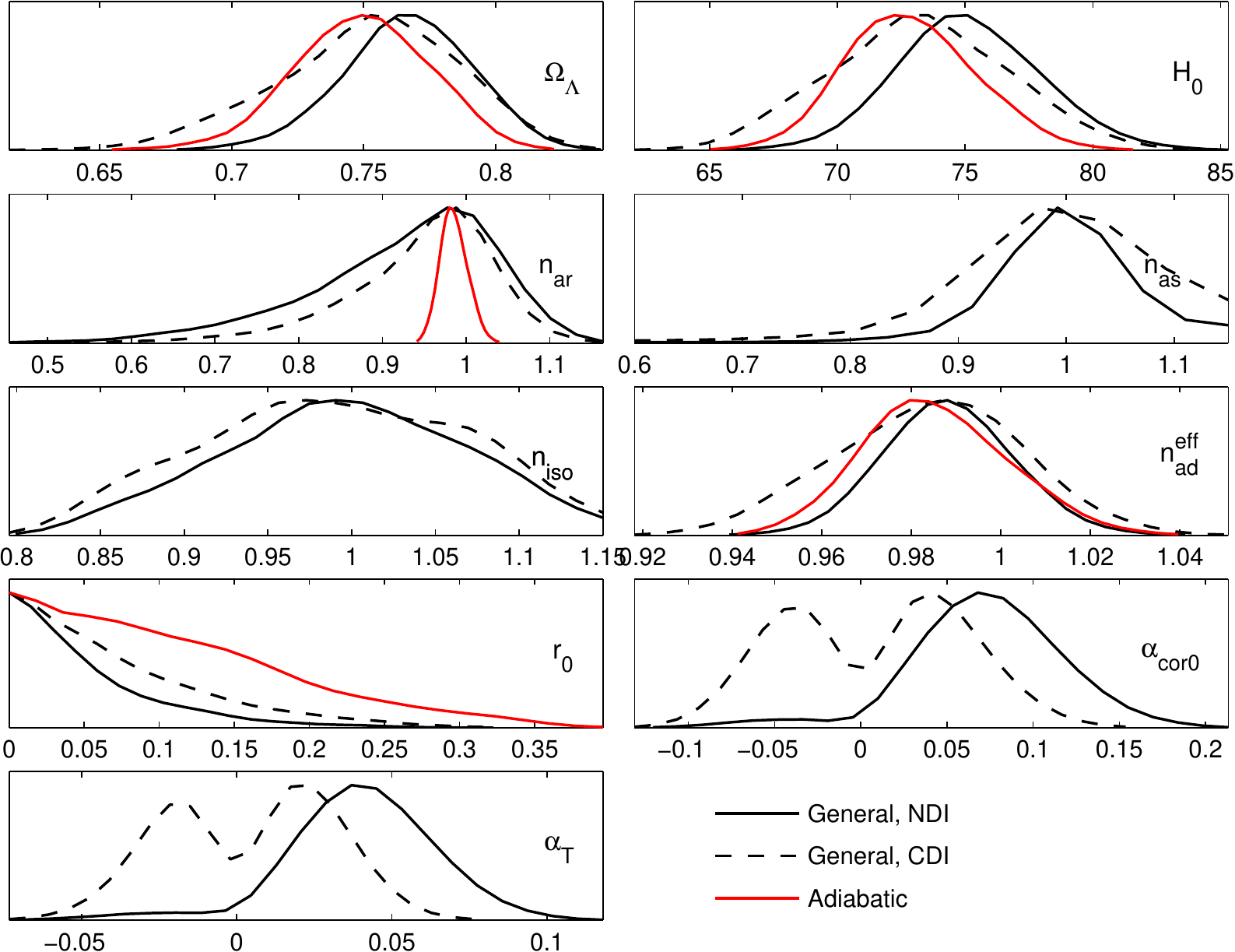}
  \caption{{\bf Slow-roll parametrization, general correlation, derived parameters.} Marginalized 1-d posterior pdfs as in Fig.~\ref{fig:InflPrimaryNeutDens}, but for selected derived parameters.
 \label{fig:InflDerivedNeutDens}
}
\end{figure}

The phase of the NVI mode is very different from the NDI and CDI. Positively correlated NVI would tend to add power to the left side of the first adiabatic acoustic peak and to reduce the relative power on the right side of the peak. In the mixed model this would move the first acoustic peak to the left compared to the pure adiabatic model. However, a negative NVI correlation works in the opposite way, and hence leads to a very similar effect as the positive correlation in the NDI and CDI cases. Then the pdfs of the primary background parameters $\omega_c$, $\theta$, and $\tau$ (Fig.~\ref{fig:AmplPrimaryNeutDens}), as well as the derived parameters $\Omega_\Lambda$ and $H_0$  (Fig.~\ref{fig:AmplDerivedNeutDens}) are very similar in all the mixed models, whereas the pdfs of $\gamma_{1,2,0}$ for the mixed NVI model are rough mirror images of those of the mixed NDI and CDI models. 

The constraints on the primordial isocurvature fraction are tightest for the CDI on large scales (see $\alpha_1$ in Fig.~\ref{fig:AmplPrimaryNeutDens}) and weakest on small scales (see $\alpha_2$). This is reflected in the derived parameter $\niso$ in Fig.~\ref{fig:AmplDerivedNeutDens}. Indeed, assuming that the data do not like any isocurvature contribution, the preferred order of the values of $\niso$ could have been guessed by seeing the left panel of Fig.~\ref{fig:Pure_Models}. Nearly scale-invariant spectrum is preferred in the NVI case, since this type of NVI mode leads to roughly constant fractional isocurvature contribution over the whole range of acoustic peaks. 

The primordial correlation amplitude $\alpha_{\mtr{cor}}$ describes best the primordial deviation from pure adiabaticity, unless the correlation parameter $\gamma$ is (nearly) zero. We find at $k_0 = 0.01\,$Mpc$^{-1}$ a constraint $-0.08 < \alpha_{\mtr{cor}0} < 0.18$ for NDI, $-0.16 < \alpha_{\mtr{cor}0} < -0.03$ for NVI, and $-0.08 < \alpha_{\mtr{cor}0} < 0.15$ for CDI at 95\% C.L.
For the primordial isocurvature fraction the corresponding numbers are $\alpha_0 <$ 0.14 (NDI), 0.10 (NVI), and 0.10 (CDI). The CMB data do not show any preference for the mixed models: all the posterior pdfs of the primordial isocurvature fraction $\alpha$ peak at zero or very near to zero. The improvement of the $\chi^2$ of the best-fitting models compared to the adiabatic model does not exceed the number of extra parameters introduced by the isocurvature modes.

From Fig.~\ref{fig:AmplDerivedNeutDens} we see that a larger nonadiabatic contribution $ \alpha_T$ to the CMB temperature variance is allowed by the data in the case of the neutrino isocurvature modes than for the CDI mode. This is because the $C_\ell$ contribution from the neutrino modes is not as much off-phase from the adiabatic contribution (and the data agrees well with this adiabatic placement of the acoustic peaks), see Fig.~\ref{fig:Pure_Models}. Another contributor to this result is that in the CDI case the correlation component $C_\ell^{\mtr{cor}}$ keeps changing its sign as a function of multipole, whereas in the neutrino isocurvature cases $C_\ell^{\mtr{cor}}$ has the same sign as the primordial correlation over the whole multipole range. Thus, in the CDI case, there are some cancellations in the summation over $\ell$ in Eq.~(\ref{eq:alphaT}). This may lead to a smaller non-adiabatic contribution to the total CMB temperature variance than to the individual multipoles in the CDI case. The 95\% C.L. constraints are
$-5\% < \alpha_T < 10\%$ (NDI), $-13\% < \alpha_T < -1\%$ (NVI), and $-3\% < \alpha_T < 5\%$ (CDI). The apparent missing of the adiabatic case from the 95\% C.L. interval of the NVI case is due to the very similar acoustic peak structures of the NVI and adiabatic modes. Thus the NVI mode is the most difficult to distinguish from the adiabatic one. However, as written above, the pdf of the primordial isocurvature fraction peaks at zero even in the NVI case.

Now we comment on the Bayesian evidences reported on the last lines of Tables \ref{tab:NDIamp}, \ref{tab:NVIamp} and \ref{tab:CDIamp}. The adiabatic model is favored. It has $-\ln \mtc{Z} \approx 3,901.17$, whereas the mixed models (with general correlation) all have $-\ln \mtc{Z} \approx 3,905$\ldots$3,906$. So the betting odds in favor of the pure adiabatic model against the mixed models are roughly $100 : 1$.

Finally, in Fig.~\ref{fig:rvstilder} we show the effect of our new $\tilde r$ parametrization on the posterior pdfs of $\gamma_{1,2,0}$ and $r_0$ (which in the $\tilde r$ parametrization is a derived parameter). The constraints of $r_0$ are not much affected, but as we expected, in $\tilde r$ parametrization the values of $\gamma$ are constrained by the data, not by the unphysical prior of the derived parameter $n_T$ (tensor spectral index). As a result our new constraints on $\gamma$ are weaker than those presented in \cite{Valiviita:2012ub} for the CDI case with WMAP-7 or those presented in \cite{Kawasaki:2007mb} for NDI, NVI, and CDI cases with WMAP-3 data. Naturally, the posterior pdfs of other parameters than $\gamma$ or $r$ are also affected to some extent. For example, we obtain slightly tighter constraints on the isocurvature fraction in the new parametrization (since larger correlation fractions are allowed and hence a fixed $\alpha$ leads to a larger non-adiabatic modification).

\subsection{Two-field inflation approach --- Slow-roll parametrization for NDI and CDI \label{sec:general_cmb_infl_par}}

Fig.~\ref{fig:InflPrimaryNeutDens} shows marginalized 1-d posterior pdfs of the primary parameters in the slow-roll parametrization, and Fig.~\ref{fig:InflDerivedNeutDens} the selected derived parameters. 
The medians of the pdfs and 68\% or 95\% C.L. intervals are provided in Appendix \ref{app:Tables1} in Tables \ref{tab:NDIslow} and \ref{tab:CDIslow}
for the mixed NDI and CDI models. (As discussed earlier, the NVI mode is hard to think of as resulting from inflationary physics, hence we do not include it in our slow-roll analysis.)

The most significant difference from the previous subsection is that the slow-roll parametrization forces the power spectra, in particular the isocurvature and correlation spectra, to be nearly scale-invariant,  see Eq.~(\ref{eq:BWresults}). In the CDI case the difference is most dramatic, since the phenomenological approach led to the median $\niso \sim 2.05$. For NDI the difference is smaller, since it gave the median $\niso \sim 1.45$. For NVI the slow-roll and phenomenological ($\niso\sim 1.15$) approaches would be almost identical. In the slow-roll parametrization the CDI and NDI modes can significantly modify only the low-$\ell$ part of the $C_\ell$ spectrum, see the left panel of Fig.~\ref{fig:Pure_Models}. We find tighter constraints than in the phenomenological approach: now $-0.04 < \alpha_{\mtr{cor}0} < 0.15$, $\alpha_0 <0.06$ (NDI), and $-0.08 < \alpha_{\mtr{cor}0} < 0.10$, $\alpha_0<0.03$ (CDI).

\begin{figure}[t]
  \centering
  \includegraphics[width=\columnwidth]{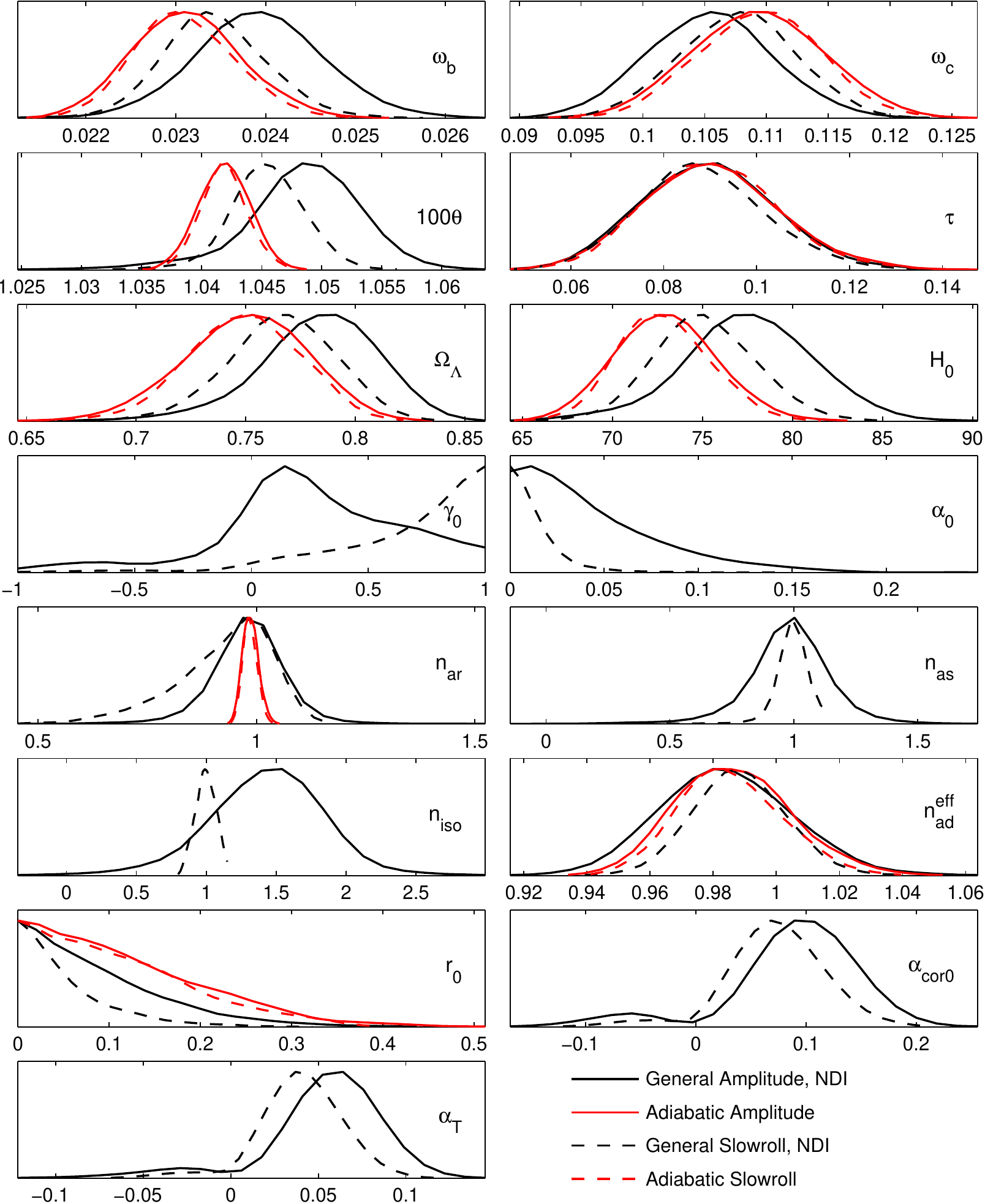}
  \caption{{\bf Comparison of amplitude and slow-roll parametrizations, general correlation.} Marginalized 1-d posterior pdfs from MultiNest runs made in the phenomenological amplitude parametrization (solid lines) and inflationary slow-roll parametrization (dashed lines) with generally correlated mixture of primordial adiabatic and NDI modes (black) and with the pure adiabatic mode (red). 
 \label{fig:AmplVsInfl}
}
\end{figure}

A comparison of the posterior pdfs in amplitude and slow-roll parametrization for the NDI case is shown in Fig.~\ref{fig:AmplVsInfl}. (A similar comparison for the CDI case can be found in \cite{Valiviita:2012ub} with WMAP-7 data, or with WMAP-9 by comparing Figs.~\ref{fig:AmplPrimaryNeutDens} and \ref{fig:AmplDerivedNeutDens} to \ref{fig:InflPrimaryNeutDens} and \ref{fig:InflDerivedNeutDens}.) In both the NDI and CDI cases, the preferred values of most of the parameters in slow roll parametrization are between preferred values of the pure adiabatic case and the amplitude parametrization. The reason is that the near scale-invariance of the primordial isocurvature spectrum prevents any significant non-adiabatic contribution to the acoustic peak structure, thus leaving the high-$\ell$ part of angular power spectrum virtually ``adiabatic''.

The data prefer positive correlation between the NDI and adiabatic mode, as happened also in the phenomenological case. The nearly scale-invariant NDI mode is able to modify the first acoustic peak almost in the same way as in the amplitude parametrization (with $\niso \sim 1.45$) if the correlation fraction is large enough, compare the left and right panels of Fig.~\ref{fig:Pure_Models}.  Thus, in the slow-roll parametrization much larger correlation fractions are favored; the pdf of $\gamma_0$ peaks at one --- at the full correlation. The situation is very different for the CDI mode, see $\gamma_0$ in Fig.~\ref{fig:InflPrimaryNeutDens}. 
While positive correlation was clearly preferred in the amplitude parametrization (due to the effects on the first acoustic peak), now any correlation fraction $\gamma_0$ between $-1$ and $+1$ is allowed. As the only effect of the correlated CDI in the slow-roll parametrization is to add or reduce some power at low-$\ell$, which is dominated by cosmic variance, the data are insensitive to the sign of correlation. Thus, for example, the parameters $\alpha_{\mtr{cor}0}$ and $\alpha_T$ just reflect the uncertainty caused by the cosmic variance, and their pdf is almost symmetric about zero. 

In particular in the NDI case, the tensor-to-scalar ratio, $r_0$, is constrained tighter in the slow-roll parametrization, since the positive correlation adds power at low-$\ell$. Thus there is less room for the tensor contribution which would also add power at low-$\ell$.

Of the four slow-roll parameters only $\veps$ is well constrained in all studied cases, while the three $\eta_{ij}$ are unconstrained or very poorly constrained, except $\etzz$ in the adiabatic case. The constraint on $\veps$ does not come only from the tensor contribution. From the first line of Eq.~(\ref{eq:BWresults}) it is obvious that if $\veps$ was near to the upper bound $(0.075)$ of our chosen prior range, this would lead to too red-tilted adiabatic spectrum, which cannot be compensated by the blue-tilted isocurvature spectrum at high-$\ell$, since the third line of Eq.~(\ref{eq:BWresults}) gives $0.70 < \niso < 1.15$.

In the slow-roll parametrization the adiabatic model has $-\ln \mtc{Z} \approx 3,898.9$, whereas the mixed models (with general correlation) have
$-\ln \mtc{Z} \approx 3,901.5$ (NDI) and $-\ln \mtc{Z} \approx 3,902.9$ (CDI). So the betting odds in favor of the adiabatic model are
13 : 1 against NDI, and 51 : 1 against CDI. In particular, the mixed NDI case is not overwhelmingly disfavored by the Bayesian model comparison when the slow-roll approach is adopted.

\section{Special cases \label{sec:specialcases}}

Now we study uncorrelated ($\gamma=0$) and maximally correlated ($\gamma=1$) or anticorrelated ($\gamma=-1$) cases. The uncorrelated model has only two extra parameters compared to the ``standard'' adiabatic $\Lambda$CDM model. In the maximally correlated cases, in the amplitude parametrization, we make an extra assumption that the adiabatic and isocurvature spectra have the same shape, which further reduces the number of parameters by one, leading to only one extra parameter. In the slow-roll parametrization the same shape of spectra follows directly from the second and third lines of Eq. (\ref{eq:BWresults}).

\subsection{Uncorrelated case for NDI, NVI, and CDI \label{sec:uncorrelated}}

In the uncorrelated case the $\Pow_{\mtr{as}}(k)$ spectrum is absent, so we have 
only nine independent parameters: the four background parameters
 \beq
 	\omega_b\,,\ \omega_c\,,\ \theta\,,\ \tau\,,
 \eeq
and five perturbation parameters
 \beq
 	\ln A_0^2\,,\ \nadI\,,\ \alpha_0\,, \niso\,,\ r_0\,,
 \eeq
where
 \bea 
 	\nadI & = & 1 - 6\veps+2\etzz \nn\\  
	\niso & = & 1 - 2\veps+2\etss \nn\\
	         r_0 & = & 16\veps\,.
\label{eq:nocor}
 \eea

The primary perturbation parameters in the amplitude parametrization are
 \beq
 	\ln A_1^2\,,\ \ln A_2^2\,,\ \alpha_1\,,\ \alpha_2\,,\ \tilde r_0 = r_0\,,
 \eeq
 and in the slow-roll parametrization
 \beq
    \ln A_0^2\,,\ \alpha_0\,,\ \etzz\,,\ \etss\,, \ \veps\,.
 \eeq
 
\begin{figure}[t]
  \centering
  \includegraphics[width=\columnwidth]{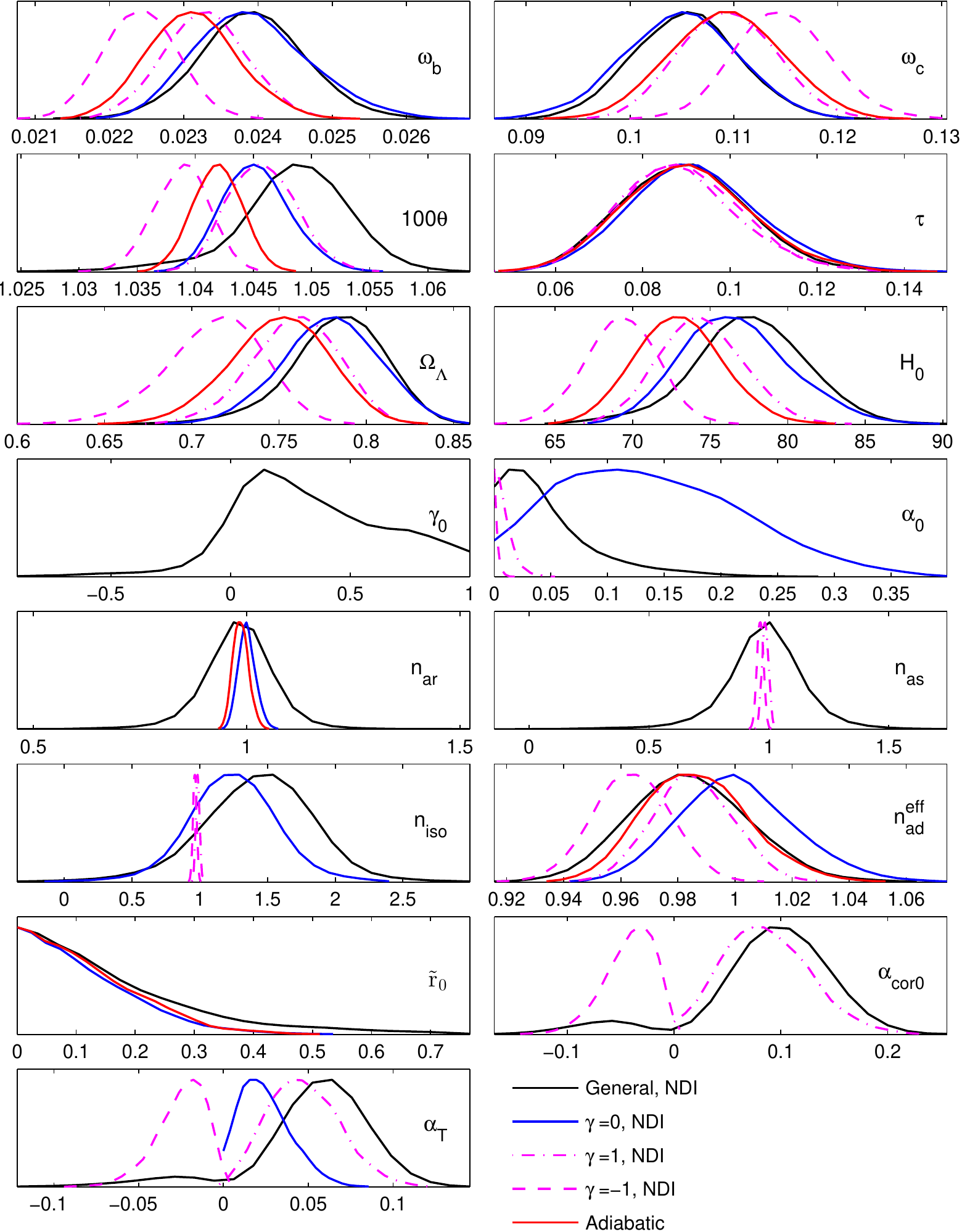}
 \caption{{\bf Amplitude parametrization, comparison of special and general NDI cases.} Marginalized 1-d pdfs for generally correlated mixture of primordial adiabatic and NDI (solid black) modes compared to the uncorrelated ($\gamma=0$, solid blue), maximally correlated ($\gamma=1$, dot-dashed magenta), and maximally anticorrelated ($\gamma=-1$, dashed magenta) cases, as well as, to the pure adiabatic model (solid red).
\label{fig:AmplSpecialCasesNeutDens}
}
\end{figure}

\begin{figure}[h]
  \centering
  \includegraphics[width=\columnwidth]{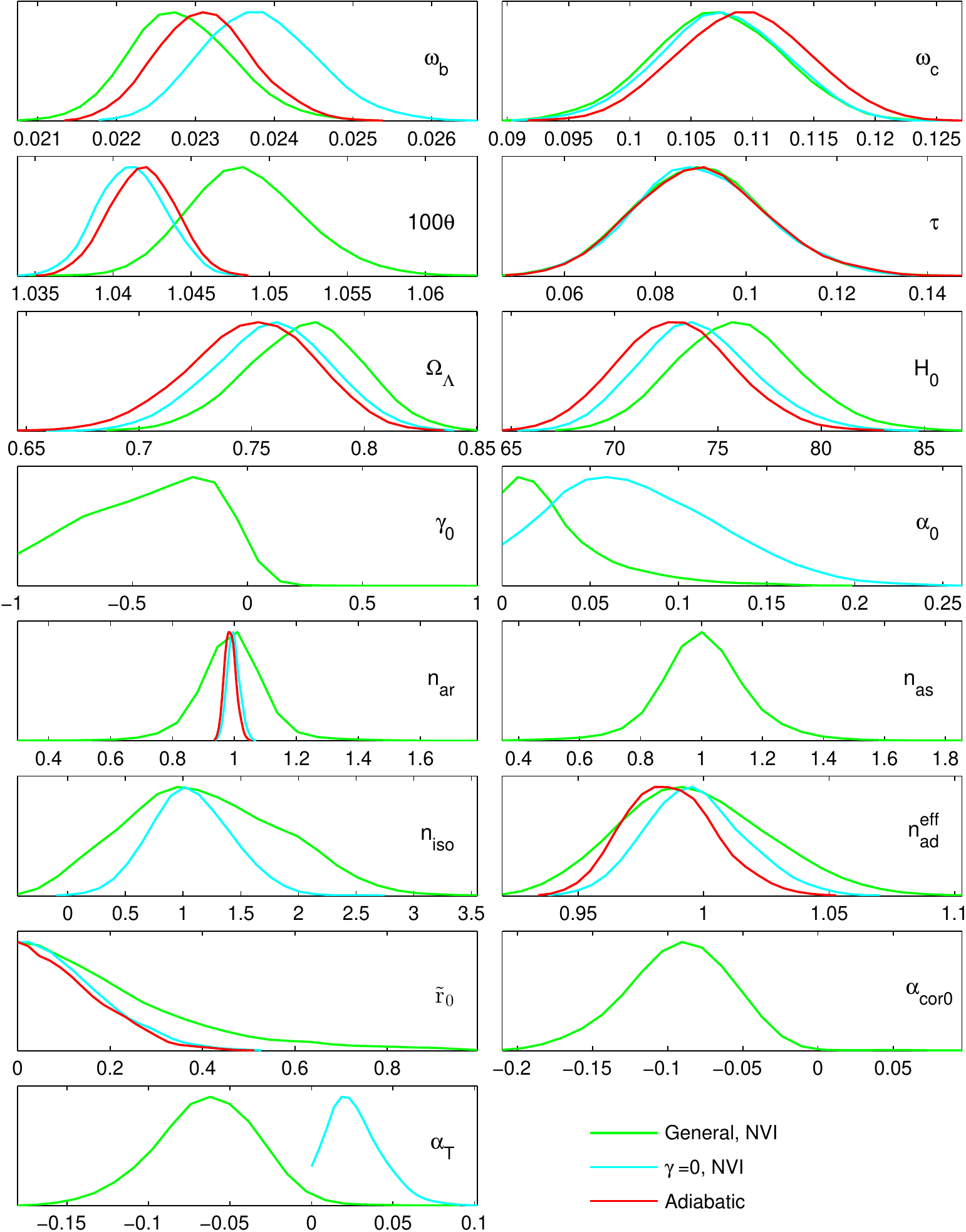}
\caption{{\bf Amplitude parametrization, comparison of special and general NVI cases.} Marginalized 1-d pdfs for generally correlated mixture of primordial adiabatic and NVI (solid green) modes compared to the uncorrelated ($\gamma=0$, solid cyan) case and to the pure adiabatic model (solid red).
\label{fig:AmplSpecialCasesNeutVel}
}
\end{figure}

The marginalized 1-d posterior pdfs are indicated in Fig.~\ref{fig:AmplSpecialCasesNeutDens} 
by solid blue $\gamma=0$ curves for NDI and in Fig.~\ref{fig:AmplSpecialCasesNeutVel} by solid cyan $\gamma=0$ curves for NVI 
in the amplitude parametrization, and in Figs.~\ref{fig:InflSpecialCases} and \ref{fig:InflSpecialCasesSlowRoll} by solid blue $\gamma=0$ curves for
NDI in the slow-roll parametrization. (Again, we drop the NVI case from the slow-roll analysis as it is hard to motivate.) 
To allow for an easy comparison, we also plot the generally correlated and adiabatic cases presented in the previous section.
Numerical results for the $\gamma=0$ case are reported in the fourth columns of Tables \ref{tab:NDIamp} -- \ref{tab:CDIslow}.

On all scales in all cases the allowed primordial isocurvature fraction ($\alpha_{1,2,0}$) is much larger in the uncorrelated case than in the generally correlated case
or in the maximally correlated cases $\gamma=\pm1$ (studied in the next subsection). In the other models the main non-adiabatic effect comes from the correlation
(whose amplitude is somewhere between the adiabatic and isocurvature contributions), but in the uncorrelated case the only disturbance to the adiabatic spectrum comes from the isocurvature itself. Thus rather large \emph{primordial} fractions can be accommodated by the CMB data. However, since the isocurvature component is more off-phase from the adiabatic one than the correlation component, the allowed non-adiabatic contribution, $|\alpha_T|$, to the observed CMB temperature variance is smaller in all uncorrelated cases than in the general cases.


In the uncorrelated NDI and CDI cases the slow-roll parameters $\etzz$ and $\epsilon$ are constrained equally well as in the adiabatic case, whereas $\etss$ remains unconstrained. A small isocurvature contribution at low-$\ell$ is allowed by the data (in particular due to cosmic variance) and the value of $\niso$ does not matter as long as the spectrum is nearly scale invariant (as it is due to our chosen priors of the slow-roll parameters). Since $r_0$ is simply 16 times $\veps$, we find a tight constraint on the tensor-to-scalar ratio. Indeed, we find $r_0 < 0.25$ both in the uncorrelated NDI and CDI cases, while the pure adiabatic case leads to $r_0 < 0.31$ at 95\% C.L. The tightening of the constraint on $r_0$ when allowing for the uncorrelated isocurvature component is natural, since adding power at low-$\ell$ eats room from the tensor contribution.

The betting odds in favor of the adiabatic model are 3.4, 16, 5.2, 3.5, and 16 : 1 when compared to the NDI (ampl. par.), NVI (ampl. par.), CDI (ampl. par.), NDI (slow-roll), CDI (slow-roll), respectively. Indeed, out of all models studied in this paper, the uncorrelated mixed NDI model in amplitude parametrization turns out to be least disfavored compared to the adiabatic model in terms of Bayesian model comparison.

\subsection{Maximally (anti)correlated NDI or CDI \label{sec:fullycorrelated}}

In the fully correlated cases the $\Pow_{\mtr{ar}}(k)$ spectrum is absent, and, following
\cite{Valiviita:2012ub}, we assume there are no tensor perturbations.
Moreover, according to Eq.~(\ref{eq:BWresults}), the two-field slow-roll inflation gives
 \beq 
 	\nadII = \niso \ = \ 1 + 2(\etss-\veps) \,.
 \eeq
No matter what $\veps$ is, these models lead to zero tensor contribution. Thus we can only constrain the combination $\etss-\veps$, not $\etss$ and $\veps$ individually.
Unlike in \cite{Valiviita:2012ub}, we assume $\nadII = \niso$ also in the amplitude parametrization in order to make comparison to slow-roll results more straightforward and to have the same number of parameters in both parametrizations, which affects the Bayesian model comparison results.

\begin{figure}[t]
  \centering
  \includegraphics[width=\columnwidth]{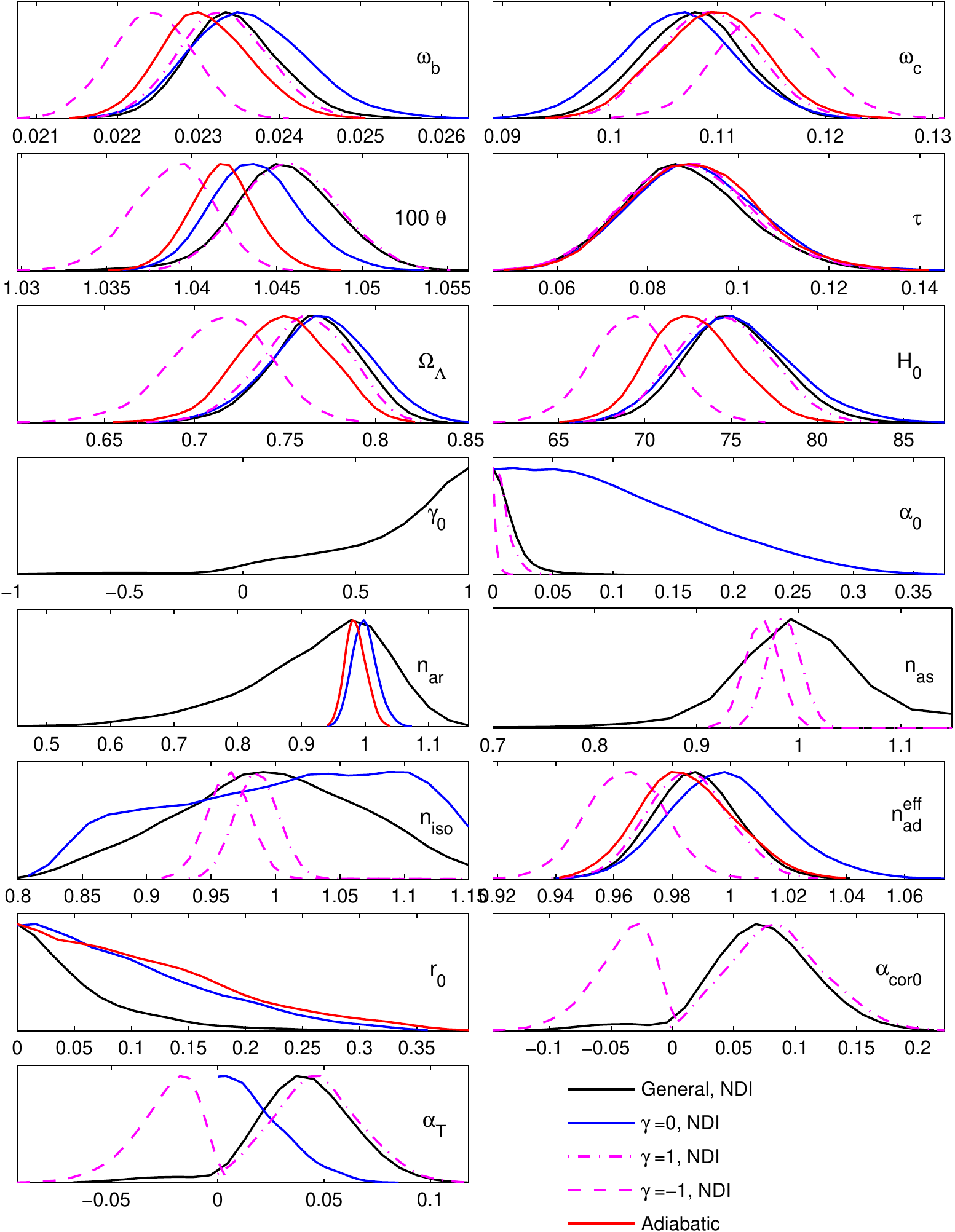}
\caption{{\bf Slow-roll parametrization, comparison of special and general NDI cases.} Marginalized 1-d pdfs for generally correlated mixture of primordial adiabatic and NDI (solid black) modes compared to the uncorrelated ($\gamma=0$, solid blue), maximally correlated ($\gamma=1$, dot-dashed magenta), and maximally anticorrelated ($\gamma=-1$, dashed magenta) cases, as well as, to the pure adiabatic model (solid red).
 \label{fig:InflSpecialCases}
}
\end{figure}
\begin{figure}[t]
  \centering
  \includegraphics[width=\columnwidth]{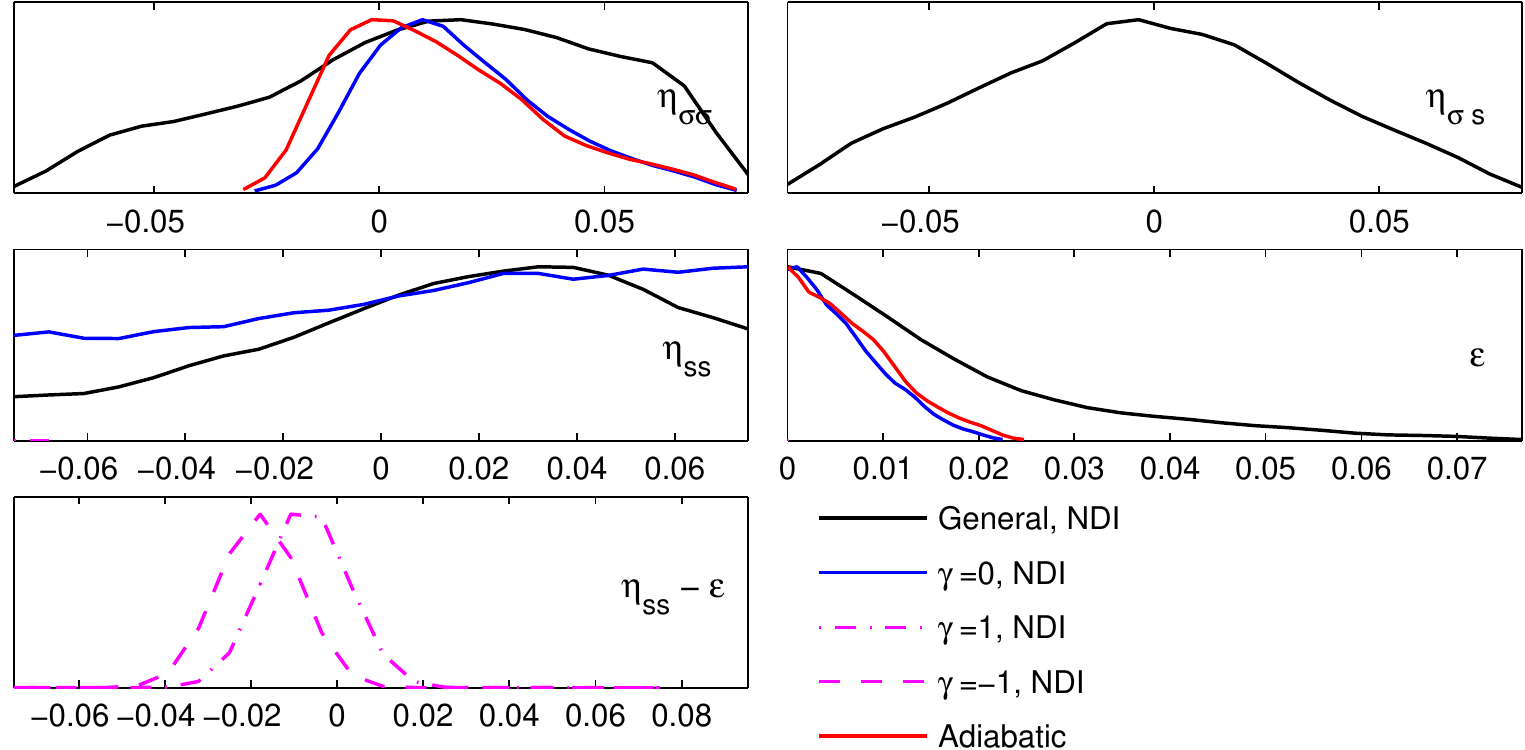}
  \caption{{\bf Slow-roll parametrization, comparison of slow-roll parameters in special and general NDI cases.} Marginalized 1-d posterior probability densities of the primary slow-roll parameters. The line styles are the same as in Fig.~\ref{fig:InflSpecialCases}.
\label{fig:InflSpecialCasesSlowRoll}
}
\end{figure}

Therefore, we have only seven  independent parameters:  background parameters
 \beq
 	\omega_b\,,\ \omega_c\,,\ \theta\,,\ \tau\,,
 \eeq
and three perturbation parameters
 \beq
 	\ln A_0^2\,,\ \niso=\nadII \,,\ \alpha_0\,. 
 \eeq
Since $\nadII = \niso$ (and the ``ar'' component is missing), the primordial isocurvature fraction is scale independent,
$\alpha = \alpha_{1,2} = \alpha_0$.

The primary perturbation parameters in the amplitude parametrization are
 \beq
 	\ln A_1^2\,,\ \ln A_2^2\,,\ \alpha_1\,(=\alpha_2),
 \eeq
and in the slow-roll parametrization
 \beq
    \ln A_0^2\,,\ \alpha_0\,,\ \etss-\veps \,.
 \eeq

The marginalized 1-d posterior pdfs of the NDI case are indicated by the dashed magenta ($\gamma=-1$, 100\% anticorrelation) and dot-dashed magenta ($\gamma=+1$, 100\% correlation)  curves in Fig.~\ref{fig:AmplSpecialCasesNeutDens} for the amplitude parametrization, and in Figs.~\ref{fig:InflSpecialCases} and \ref{fig:InflSpecialCasesSlowRoll} for the slow-roll parametrization. The numerical values are tabulated in the
fifth and sixth columns of Tables \ref{tab:NDIamp} (NDI) and \ref{tab:CDIamp} (CDI) in the amplitude parametrization, and in Tables \ref{tab:NDIslow} (NDI) and \ref{tab:CDIslow} (CDI) in the slow-roll parametrization. Note that we do not study the mixed NVI model with maximal (anti)correlation, since it is hard to think any physical mechanism that would lead to a correlation between NVI and adiabatic perturbations, since they may originate from very different epoch of the evolution of the universe. 
 
In the amplitude parametrization, we obtain very tightly constrained isocurvature fraction $\alpha$ for both the NDI and CDI cases, $\gamma=\pm1$. The $95\%$ C.L. limits for NDI are  $\alpha$ $<0.0303$ and  $\alpha$  $<0.0093$ for $\gamma=+1$ and $\gamma=-1$, respectively. Since the data force the adiabatic spectrum to be nearly scale invariant, and $\niso=\nadII$, we would expect very little difference to the slow-roll case, and indeed we find in the slow-roll parametrization very similar results, $\alpha<0.0280$ and $\alpha<0.0104$ for $\gamma=+1$ and $\gamma=-1$, respectively. The reason for these tight constraints is that in the maximally correlated cases a fixed value of $\alpha$ leads to much larger non-adiabatic contribution than in the partially correlated cases. As we would expect, in the cases where the maximal correlation has the same sign that was preferred in the general case, the constraints on $\alpha_{\mtr{cor}0}$ and $\alpha_T$ are very similar between the maximally correlated and generally correlated models.

The only ``slow-roll parameter'' of the maximally correlated cases, the combination $\etss - \veps$, is well constrained.

The Bayesian evidences for the maximally correlated models are in all cases worse than for the uncorrelated models where we found the best evidences compared to the adiabatic model.

\section{Discussion \label{sec:discussion}}

We constrained the primordial fraction of all regular isocurvature modes, one at a time
(matter density, neutrino density, and neutrino velocity isocurvature),
as well as the non-adiabatic contribution to the observed CMB temperature variance.
As the matter density (or CDI) mode has been extensively studied previously,
our focus was on the neutrino isocurvature modes (NDI, NVI), but we also updated
the constraints on CDI. Since primordial tensor perturbations are produced in
typical inflationary models, we included tensor perturbations throughout the analysis.

In the phenomenological approach the power law spectra of the curvature and
isocurvature mode, and the correlation between them, had independent amplitudes
and tilts. This added four independent perturbation parameters to the
standard adiabatic flat $\Lambda$CDM scenario. If any clear non-adiabatic features were present in the CMB data used
(WMAP-9, ACBAR, and QUaD), this approach should have found them. Neither frequentist
nor Bayesian methods indicated any preference for any of the isocurvature modes: the CMB data
set tight upper bounds on non-adiabatic contribution to the observed temperature variance.  

Using Bayesian evidences calculated by MultiNest we established the
betting odds for the models studied. For the generally
correlated mixture of the adiabatic and one isocurvature mode (either
NDI, NVI, or CDI) compared to the pure adiabatic primordial
perturbation mode we found the odds to be as small as 1 : 100.
However, in the special cases where we imposed
restrictions to the correlation component, the betting
odds were higher. In particular, for an uncorrelated
mixed NDI model (which had two non-adiabatic extra parameters) the betting odds 
were 1 : 3.4 compared to the pure adiabatic model.

In the phenomenological set-up, with generally correlated mixed adiabatic and isocurvature perturbations,
the tensor perturbations have been included at least in two different previous publications,
in \cite{Kawasaki:2007mb} for NDI, NVI, and CDI, and in \cite{Valiviita:2012ub} for CDI. In both of these, the first inflationary consistency relation was used in order to reduce the number of extra parameters. Namely, the tensor spectral index was determined from the consistency relation, $n_T = -r/[8(1-|\gamma|)]$, where $r$ was the tensor-to-scalar ratio (the ratio of tensor and total curvature perturbation power at the primordial time) and $\gamma$ the correlation fraction. In both  \cite{Kawasaki:2007mb}  and \cite{Valiviita:2012ub}, assigning uniform priors on $r$ and $\gamma$ (or $\sqrt{1-|\gamma|}$) led to ``tight'' constraints on $\gamma$. This was due to an unphysical prior of $n_T$: whenever $|\gamma|$ was near to one, the tensor spectral index was very negative. The huge tensor contribution thus induced was disfavored by the low-$\ell$ data. However, the original idea behind using the consistency relation was to obtain physically motivated (near to zero) values for $n_T$ as predicted by generic inflationary models, not huge negative values. In this paper we, for the first time, addressed these problems. We introduced the tensor-to-scalar ratio at horizon exit during inflation, $\tilde r$. This is related to the above definition by $r = \tilde r (1-|\gamma|)$. The inflationary consistency relation (derived to first order in slow-roll parameters) now read $n_T=-\tilde r / 8$, and this led to a uniform prior on $n_T$ between $\min(\tilde r)$ and zero, avoiding unphysical values, and most importantly avoiding the interference of the use of consistency relation with the constraints on $\gamma$. The difference between the ``old'' and new approaches was presented in Fig.~\ref{fig:rvstilder}.

We studied the matter and neutrino density
modes also in the two-field slow-roll inflation context, where we
assumed uniform priors on the four first order slow-roll parameters ($\veps$, $\etzz$, $\etzs$, $\etss$)
and assumed their magnitude to be small, i.e., less than $0.075$, so that the slow-roll approximation
was accurate enough. The main difference to the phenomenological
approach came from the fact that the choice of prior ranges of
the slow-roll parameters restricted all the primordial spectra
to be nearly scale invariant. (In the phenomenological approach
the data favored blue tilted isocurvature spectra with spectral indices $\niso \sim 1.45$
for NDI or $\niso \sim 2.05$ for CDI, which in the slow-roll approach were excluded by the prior.)

In all those slow-roll cases, where tensor perturbations were
produced, the posterior probability density of $\veps$ was much
narrower than its prior. In the models with generally correlated
primordial adiabatic and CDI or NDI mode, the constraint on $\veps$
was weaker than in the pure adiabatic model, but in the models with
uncorrelated adiabatic and CDI or NDI mode the constraint on $\veps$
was tighter since the only possible effect (on the temperature angular
power) of the nearly scale-invariant uncorrelated isocurvature
component is to add power to the low multipoles, where also the tensor
contribution would add power. Unlike $\veps$, all three $\eta_{ij}$ parameters were weakly constrained or
unconstrained: since our slow-roll approach led to almost
scale-invariant isocurvature and correlation spectra, the ``non-adiabatic''
modifications to $C_\ell$ appeared only in the low-$\ell$ region which
is cosmic variance dominated and hence insensitive to the small tilts of
the isocurvature and correlation components.

In Sec.~\ref{sec:mechanisms}, assuming a curvaton-type model with inhomogeneous lepton asymmetry, and taking into
account the big bang nucleosynthesis constraint on the neutrino asymmetry and the Planck constraint on
non-Gaussianity (and converting this to a constraint of the curvaton
inertia fraction $R$ at its decay time), we derived an upper limit for the
primordial isocurvature fraction, $\alpha<0.0045$, within this specific
maximally correlated neutrino density isocurvature model. The direct constraint from the CMB, $\alpha < 0.0256$
(see Sec.~\ref{sec:fullycorrelated} for the constraints on NDI $\gamma=1$ case in the amplitude parametrization),
is weaker by a factor six. 

In the recent literature, the CDI modes have been extensively contrasted against observations, but the observational constraints
on neutrino isocurvature have been studied less, although theoretical work and future forecasts can be found in many publications, see e.g. \cite{Bucher:2000kb,Bucher:2000hy,Kasanda:2011np,Zunckel:2010mm,Langlois:2012tm}.

The most recent constraints on the phenomenological mixed CDI, NDI, and
NVI models come from the Planck temperature anisotropy data, see
\cite{Ade:2013uln}. Perhaps surprisingly, the Planck constraints are
weaker than what we find here with WMAP-9, ACBAR and QUaD. The Planck
(our) upper bounds on the primordial isocurvature fraction on large scales, $\alpha_1$ (in \cite{Ade:2013uln} called $\beta_{\mtr{iso}}(k_{\mtr{low}})$), are 0.075 (0.045) for CDI, 0.27 (0.10) for NDI, and 0.18 (0.12) for NVI. The Planck parameter $\beta_{\mtr{iso}}(k_{\mtr{mid}})$ corresponds to our $\alpha_2$. The upper bounds on this are 0.39 (0.38) for CDI, 0.27 (0.27) for NDI, and 0.14 (0.13) for NVI. Finally, the 95\% C.L. intervals for the non-adiabatic contribution to the observed CMB temperature variance $\alpha_T$ (which is $1 - \alpha_{\mtc{RR}}^{(2,2500)}$ in \cite{Ade:2013uln}) are $-0.07$\ldots$ 0.02$ ($-0.03$\ldots$0.05$) for CDI, $-0.09$\ldots$0.01$ ($-0.05$\ldots$0.10$) for NDI, and $-0.05$\ldots$0.04$ ($-0.13$\ldots$-0.01$) for NVI.

The reason for such an unexpected difference between WMAP-9 and Planck is that the Planck data seem to prefer a negative correlation due to a relatively low power at low multipoles $\ell \sim 2$\ldots$40$ compared to the higher multipoles.
The adiabatic $\Lambda$CDM model fits  the acoustic peak structure of the Planck data with high precision, but even the best-fitting adiabatic model leads to more power at low-$\ell$ than seen in the data. This leads to a ``demand'' of some power-reducing mechanism at low-$\ell$; a negatively correlated isocurvature can provide such an effect. This explains why the Planck constraints are weaker than the WMAP-9 constraints in particular at large scales (i.e., on $\alpha_1$) and why Planck prefers negative correlation for all three cases (CDI, NDI, NVI), whereas WMAP-9 prefers a positive correlation in the CDI and NDI cases and a negative correlation in the NVI case (since with WMAP-9 the main non-adiabatic effects come from the first acoustic peak region). 

Another crucial difference between WMAP-9 and Planck is that the Planck data prefer smaller $H_0$ and $\Omega_\Lambda$, and constrain the background parameters much tighter, thus leaving less freedom to play with their values. The WMAP-9 data preferred very large $H_0$ and $\Omega_\Lambda$ in the phenomenological mixed models in order to compensate the shift of the first acoustic peak to right caused by the non-adiabatic component.

Some of the differences between Planck and our WMAP-9 results may come from the different parametrizations and assumptions. In \cite{Ade:2013uln} the curvature (i.e. adiabatic) spectrum was described by one power law, the isocurvature by one, and the correlation by one power law, which had a kink either at the low-$k$ or at the high-$k$ region to keep $|\mtc{C}_{\mtc{R}S}(k)| \le \sqrt{\Pow_{\mtc{R}}(k)\Pow_S(k)}$. Instead, we assumed two power-law components for the curvature perturbation, of which the other was fully correlated with the isocurvature power spectrum. This led our curvature spectrum to ``auto-adjust'' (run) in such a way that the above-mentioned mathematically necessary condition was always automatically satisfied without introducing kinks to any of the spectra. Moreover, we used the relative amplitudes $\alpha_{1,2}$ and $\gamma_{1,2}$ and logarithm of the ``total'' amplitudes $\ln(10^{10}A_{1,2}^2)$ as primary parameters, but in \cite{Ade:2013uln} the actual amplitudes of the three power spectra at two scales were primary parameters. We included also tensor perturbations in the analysis while in \cite{Ade:2013uln} the isocurvature analysis was done without tensor perturbations. However, based on \cite{Valiviita:2012ub} we do not expect this to cause major differences.

In \cite{Valiviita:2012ub} we studied the CDI mode in a similar set-up as NDI, NVI, and CDI here. We have checked the consistency of our new CDI results against the old ones. The differences can be traced to the following three points: different parametrization for the tensor-to-scalar ratio (in the phenomenological approach), different data, i.e., WMAP-9 versus WMAP-7, and an updated recombination code RECFAST in CAMB.

\begin{acknowledgments}
This work was supported by the Academy of Finland grant 257989.
This work was granted access to the HPC resources of CSC made
available within the  Distributed European Computing Initiative by
the PRACE-2IP, receiving funding from the European Community's Seventh
Framework Programme (FP7/2007-2013) under grant agreement RI-283493.
We thank the CSC - Scientific Computing Ltd (Finland) for computational
resources.
MS was supported by the Magnus Ehrnrooth Foundation, and
SR by the V\"{a}is\"{a}l\"{a}  Foundation.
\end{acknowledgments}

\bibliography{neutrino_isocurvature_210713}

\begin{thebibliography}{93}%
\makeatletter
\providecommand \@ifxundefined [1]{%
 \@ifx{#1\undefined}
}%
\providecommand \@ifnum [1]{%
 \ifnum #1\expandafter \@firstoftwo
 \else \expandafter \@secondoftwo
 \fi
}%
\providecommand \@ifx [1]{%
 \ifx #1\expandafter \@firstoftwo
 \else \expandafter \@secondoftwo
 \fi
}%
\providecommand \natexlab [1]{#1}%
\providecommand \enquote  [1]{``#1''}%
\providecommand \bibnamefont  [1]{#1}%
\providecommand \bibfnamefont [1]{#1}%
\providecommand \citenamefont [1]{#1}%
\providecommand \href@noop [0]{\@secondoftwo}%
\providecommand \href [0]{\begingroup \@sanitize@url \@href}%
\providecommand \@href[1]{\@@startlink{#1}\@@href}%
\providecommand \@@href[1]{\endgroup#1\@@endlink}%
\providecommand \@sanitize@url [0]{\catcode `\\12\catcode `\$12\catcode
  `\&12\catcode `\#12\catcode `\^12\catcode `\_12\catcode `\%12\relax}%
\providecommand \@@startlink[1]{}%
\providecommand \@@endlink[0]{}%
\providecommand \url  [0]{\begingroup\@sanitize@url \@url }%
\providecommand \@url [1]{\endgroup\@href {#1}{\urlprefix }}%
\providecommand \urlprefix  [0]{URL }%
\providecommand \Eprint [0]{\href }%
\providecommand \doibase [0]{http://dx.doi.org/}%
\providecommand \selectlanguage [0]{\@gobble}%
\providecommand \bibinfo  [0]{\@secondoftwo}%
\providecommand \bibfield  [0]{\@secondoftwo}%
\providecommand \translation [1]{[#1]}%
\providecommand \BibitemOpen [0]{}%
\providecommand \bibitemStop [0]{}%
\providecommand \bibitemNoStop [0]{.\EOS\space}%
\providecommand \EOS [0]{\spacefactor3000\relax}%
\providecommand \BibitemShut  [1]{\csname bibitem#1\endcsname}%
\let\auto@bib@innerbib\@empty
\bibitem [{\citenamefont {Valiviita}\ \emph {et~al.}(2012)\citenamefont
  {Valiviita}, \citenamefont {Savelainen}, \citenamefont {Talvitie},
  \citenamefont {Kurki-Suonio},\ and\ \citenamefont
  {Rusak}}]{Valiviita:2012ub}%
  \BibitemOpen
  \bibfield  {author} {\bibinfo {author} {\bibfnamefont {J.}~\bibnamefont
  {Valiviita}}, \bibinfo {author} {\bibfnamefont {M.}~\bibnamefont
  {Savelainen}}, \bibinfo {author} {\bibfnamefont {M.}~\bibnamefont
  {Talvitie}}, \bibinfo {author} {\bibfnamefont {H.}~\bibnamefont
  {Kurki-Suonio}}, \ and\ \bibinfo {author} {\bibfnamefont {S.}~\bibnamefont
  {Rusak}},\ }\href {\doibase 10.1088/0004-637X/753/2/151} {\bibfield
  {journal} {\bibinfo  {journal} {Astrophys.J.}\ }\textbf {\bibinfo {volume}
  {753}},\ \bibinfo {pages} {151} (\bibinfo {year} {2012})},\ \Eprint
  {http://arxiv.org/abs/1202.2852} {arXiv:1202.2852 [astro-ph.CO]} \BibitemShut
  {NoStop}%
\bibitem [{\citenamefont {Bennett}\ \emph {et~al.}(2012)\citenamefont {Bennett}
  \emph {et~al.}}]{Bennett:2012zja}%
  \BibitemOpen
  \bibfield  {author} {\bibinfo {author} {\bibfnamefont {C.}~\bibnamefont
  {Bennett}} \emph {et~al.} (\bibinfo {collaboration} {WMAP Collaboration}),\
  }\href@noop {} {\  (\bibinfo {year} {2012})},\ \Eprint
  {http://arxiv.org/abs/1212.5225} {arXiv:1212.5225 [astro-ph.CO]} \BibitemShut
  {NoStop}%
\bibitem [{\citenamefont {Larson}\ \emph {et~al.}(2011)\citenamefont {Larson},
  \citenamefont {Dunkley}, \citenamefont {Hinshaw}, \citenamefont {Komatsu},
  \citenamefont {Nolta} \emph {et~al.}}]{Larson:2010gs}%
  \BibitemOpen
  \bibfield  {author} {\bibinfo {author} {\bibfnamefont {D.}~\bibnamefont
  {Larson}}, \bibinfo {author} {\bibfnamefont {J.}~\bibnamefont {Dunkley}},
  \bibinfo {author} {\bibfnamefont {G.}~\bibnamefont {Hinshaw}}, \bibinfo
  {author} {\bibfnamefont {E.}~\bibnamefont {Komatsu}}, \bibinfo {author}
  {\bibfnamefont {M.}~\bibnamefont {Nolta}},  \emph {et~al.},\ }\href {\doibase
  10.1088/0067-0049/192/2/16} {\bibfield  {journal} {\bibinfo  {journal}
  {Astrophys.J.Suppl.}\ }\textbf {\bibinfo {volume} {192}},\ \bibinfo {pages}
  {16} (\bibinfo {year} {2011})},\ \Eprint {http://arxiv.org/abs/1001.4635}
  {arXiv:1001.4635 [astro-ph.CO]} \BibitemShut {NoStop}%
\bibitem [{\citenamefont {Reichardt}\ \emph {et~al.}(2009)\citenamefont
  {Reichardt} \emph {et~al.}}]{Reichardt:2008ay}%
  \BibitemOpen
  \bibfield  {author} {\bibinfo {author} {\bibfnamefont {C.~L.}\ \bibnamefont
  {Reichardt}} \emph {et~al.},\ }\href {\doibase 10.1088/0004-637X/694/2/1200}
  {\bibfield  {journal} {\bibinfo  {journal} {Astrophys. J.}\ }\textbf
  {\bibinfo {volume} {694}},\ \bibinfo {pages} {1200} (\bibinfo {year}
  {2009})},\ \Eprint {http://arxiv.org/abs/0801.1491} {arXiv:0801.1491
  [astro-ph]} \BibitemShut {NoStop}%
\bibitem [{\citenamefont {Brown}\ \emph {et~al.}(2009)\citenamefont {Brown}
  \emph {et~al.}}]{Brown:2009uy}%
  \BibitemOpen
  \bibfield  {author} {\bibinfo {author} {\bibfnamefont {M.}~\bibnamefont
  {Brown}} \emph {et~al.} (\bibinfo {collaboration} {QUaD collaboration}),\
  }\href {\doibase 10.1088/0004-637X/705/1/978} {\bibfield  {journal} {\bibinfo
   {journal} {Astrophys.J.}\ }\textbf {\bibinfo {volume} {705}},\ \bibinfo
  {pages} {978} (\bibinfo {year} {2009})},\ \Eprint
  {http://arxiv.org/abs/0906.1003} {arXiv:0906.1003 [astro-ph.CO]} \BibitemShut
  {NoStop}%
\bibitem [{\citenamefont {Ade}\ \emph {et~al.}(2013{\natexlab{a}})\citenamefont
  {Ade} \emph {et~al.}}]{Ade:2013uln}%
  \BibitemOpen
  \bibfield  {author} {\bibinfo {author} {\bibfnamefont {P.}~\bibnamefont
  {Ade}} \emph {et~al.} (\bibinfo {collaboration} {Planck Collaboration}),\
  }\href@noop {} {\  (\bibinfo {year} {2013}{\natexlab{a}})},\ \Eprint
  {http://arxiv.org/abs/1303.5082} {arXiv:1303.5082 [astro-ph.CO]} \BibitemShut
  {NoStop}%
\bibitem [{\citenamefont {Bucher}\ \emph {et~al.}(2000)\citenamefont {Bucher},
  \citenamefont {Moodley},\ and\ \citenamefont {Turok}}]{Bucher:1999re}%
  \BibitemOpen
  \bibfield  {author} {\bibinfo {author} {\bibfnamefont {M.}~\bibnamefont
  {Bucher}}, \bibinfo {author} {\bibfnamefont {K.}~\bibnamefont {Moodley}}, \
  and\ \bibinfo {author} {\bibfnamefont {N.}~\bibnamefont {Turok}},\ }\href
  {\doibase 10.1103/PhysRevD.62.083508} {\bibfield  {journal} {\bibinfo
  {journal} {Phys. Rev.}\ }\textbf {\bibinfo {volume} {D62}},\ \bibinfo {pages}
  {083508} (\bibinfo {year} {2000})},\ \Eprint
  {http://arxiv.org/abs/astro-ph/9904231} {arXiv:astro-ph/9904231} \BibitemShut
  {NoStop}%
\bibitem [{\citenamefont {Grin}\ \emph {et~al.}(2013)\citenamefont {Grin},
  \citenamefont {Hanson}, \citenamefont {Holder}, \citenamefont {Dor{\'e}},\
  and\ \citenamefont {Kamionkowski}}]{Grin:2013uya}%
  \BibitemOpen
  \bibfield  {author} {\bibinfo {author} {\bibfnamefont {D.}~\bibnamefont
  {Grin}}, \bibinfo {author} {\bibfnamefont {D.}~\bibnamefont {Hanson}},
  \bibinfo {author} {\bibfnamefont {G.}~\bibnamefont {Holder}}, \bibinfo
  {author} {\bibfnamefont {O.}~\bibnamefont {Dor{\'e}}}, \ and\ \bibinfo
  {author} {\bibfnamefont {M.}~\bibnamefont {Kamionkowski}},\ }\href@noop {} {\
   (\bibinfo {year} {2013})},\ \Eprint {http://arxiv.org/abs/1306.4319}
  {arXiv:1306.4319 [astro-ph.CO]} \BibitemShut {NoStop}%
\bibitem [{\citenamefont {Grin}\ \emph {et~al.}(2011)\citenamefont {Grin},
  \citenamefont {Dor{\'e}},\ and\ \citenamefont {Kamionkowski}}]{Grin:2011tf}%
  \BibitemOpen
  \bibfield  {author} {\bibinfo {author} {\bibfnamefont {D.}~\bibnamefont
  {Grin}}, \bibinfo {author} {\bibfnamefont {O.}~\bibnamefont {Dor{\'e}}}, \
  and\ \bibinfo {author} {\bibfnamefont {M.}~\bibnamefont {Kamionkowski}},\
  }\href {\doibase 10.1103/PhysRevD.84.123003} {\bibfield  {journal} {\bibinfo
  {journal} {Phys.Rev.}\ }\textbf {\bibinfo {volume} {D84}},\ \bibinfo {pages}
  {123003} (\bibinfo {year} {2011})},\ \Eprint {http://arxiv.org/abs/1107.5047}
  {arXiv:1107.5047 [astro-ph.CO]} \BibitemShut {NoStop}%
\bibitem [{\citenamefont {Lewis}(2011)}]{cambnote}%
  \BibitemOpen
  \bibfield  {author} {\bibinfo {author} {\bibfnamefont {A.}~\bibnamefont
  {Lewis}},\ }\href@noop {} {\enquote {\bibinfo {title} {Camb notes},}\ }
  (\bibinfo {year} {2011}),\ \bibinfo {note} {\\
  {h}ttp://cosmologist.info/notes/CAMB.pdf}\BibitemShut {NoStop}%
\bibitem [{\citenamefont {Langlois}\ and\ \citenamefont {van
  Tent}(2012)}]{Langlois:2012tm}%
  \BibitemOpen
  \bibfield  {author} {\bibinfo {author} {\bibfnamefont {D.}~\bibnamefont
  {Langlois}}\ and\ \bibinfo {author} {\bibfnamefont {B.}~\bibnamefont {van
  Tent}},\ }\href {\doibase 10.1088/1475-7516/2012/07/040} {\bibfield
  {journal} {\bibinfo  {journal} {JCAP}\ }\textbf {\bibinfo {volume} {1207}},\
  \bibinfo {pages} {040} (\bibinfo {year} {2012})},\ \Eprint
  {http://arxiv.org/abs/1204.5042} {arXiv:1204.5042 [astro-ph.CO]} \BibitemShut
  {NoStop}%
\bibitem [{\citenamefont {Enqvist}\ \emph {et~al.}(2000)\citenamefont
  {Enqvist}, \citenamefont {Kurki-Suonio},\ and\ \citenamefont
  {Valiviita}}]{Enqvist:2000hp}%
  \BibitemOpen
  \bibfield  {author} {\bibinfo {author} {\bibfnamefont {K.}~\bibnamefont
  {Enqvist}}, \bibinfo {author} {\bibfnamefont {H.}~\bibnamefont
  {Kurki-Suonio}}, \ and\ \bibinfo {author} {\bibfnamefont {J.}~\bibnamefont
  {Valiviita}},\ }\href {\doibase 10.1103/PhysRevD.62.103003} {\bibfield
  {journal} {\bibinfo  {journal} {Phys. Rev.}\ }\textbf {\bibinfo {volume}
  {D62}},\ \bibinfo {pages} {103003} (\bibinfo {year} {2000})},\ \Eprint
  {http://arxiv.org/abs/astro-ph/0006429} {arXiv:astro-ph/0006429} \BibitemShut
  {NoStop}%
\bibitem [{\citenamefont {Enqvist}\ \emph {et~al.}(2002)\citenamefont
  {Enqvist}, \citenamefont {Kurki-Suonio},\ and\ \citenamefont
  {Valiviita}}]{Enqvist:2001fu}%
  \BibitemOpen
  \bibfield  {author} {\bibinfo {author} {\bibfnamefont {K.}~\bibnamefont
  {Enqvist}}, \bibinfo {author} {\bibfnamefont {H.}~\bibnamefont
  {Kurki-Suonio}}, \ and\ \bibinfo {author} {\bibfnamefont {J.}~\bibnamefont
  {Valiviita}},\ }\href {\doibase 10.1103/PhysRevD.65.043002} {\bibfield
  {journal} {\bibinfo  {journal} {Phys. Rev.}\ }\textbf {\bibinfo {volume}
  {D65}},\ \bibinfo {pages} {043002} (\bibinfo {year} {2002})},\ \Eprint
  {http://arxiv.org/abs/astro-ph/0108422} {arXiv:astro-ph/0108422} \BibitemShut
  {NoStop}%
\bibitem [{\citenamefont {Beltran}\ \emph
  {et~al.}(2005{\natexlab{a}})\citenamefont {Beltran}, \citenamefont
  {Garcia-Bellido}, \citenamefont {Lesgourgues}, \citenamefont {Liddle},\ and\
  \citenamefont {Slosar}}]{Beltran:2005xd}%
  \BibitemOpen
  \bibfield  {author} {\bibinfo {author} {\bibfnamefont {M.}~\bibnamefont
  {Beltran}}, \bibinfo {author} {\bibfnamefont {J.}~\bibnamefont
  {Garcia-Bellido}}, \bibinfo {author} {\bibfnamefont {J.}~\bibnamefont
  {Lesgourgues}}, \bibinfo {author} {\bibfnamefont {A.~R.}\ \bibnamefont
  {Liddle}}, \ and\ \bibinfo {author} {\bibfnamefont {A.}~\bibnamefont
  {Slosar}},\ }\href {\doibase 10.1103/PhysRevD.71.063532} {\bibfield
  {journal} {\bibinfo  {journal} {Phys. Rev.}\ }\textbf {\bibinfo {volume}
  {D71}},\ \bibinfo {pages} {063532} (\bibinfo {year} {2005}{\natexlab{a}})},\
  \Eprint {http://arxiv.org/abs/astro-ph/0501477} {arXiv:astro-ph/0501477}
  \BibitemShut {NoStop}%
\bibitem [{\citenamefont {Beltran}\ \emph
  {et~al.}(2005{\natexlab{b}})\citenamefont {Beltran}, \citenamefont
  {Garcia-Bellido}, \citenamefont {Lesgourgues},\ and\ \citenamefont
  {Viel}}]{Beltran:2005gr}%
  \BibitemOpen
  \bibfield  {author} {\bibinfo {author} {\bibfnamefont {M.}~\bibnamefont
  {Beltran}}, \bibinfo {author} {\bibfnamefont {J.}~\bibnamefont
  {Garcia-Bellido}}, \bibinfo {author} {\bibfnamefont {J.}~\bibnamefont
  {Lesgourgues}}, \ and\ \bibinfo {author} {\bibfnamefont {M.}~\bibnamefont
  {Viel}},\ }\href {\doibase 10.1103/PhysRevD.72.103515} {\bibfield  {journal}
  {\bibinfo  {journal} {Phys. Rev.}\ }\textbf {\bibinfo {volume} {D72}},\
  \bibinfo {pages} {103515} (\bibinfo {year} {2005}{\natexlab{b}})},\ \Eprint
  {http://arxiv.org/abs/astro-ph/0509209} {arXiv:astro-ph/0509209} \BibitemShut
  {NoStop}%
\bibitem [{\citenamefont {Keskitalo}\ \emph {et~al.}(2007)\citenamefont
  {Keskitalo}, \citenamefont {Kurki-Suonio}, \citenamefont {Muhonen},\ and\
  \citenamefont {Valiviita}}]{Keskitalo:2006qv}%
  \BibitemOpen
  \bibfield  {author} {\bibinfo {author} {\bibfnamefont {R.}~\bibnamefont
  {Keskitalo}}, \bibinfo {author} {\bibfnamefont {H.}~\bibnamefont
  {Kurki-Suonio}}, \bibinfo {author} {\bibfnamefont {V.}~\bibnamefont
  {Muhonen}}, \ and\ \bibinfo {author} {\bibfnamefont {J.}~\bibnamefont
  {Valiviita}},\ }\href {\doibase 10.1088/1475-7516/2007/09/008} {\bibfield
  {journal} {\bibinfo  {journal} {JCAP}\ }\textbf {\bibinfo {volume} {0709}},\
  \bibinfo {pages} {008} (\bibinfo {year} {2007})},\ \Eprint
  {http://arxiv.org/abs/astro-ph/0611917} {arXiv:astro-ph/0611917} \BibitemShut
  {NoStop}%
\bibitem [{\citenamefont {Trotta}(2007)}]{Trotta:2006ww}%
  \BibitemOpen
  \bibfield  {author} {\bibinfo {author} {\bibfnamefont {R.}~\bibnamefont
  {Trotta}},\ }\href {\doibase 10.1111/j.1745-3933.2006.00268.x} {\bibfield
  {journal} {\bibinfo  {journal} {Mon. Not. Roy. Astron. Soc. Lett.}\ }\textbf
  {\bibinfo {volume} {375}},\ \bibinfo {pages} {L26} (\bibinfo {year}
  {2007})},\ \Eprint {http://arxiv.org/abs/astro-ph/0608116}
  {arXiv:astro-ph/0608116} \BibitemShut {NoStop}%
\bibitem [{\citenamefont {Seljak}\ \emph {et~al.}(2006)\citenamefont {Seljak},
  \citenamefont {Slosar},\ and\ \citenamefont {McDonald}}]{Seljak:2006bg}%
  \BibitemOpen
  \bibfield  {author} {\bibinfo {author} {\bibfnamefont {U.}~\bibnamefont
  {Seljak}}, \bibinfo {author} {\bibfnamefont {A.}~\bibnamefont {Slosar}}, \
  and\ \bibinfo {author} {\bibfnamefont {P.}~\bibnamefont {McDonald}},\
  }\href@noop {} {\bibfield  {journal} {\bibinfo  {journal} {JCAP}\ }\textbf
  {\bibinfo {volume} {0610}},\ \bibinfo {pages} {014} (\bibinfo {year}
  {2006})},\ \Eprint {http://arxiv.org/abs/astro-ph/0604335}
  {arXiv:astro-ph/0604335} \BibitemShut {NoStop}%
\bibitem [{\citenamefont {Lewis}(2006)}]{Lewis:2006ma}%
  \BibitemOpen
  \bibfield  {author} {\bibinfo {author} {\bibfnamefont {A.}~\bibnamefont
  {Lewis}},\ }\href@noop {} {\  (\bibinfo {year} {2006})},\ \Eprint
  {http://arxiv.org/abs/astro-ph/0603753} {arXiv:astro-ph/0603753} \BibitemShut
  {NoStop}%
\bibitem [{\citenamefont {Bean}\ \emph {et~al.}(2006)\citenamefont {Bean},
  \citenamefont {Dunkley},\ and\ \citenamefont {Pierpaoli}}]{Bean:2006qz}%
  \BibitemOpen
  \bibfield  {author} {\bibinfo {author} {\bibfnamefont {R.}~\bibnamefont
  {Bean}}, \bibinfo {author} {\bibfnamefont {J.}~\bibnamefont {Dunkley}}, \
  and\ \bibinfo {author} {\bibfnamefont {E.}~\bibnamefont {Pierpaoli}},\ }\href
  {\doibase 10.1103/PhysRevD.74.063503} {\bibfield  {journal} {\bibinfo
  {journal} {Phys. Rev.}\ }\textbf {\bibinfo {volume} {D74}},\ \bibinfo {pages}
  {063503} (\bibinfo {year} {2006})},\ \Eprint
  {http://arxiv.org/abs/astro-ph/0606685} {arXiv:astro-ph/0606685} \BibitemShut
  {NoStop}%
\bibitem [{\citenamefont {Kawasaki}\ and\ \citenamefont
  {Sekiguchi}(2008)}]{Kawasaki:2007mb}%
  \BibitemOpen
  \bibfield  {author} {\bibinfo {author} {\bibfnamefont {M.}~\bibnamefont
  {Kawasaki}}\ and\ \bibinfo {author} {\bibfnamefont {T.}~\bibnamefont
  {Sekiguchi}},\ }\href {\doibase 10.1143/PTP.120.995} {\bibfield  {journal}
  {\bibinfo  {journal} {Prog. Theor. Phys.}\ }\textbf {\bibinfo {volume}
  {120}},\ \bibinfo {pages} {995} (\bibinfo {year} {2008})},\ \Eprint
  {http://arxiv.org/abs/0705.2853} {arXiv:0705.2853 [astro-ph]} \BibitemShut
  {NoStop}%
\bibitem [{\citenamefont {Beltran}(2008)}]{Beltran:2008ei}%
  \BibitemOpen
  \bibfield  {author} {\bibinfo {author} {\bibfnamefont {M.}~\bibnamefont
  {Beltran}},\ }\href {\doibase 10.1103/PhysRevD.78.023530} {\bibfield
  {journal} {\bibinfo  {journal} {Phys. Rev.}\ }\textbf {\bibinfo {volume}
  {D78}},\ \bibinfo {pages} {023530} (\bibinfo {year} {2008})},\ \Eprint
  {http://arxiv.org/abs/0804.1097} {arXiv:0804.1097 [astro-ph]} \BibitemShut
  {NoStop}%
\bibitem [{\citenamefont {Sollom}\ \emph {et~al.}(2009)\citenamefont {Sollom},
  \citenamefont {Challinor},\ and\ \citenamefont {Hobson}}]{Sollom:2009vd}%
  \BibitemOpen
  \bibfield  {author} {\bibinfo {author} {\bibfnamefont {I.}~\bibnamefont
  {Sollom}}, \bibinfo {author} {\bibfnamefont {A.}~\bibnamefont {Challinor}}, \
  and\ \bibinfo {author} {\bibfnamefont {M.~P.}\ \bibnamefont {Hobson}},\
  }\href {\doibase 10.1103/PhysRevD.79.123521} {\bibfield  {journal} {\bibinfo
  {journal} {Phys. Rev.}\ }\textbf {\bibinfo {volume} {D79}},\ \bibinfo {pages}
  {123521} (\bibinfo {year} {2009})},\ \Eprint {http://arxiv.org/abs/0903.5257}
  {arXiv:0903.5257 [astro-ph.CO]} \BibitemShut {NoStop}%
\bibitem [{\citenamefont {Valiviita}\ and\ \citenamefont
  {Giannantonio}(2009)}]{Valiviita:2009bp}%
  \BibitemOpen
  \bibfield  {author} {\bibinfo {author} {\bibfnamefont {J.}~\bibnamefont
  {Valiviita}}\ and\ \bibinfo {author} {\bibfnamefont {T.}~\bibnamefont
  {Giannantonio}},\ }\href {\doibase 10.1103/PhysRevD.80.123516} {\bibfield
  {journal} {\bibinfo  {journal} {Phys.Rev.}\ }\textbf {\bibinfo {volume}
  {D80}},\ \bibinfo {pages} {123516} (\bibinfo {year} {2009})},\ \Eprint
  {http://arxiv.org/abs/0909.5190} {arXiv:0909.5190 [astro-ph.CO]} \BibitemShut
  {NoStop}%
\bibitem [{\citenamefont {Castro}\ \emph {et~al.}(2009)\citenamefont {Castro}
  \emph {et~al.}}]{Castro:2009ej}%
  \BibitemOpen
  \bibfield  {author} {\bibinfo {author} {\bibfnamefont {P.}~\bibnamefont
  {Castro}} \emph {et~al.} (\bibinfo {collaboration} {QUaD collaboration}),\
  }\href {\doibase 10.1088/0004-637X/701/2/857} {\bibfield  {journal} {\bibinfo
   {journal} {Astrophys.J.}\ }\textbf {\bibinfo {volume} {701}},\ \bibinfo
  {pages} {857} (\bibinfo {year} {2009})},\ \Eprint
  {http://arxiv.org/abs/0901.0810} {arXiv:0901.0810 [astro-ph.CO]} \BibitemShut
  {NoStop}%
\bibitem [{\citenamefont {Li}\ \emph {et~al.}(2011)\citenamefont {Li},
  \citenamefont {Liu}, \citenamefont {Xia},\ and\ \citenamefont
  {Cai}}]{Li:2010yb}%
  \BibitemOpen
  \bibfield  {author} {\bibinfo {author} {\bibfnamefont {H.}~\bibnamefont
  {Li}}, \bibinfo {author} {\bibfnamefont {J.}~\bibnamefont {Liu}}, \bibinfo
  {author} {\bibfnamefont {J.-Q.}\ \bibnamefont {Xia}}, \ and\ \bibinfo
  {author} {\bibfnamefont {Y.-F.}\ \bibnamefont {Cai}},\ }\href {\doibase
  10.1103/PhysRevD.83.123517} {\bibfield  {journal} {\bibinfo  {journal}
  {Phys.Rev.}\ }\textbf {\bibinfo {volume} {D83}},\ \bibinfo {pages} {123517}
  (\bibinfo {year} {2011})},\ \Eprint {http://arxiv.org/abs/1012.2511}
  {arXiv:1012.2511 [astro-ph.CO]} \BibitemShut {NoStop}%
\bibitem [{\citenamefont {Peiris}\ \emph {et~al.}(2003)\citenamefont {Peiris}
  \emph {et~al.}}]{Peiris:2003ff}%
  \BibitemOpen
  \bibfield  {author} {\bibinfo {author} {\bibfnamefont {H.}~\bibnamefont
  {Peiris}} \emph {et~al.} (\bibinfo {collaboration} {WMAP Collaboration}),\
  }\href {\doibase 10.1086/377228} {\bibfield  {journal} {\bibinfo  {journal}
  {Astrophys.J.Suppl.}\ }\textbf {\bibinfo {volume} {148}},\ \bibinfo {pages}
  {213} (\bibinfo {year} {2003})},\ \Eprint
  {http://arxiv.org/abs/astro-ph/0302225} {arXiv:astro-ph/0302225 [astro-ph]}
  \BibitemShut {NoStop}%
\bibitem [{\citenamefont {Valiviita}\ and\ \citenamefont
  {Muhonen}(2003)}]{Valiviita:2003ty}%
  \BibitemOpen
  \bibfield  {author} {\bibinfo {author} {\bibfnamefont {J.}~\bibnamefont
  {Valiviita}}\ and\ \bibinfo {author} {\bibfnamefont {V.}~\bibnamefont
  {Muhonen}},\ }\href {\doibase 10.1103/PhysRevLett.91.131302} {\bibfield
  {journal} {\bibinfo  {journal} {Phys. Rev. Lett.}\ }\textbf {\bibinfo
  {volume} {91}},\ \bibinfo {pages} {131302} (\bibinfo {year} {2003})},\
  \Eprint {http://arxiv.org/abs/astro-ph/0304175} {arXiv:astro-ph/0304175}
  \BibitemShut {NoStop}%
\bibitem [{\citenamefont {Kurki-Suonio}\ \emph {et~al.}(2005)\citenamefont
  {Kurki-Suonio}, \citenamefont {Muhonen},\ and\ \citenamefont
  {Valiviita}}]{KurkiSuonio:2004mn}%
  \BibitemOpen
  \bibfield  {author} {\bibinfo {author} {\bibfnamefont {H.}~\bibnamefont
  {Kurki-Suonio}}, \bibinfo {author} {\bibfnamefont {V.}~\bibnamefont
  {Muhonen}}, \ and\ \bibinfo {author} {\bibfnamefont {J.}~\bibnamefont
  {Valiviita}},\ }\href {\doibase 10.1103/PhysRevD.71.063005} {\bibfield
  {journal} {\bibinfo  {journal} {Phys. Rev.}\ }\textbf {\bibinfo {volume}
  {D71}},\ \bibinfo {pages} {063005} (\bibinfo {year} {2005})},\ \Eprint
  {http://arxiv.org/abs/astro-ph/0412439} {arXiv:astro-ph/0412439} \BibitemShut
  {NoStop}%
\bibitem [{\citenamefont {Lidsey}\ \emph {et~al.}(1997)\citenamefont {Lidsey},
  \citenamefont {Liddle}, \citenamefont {Kolb}, \citenamefont {Copeland},
  \citenamefont {Barreiro} \emph {et~al.}}]{Lidsey:1995np}%
  \BibitemOpen
  \bibfield  {author} {\bibinfo {author} {\bibfnamefont {J.~E.}\ \bibnamefont
  {Lidsey}}, \bibinfo {author} {\bibfnamefont {A.~R.}\ \bibnamefont {Liddle}},
  \bibinfo {author} {\bibfnamefont {E.~W.}\ \bibnamefont {Kolb}}, \bibinfo
  {author} {\bibfnamefont {E.~J.}\ \bibnamefont {Copeland}}, \bibinfo {author}
  {\bibfnamefont {T.}~\bibnamefont {Barreiro}},  \emph {et~al.},\ }\href
  {\doibase 10.1103/RevModPhys.69.373} {\bibfield  {journal} {\bibinfo
  {journal} {Rev.Mod.Phys.}\ }\textbf {\bibinfo {volume} {69}},\ \bibinfo
  {pages} {373} (\bibinfo {year} {1997})},\ \Eprint
  {http://arxiv.org/abs/astro-ph/9508078} {arXiv:astro-ph/9508078 [astro-ph]}
  \BibitemShut {NoStop}%
\bibitem [{\citenamefont {Bartolo}\ \emph
  {et~al.}(2001{\natexlab{a}})\citenamefont {Bartolo}, \citenamefont
  {Matarrese},\ and\ \citenamefont {Riotto}}]{Bartolo:2001rt}%
  \BibitemOpen
  \bibfield  {author} {\bibinfo {author} {\bibfnamefont {N.}~\bibnamefont
  {Bartolo}}, \bibinfo {author} {\bibfnamefont {S.}~\bibnamefont {Matarrese}},
  \ and\ \bibinfo {author} {\bibfnamefont {A.}~\bibnamefont {Riotto}},\ }\href
  {\doibase 10.1103/PhysRevD.64.123504} {\bibfield  {journal} {\bibinfo
  {journal} {Phys. Rev.}\ }\textbf {\bibinfo {volume} {D64}},\ \bibinfo {pages}
  {123504} (\bibinfo {year} {2001}{\natexlab{a}})},\ \Eprint
  {http://arxiv.org/abs/astro-ph/0107502} {arXiv:astro-ph/0107502} \BibitemShut
  {NoStop}%
\bibitem [{\citenamefont {Wands}\ \emph {et~al.}(2002)\citenamefont {Wands},
  \citenamefont {Bartolo}, \citenamefont {Matarrese},\ and\ \citenamefont
  {Riotto}}]{Wands:2002bn}%
  \BibitemOpen
  \bibfield  {author} {\bibinfo {author} {\bibfnamefont {D.}~\bibnamefont
  {Wands}}, \bibinfo {author} {\bibfnamefont {N.}~\bibnamefont {Bartolo}},
  \bibinfo {author} {\bibfnamefont {S.}~\bibnamefont {Matarrese}}, \ and\
  \bibinfo {author} {\bibfnamefont {A.}~\bibnamefont {Riotto}},\ }\href
  {\doibase 10.1103/PhysRevD.66.043520} {\bibfield  {journal} {\bibinfo
  {journal} {Phys. Rev.}\ }\textbf {\bibinfo {volume} {D66}},\ \bibinfo {pages}
  {043520} (\bibinfo {year} {2002})},\ \Eprint
  {http://arxiv.org/abs/astro-ph/0205253} {arXiv:astro-ph/0205253} \BibitemShut
  {NoStop}%
\bibitem [{\citenamefont {Cortes}\ and\ \citenamefont
  {Liddle}(2006)}]{Cortes:2006ap}%
  \BibitemOpen
  \bibfield  {author} {\bibinfo {author} {\bibfnamefont {M.}~\bibnamefont
  {Cortes}}\ and\ \bibinfo {author} {\bibfnamefont {A.~R.}\ \bibnamefont
  {Liddle}},\ }\href {\doibase 10.1103/PhysRevD.73.083523} {\bibfield
  {journal} {\bibinfo  {journal} {Phys. Rev.}\ }\textbf {\bibinfo {volume}
  {D73}},\ \bibinfo {pages} {083523} (\bibinfo {year} {2006})},\ \Eprint
  {http://arxiv.org/abs/astro-ph/0603016} {arXiv:astro-ph/0603016} \BibitemShut
  {NoStop}%
\bibitem [{\citenamefont {Cortes}\ \emph {et~al.}(2007)\citenamefont {Cortes},
  \citenamefont {Liddle},\ and\ \citenamefont {Mukherjee}}]{Cortes:2007ak}%
  \BibitemOpen
  \bibfield  {author} {\bibinfo {author} {\bibfnamefont {M.}~\bibnamefont
  {Cortes}}, \bibinfo {author} {\bibfnamefont {A.~R.}\ \bibnamefont {Liddle}},
  \ and\ \bibinfo {author} {\bibfnamefont {P.}~\bibnamefont {Mukherjee}},\
  }\href {\doibase 10.1103/PhysRevD.75.083520} {\bibfield  {journal} {\bibinfo
  {journal} {Phys. Rev.}\ }\textbf {\bibinfo {volume} {D75}},\ \bibinfo {pages}
  {083520} (\bibinfo {year} {2007})},\ \Eprint
  {http://arxiv.org/abs/astro-ph/0702170} {arXiv:astro-ph/0702170} \BibitemShut
  {NoStop}%
\bibitem [{\citenamefont {Byrnes}\ and\ \citenamefont
  {Wands}(2006)}]{Byrnes:2006fr}%
  \BibitemOpen
  \bibfield  {author} {\bibinfo {author} {\bibfnamefont {C.~T.}\ \bibnamefont
  {Byrnes}}\ and\ \bibinfo {author} {\bibfnamefont {D.}~\bibnamefont {Wands}},\
  }\href {\doibase 10.1103/PhysRevD.74.043529} {\bibfield  {journal} {\bibinfo
  {journal} {Phys. Rev.}\ }\textbf {\bibinfo {volume} {D74}},\ \bibinfo {pages}
  {043529} (\bibinfo {year} {2006})},\ \Eprint
  {http://arxiv.org/abs/astro-ph/0605679} {arXiv:astro-ph/0605679} \BibitemShut
  {NoStop}%
\bibitem [{\citenamefont {van Tent}(2004)}]{vanTent:2003mn}%
  \BibitemOpen
  \bibfield  {author} {\bibinfo {author} {\bibfnamefont {B.}~\bibnamefont {van
  Tent}},\ }\href {\doibase 10.1088/0264-9381/21/2/002} {\bibfield  {journal}
  {\bibinfo  {journal} {Class.Quant.Grav.}\ }\textbf {\bibinfo {volume} {21}},\
  \bibinfo {pages} {349} (\bibinfo {year} {2004})},\ \Eprint
  {http://arxiv.org/abs/astro-ph/0307048} {arXiv:astro-ph/0307048 [astro-ph]}
  \BibitemShut {NoStop}%
\bibitem [{\citenamefont {Peterson}\ and\ \citenamefont
  {Tegmark}(2011)}]{Peterson:2010np}%
  \BibitemOpen
  \bibfield  {author} {\bibinfo {author} {\bibfnamefont {C.~M.}\ \bibnamefont
  {Peterson}}\ and\ \bibinfo {author} {\bibfnamefont {M.}~\bibnamefont
  {Tegmark}},\ }\href {\doibase 10.1103/PhysRevD.83.023522} {\bibfield
  {journal} {\bibinfo  {journal} {Phys.Rev.}\ }\textbf {\bibinfo {volume}
  {D83}},\ \bibinfo {pages} {023522} (\bibinfo {year} {2011})},\ \Eprint
  {http://arxiv.org/abs/1005.4056} {arXiv:1005.4056 [astro-ph.CO]} \BibitemShut
  {NoStop}%
\bibitem [{\citenamefont {Norena}\ \emph {et~al.}(2012)\citenamefont {Norena},
  \citenamefont {Wagner}, \citenamefont {Verde}, \citenamefont {Peiris},\ and\
  \citenamefont {Easther}}]{Norena:2012rs}%
  \BibitemOpen
  \bibfield  {author} {\bibinfo {author} {\bibfnamefont {J.}~\bibnamefont
  {Norena}}, \bibinfo {author} {\bibfnamefont {C.}~\bibnamefont {Wagner}},
  \bibinfo {author} {\bibfnamefont {L.}~\bibnamefont {Verde}}, \bibinfo
  {author} {\bibfnamefont {H.~V.}\ \bibnamefont {Peiris}}, \ and\ \bibinfo
  {author} {\bibfnamefont {R.}~\bibnamefont {Easther}},\ }\href {\doibase
  10.1103/PhysRevD.86.023505} {\bibfield  {journal} {\bibinfo  {journal}
  {Phys.Rev.}\ }\textbf {\bibinfo {volume} {D86}},\ \bibinfo {pages} {023505}
  (\bibinfo {year} {2012})},\ \Eprint {http://arxiv.org/abs/1202.0304}
  {arXiv:1202.0304 [astro-ph.CO]} \BibitemShut {NoStop}%
\bibitem [{\citenamefont {Choi}\ \emph {et~al.}(2009)\citenamefont {Choi},
  \citenamefont {Gong},\ and\ \citenamefont {Jeong}}]{Choi:2008et}%
  \BibitemOpen
  \bibfield  {author} {\bibinfo {author} {\bibfnamefont {K.-Y.}\ \bibnamefont
  {Choi}}, \bibinfo {author} {\bibfnamefont {J.-O.}\ \bibnamefont {Gong}}, \
  and\ \bibinfo {author} {\bibfnamefont {D.}~\bibnamefont {Jeong}},\ }\href
  {\doibase 10.1088/1475-7516/2009/02/032} {\bibfield  {journal} {\bibinfo
  {journal} {JCAP}\ }\textbf {\bibinfo {volume} {0902}},\ \bibinfo {pages}
  {032} (\bibinfo {year} {2009})},\ \Eprint {http://arxiv.org/abs/0810.2299}
  {arXiv:0810.2299 [hep-ph]} \BibitemShut {NoStop}%
\bibitem [{\citenamefont {Hamazaki}(2008)}]{Hamazaki:2007eq}%
  \BibitemOpen
  \bibfield  {author} {\bibinfo {author} {\bibfnamefont {T.}~\bibnamefont
  {Hamazaki}},\ }\href {\doibase 10.1016/j.nuclphysb.2007.09.028} {\bibfield
  {journal} {\bibinfo  {journal} {Nucl. Phys.}\ }\textbf {\bibinfo {volume}
  {B791}},\ \bibinfo {pages} {20} (\bibinfo {year} {2008})},\ \Eprint
  {http://arxiv.org/abs/0711.2954} {arXiv:0711.2954 [astro-ph]} \BibitemShut
  {NoStop}%
\bibitem [{\citenamefont {Di~Marco}\ \emph {et~al.}(2007)\citenamefont
  {Di~Marco}, \citenamefont {Finelli},\ and\ \citenamefont
  {Gruppuso}}]{DiMarco:2007pb}%
  \BibitemOpen
  \bibfield  {author} {\bibinfo {author} {\bibfnamefont {F.}~\bibnamefont
  {Di~Marco}}, \bibinfo {author} {\bibfnamefont {F.}~\bibnamefont {Finelli}}, \
  and\ \bibinfo {author} {\bibfnamefont {A.}~\bibnamefont {Gruppuso}},\ }\href
  {\doibase 10.1103/PhysRevD.76.043530} {\bibfield  {journal} {\bibinfo
  {journal} {Phys. Rev.}\ }\textbf {\bibinfo {volume} {D76}},\ \bibinfo {pages}
  {043530} (\bibinfo {year} {2007})},\ \Eprint {http://arxiv.org/abs/0705.2185}
  {arXiv:0705.2185 [astro-ph]} \BibitemShut {NoStop}%
\bibitem [{\citenamefont {Lalak}\ \emph {et~al.}(2007)\citenamefont {Lalak},
  \citenamefont {Langlois}, \citenamefont {Pokorski},\ and\ \citenamefont
  {Turzynski}}]{Lalak:2007vi}%
  \BibitemOpen
  \bibfield  {author} {\bibinfo {author} {\bibfnamefont {Z.}~\bibnamefont
  {Lalak}}, \bibinfo {author} {\bibfnamefont {D.}~\bibnamefont {Langlois}},
  \bibinfo {author} {\bibfnamefont {S.}~\bibnamefont {Pokorski}}, \ and\
  \bibinfo {author} {\bibfnamefont {K.}~\bibnamefont {Turzynski}},\ }\href@noop
  {} {\bibfield  {journal} {\bibinfo  {journal} {JCAP}\ }\textbf {\bibinfo
  {volume} {0707}},\ \bibinfo {pages} {014} (\bibinfo {year} {2007})},\ \Eprint
  {http://arxiv.org/abs/0704.0212} {arXiv:0704.0212 [hep-th]} \BibitemShut
  {NoStop}%
\bibitem [{\citenamefont {Choi}\ \emph {et~al.}(2007)\citenamefont {Choi},
  \citenamefont {Hall},\ and\ \citenamefont {van~de Bruck}}]{Choi:2007su}%
  \BibitemOpen
  \bibfield  {author} {\bibinfo {author} {\bibfnamefont {K.-Y.}\ \bibnamefont
  {Choi}}, \bibinfo {author} {\bibfnamefont {L.~M.~H.}\ \bibnamefont {Hall}}, \
  and\ \bibinfo {author} {\bibfnamefont {C.}~\bibnamefont {van~de Bruck}},\
  }\href@noop {} {\bibfield  {journal} {\bibinfo  {journal} {JCAP}\ }\textbf
  {\bibinfo {volume} {0702}},\ \bibinfo {pages} {029} (\bibinfo {year}
  {2007})},\ \Eprint {http://arxiv.org/abs/astro-ph/0701247}
  {arXiv:astro-ph/0701247} \BibitemShut {NoStop}%
\bibitem [{\citenamefont {Rigopoulos}\ \emph {et~al.}(2007)\citenamefont
  {Rigopoulos}, \citenamefont {Shellard},\ and\ \citenamefont {van
  Tent}}]{Rigopoulos:2005us}%
  \BibitemOpen
  \bibfield  {author} {\bibinfo {author} {\bibfnamefont {G.~I.}\ \bibnamefont
  {Rigopoulos}}, \bibinfo {author} {\bibfnamefont {E.~P.~S.}\ \bibnamefont
  {Shellard}}, \ and\ \bibinfo {author} {\bibfnamefont {B.~J.~W.}\ \bibnamefont
  {van Tent}},\ }\href {\doibase 10.1103/PhysRevD.76.083512} {\bibfield
  {journal} {\bibinfo  {journal} {Phys. Rev.}\ }\textbf {\bibinfo {volume}
  {D76}},\ \bibinfo {pages} {083512} (\bibinfo {year} {2007})},\ \Eprint
  {http://arxiv.org/abs/astro-ph/0511041} {arXiv:astro-ph/0511041} \BibitemShut
  {NoStop}%
\bibitem [{\citenamefont {Bassett}\ \emph {et~al.}(2006)\citenamefont
  {Bassett}, \citenamefont {Tsujikawa},\ and\ \citenamefont
  {Wands}}]{Bassett:2005xm}%
  \BibitemOpen
  \bibfield  {author} {\bibinfo {author} {\bibfnamefont {B.~A.}\ \bibnamefont
  {Bassett}}, \bibinfo {author} {\bibfnamefont {S.}~\bibnamefont {Tsujikawa}},
  \ and\ \bibinfo {author} {\bibfnamefont {D.}~\bibnamefont {Wands}},\ }\href
  {\doibase 10.1103/RevModPhys.78.537} {\bibfield  {journal} {\bibinfo
  {journal} {Rev. Mod. Phys.}\ }\textbf {\bibinfo {volume} {78}},\ \bibinfo
  {pages} {537} (\bibinfo {year} {2006})},\ \Eprint
  {http://arxiv.org/abs/astro-ph/0507632} {arXiv:astro-ph/0507632} \BibitemShut
  {NoStop}%
\bibitem [{\citenamefont {Hattori}\ and\ \citenamefont
  {Yamamoto}(2005)}]{Hattori:2005ac}%
  \BibitemOpen
  \bibfield  {author} {\bibinfo {author} {\bibfnamefont {T.}~\bibnamefont
  {Hattori}}\ and\ \bibinfo {author} {\bibfnamefont {K.}~\bibnamefont
  {Yamamoto}},\ }\href {\doibase 10.1088/1475-7516/2005/07/005} {\bibfield
  {journal} {\bibinfo  {journal} {JCAP}\ }\textbf {\bibinfo {volume} {0507}},\
  \bibinfo {pages} {005} (\bibinfo {year} {2005})},\ \Eprint
  {http://arxiv.org/abs/astro-ph/0506373} {arXiv:astro-ph/0506373} \BibitemShut
  {NoStop}%
\bibitem [{\citenamefont {Di~Marco}\ and\ \citenamefont
  {Finelli}(2005)}]{DiMarco:2005nq}%
  \BibitemOpen
  \bibfield  {author} {\bibinfo {author} {\bibfnamefont {F.}~\bibnamefont
  {Di~Marco}}\ and\ \bibinfo {author} {\bibfnamefont {F.}~\bibnamefont
  {Finelli}},\ }\href {\doibase 10.1103/PhysRevD.71.123502} {\bibfield
  {journal} {\bibinfo  {journal} {Phys. Rev.}\ }\textbf {\bibinfo {volume}
  {D71}},\ \bibinfo {pages} {123502} (\bibinfo {year} {2005})},\ \Eprint
  {http://arxiv.org/abs/astro-ph/0505198} {arXiv:astro-ph/0505198} \BibitemShut
  {NoStop}%
\bibitem [{\citenamefont {Parkinson}\ \emph {et~al.}(2005)\citenamefont
  {Parkinson}, \citenamefont {Tsujikawa}, \citenamefont {Bassett},\ and\
  \citenamefont {Amendola}}]{Parkinson:2004yx}%
  \BibitemOpen
  \bibfield  {author} {\bibinfo {author} {\bibfnamefont {D.}~\bibnamefont
  {Parkinson}}, \bibinfo {author} {\bibfnamefont {S.}~\bibnamefont
  {Tsujikawa}}, \bibinfo {author} {\bibfnamefont {B.~A.}\ \bibnamefont
  {Bassett}}, \ and\ \bibinfo {author} {\bibfnamefont {L.}~\bibnamefont
  {Amendola}},\ }\href {\doibase 10.1103/PhysRevD.71.063524} {\bibfield
  {journal} {\bibinfo  {journal} {Phys. Rev.}\ }\textbf {\bibinfo {volume}
  {D71}},\ \bibinfo {pages} {063524} (\bibinfo {year} {2005})},\ \Eprint
  {http://arxiv.org/abs/astro-ph/0409071} {arXiv:astro-ph/0409071} \BibitemShut
  {NoStop}%
\bibitem [{\citenamefont {Gruzinov}(2004)}]{Gruzinov:2004ad}%
  \BibitemOpen
  \bibfield  {author} {\bibinfo {author} {\bibfnamefont {A.}~\bibnamefont
  {Gruzinov}},\ }\href@noop {} {\  (\bibinfo {year} {2004})},\ \Eprint
  {http://arxiv.org/abs/astro-ph/0401407} {arXiv:astro-ph/0401407} \BibitemShut
  {NoStop}%
\bibitem [{\citenamefont {Bartolo}\ \emph {et~al.}(2004)\citenamefont
  {Bartolo}, \citenamefont {Corasaniti}, \citenamefont {Liddle},\ and\
  \citenamefont {Malquarti}}]{Bartolo:2003ad}%
  \BibitemOpen
  \bibfield  {author} {\bibinfo {author} {\bibfnamefont {N.}~\bibnamefont
  {Bartolo}}, \bibinfo {author} {\bibfnamefont {P.~S.}\ \bibnamefont
  {Corasaniti}}, \bibinfo {author} {\bibfnamefont {A.~R.}\ \bibnamefont
  {Liddle}}, \ and\ \bibinfo {author} {\bibfnamefont {M.}~\bibnamefont
  {Malquarti}},\ }\href {\doibase 10.1103/PhysRevD.70.043532} {\bibfield
  {journal} {\bibinfo  {journal} {Phys. Rev.}\ }\textbf {\bibinfo {volume}
  {D70}},\ \bibinfo {pages} {043532} (\bibinfo {year} {2004})},\ \Eprint
  {http://arxiv.org/abs/astro-ph/0311503} {arXiv:astro-ph/0311503} \BibitemShut
  {NoStop}%
\bibitem [{\citenamefont {Vernizzi}(2004)}]{Vernizzi:2003vs}%
  \BibitemOpen
  \bibfield  {author} {\bibinfo {author} {\bibfnamefont {F.}~\bibnamefont
  {Vernizzi}},\ }\href {\doibase 10.1103/PhysRevD.69.083526} {\bibfield
  {journal} {\bibinfo  {journal} {Phys. Rev.}\ }\textbf {\bibinfo {volume}
  {D69}},\ \bibinfo {pages} {083526} (\bibinfo {year} {2004})},\ \Eprint
  {http://arxiv.org/abs/astro-ph/0311167} {arXiv:astro-ph/0311167} \BibitemShut
  {NoStop}%
\bibitem [{\citenamefont {Lee}\ and\ \citenamefont {Fang}(2004)}]{Lee:2003ed}%
  \BibitemOpen
  \bibfield  {author} {\bibinfo {author} {\bibfnamefont {W.}~\bibnamefont
  {Lee}}\ and\ \bibinfo {author} {\bibfnamefont {L.-Z.}\ \bibnamefont {Fang}},\
  }\href {\doibase 10.1103/PhysRevD.69.023514} {\bibfield  {journal} {\bibinfo
  {journal} {Phys. Rev.}\ }\textbf {\bibinfo {volume} {D69}},\ \bibinfo {pages}
  {023514} (\bibinfo {year} {2004})},\ \Eprint
  {http://arxiv.org/abs/astro-ph/0310856} {arXiv:astro-ph/0310856} \BibitemShut
  {NoStop}%
\bibitem [{\citenamefont {Mazumdar}\ and\ \citenamefont
  {Postma}(2003)}]{Mazumdar:2003iy}%
  \BibitemOpen
  \bibfield  {author} {\bibinfo {author} {\bibfnamefont {A.}~\bibnamefont
  {Mazumdar}}\ and\ \bibinfo {author} {\bibfnamefont {M.}~\bibnamefont
  {Postma}},\ }\href {\doibase 10.1016/j.physletb.2003.08.053} {\bibfield
  {journal} {\bibinfo  {journal} {Phys. Lett.}\ }\textbf {\bibinfo {volume}
  {B573}},\ \bibinfo {pages} {5} (\bibinfo {year} {2003})},\ \Eprint
  {http://arxiv.org/abs/astro-ph/0306509} {arXiv:astro-ph/0306509} \BibitemShut
  {NoStop}%
\bibitem [{\citenamefont {Malik}\ \emph {et~al.}(2003)\citenamefont {Malik},
  \citenamefont {Wands},\ and\ \citenamefont {Ungarelli}}]{Malik:2002jb}%
  \BibitemOpen
  \bibfield  {author} {\bibinfo {author} {\bibfnamefont {K.~A.}\ \bibnamefont
  {Malik}}, \bibinfo {author} {\bibfnamefont {D.}~\bibnamefont {Wands}}, \ and\
  \bibinfo {author} {\bibfnamefont {C.}~\bibnamefont {Ungarelli}},\ }\href
  {\doibase 10.1103/PhysRevD.67.063516} {\bibfield  {journal} {\bibinfo
  {journal} {Phys. Rev.}\ }\textbf {\bibinfo {volume} {D67}},\ \bibinfo {pages}
  {063516} (\bibinfo {year} {2003})},\ \Eprint
  {http://arxiv.org/abs/astro-ph/0211602} {arXiv:astro-ph/0211602} \BibitemShut
  {NoStop}%
\bibitem [{\citenamefont {Di~Marco}\ \emph {et~al.}(2003)\citenamefont
  {Di~Marco}, \citenamefont {Finelli},\ and\ \citenamefont
  {Brandenberger}}]{DiMarco:2002eb}%
  \BibitemOpen
  \bibfield  {author} {\bibinfo {author} {\bibfnamefont {F.}~\bibnamefont
  {Di~Marco}}, \bibinfo {author} {\bibfnamefont {F.}~\bibnamefont {Finelli}}, \
  and\ \bibinfo {author} {\bibfnamefont {R.}~\bibnamefont {Brandenberger}},\
  }\href {\doibase 10.1103/PhysRevD.67.063512} {\bibfield  {journal} {\bibinfo
  {journal} {Phys. Rev.}\ }\textbf {\bibinfo {volume} {D67}},\ \bibinfo {pages}
  {063512} (\bibinfo {year} {2003})},\ \Eprint
  {http://arxiv.org/abs/astro-ph/0211276} {arXiv:astro-ph/0211276} \BibitemShut
  {NoStop}%
\bibitem [{\citenamefont {Ashcroft}\ \emph {et~al.}(2004)\citenamefont
  {Ashcroft}, \citenamefont {van~de Bruck},\ and\ \citenamefont
  {Davis}}]{Ashcroft:2002vj}%
  \BibitemOpen
  \bibfield  {author} {\bibinfo {author} {\bibfnamefont {P.~R.}\ \bibnamefont
  {Ashcroft}}, \bibinfo {author} {\bibfnamefont {C.}~\bibnamefont {van~de
  Bruck}}, \ and\ \bibinfo {author} {\bibfnamefont {A.~C.}\ \bibnamefont
  {Davis}},\ }\href {\doibase 10.1103/PhysRevD.69.083516} {\bibfield  {journal}
  {\bibinfo  {journal} {Phys. Rev.}\ }\textbf {\bibinfo {volume} {D69}},\
  \bibinfo {pages} {083516} (\bibinfo {year} {2004})},\ \Eprint
  {http://arxiv.org/abs/astro-ph/0210597} {arXiv:astro-ph/0210597} \BibitemShut
  {NoStop}%
\bibitem [{\citenamefont {Bernardeau}\ and\ \citenamefont
  {Uzan}(2002)}]{Bernardeau:2002jy}%
  \BibitemOpen
  \bibfield  {author} {\bibinfo {author} {\bibfnamefont {F.}~\bibnamefont
  {Bernardeau}}\ and\ \bibinfo {author} {\bibfnamefont {J.-P.}\ \bibnamefont
  {Uzan}},\ }\href {\doibase 10.1103/PhysRevD.66.103506} {\bibfield  {journal}
  {\bibinfo  {journal} {Phys. Rev.}\ }\textbf {\bibinfo {volume} {D66}},\
  \bibinfo {pages} {103506} (\bibinfo {year} {2002})},\ \Eprint
  {http://arxiv.org/abs/hep-ph/0207295} {arXiv:hep-ph/0207295} \BibitemShut
  {NoStop}%
\bibitem [{\citenamefont {Tsujikawa}\ and\ \citenamefont
  {Bassett}(2002)}]{Tsujikawa:2002nf}%
  \BibitemOpen
  \bibfield  {author} {\bibinfo {author} {\bibfnamefont {S.}~\bibnamefont
  {Tsujikawa}}\ and\ \bibinfo {author} {\bibfnamefont {B.~A.}\ \bibnamefont
  {Bassett}},\ }\href {\doibase 10.1016/S0370-2693(02)01813-0} {\bibfield
  {journal} {\bibinfo  {journal} {Phys. Lett.}\ }\textbf {\bibinfo {volume}
  {B536}},\ \bibinfo {pages} {9} (\bibinfo {year} {2002})},\ \Eprint
  {http://arxiv.org/abs/astro-ph/0204031} {arXiv:astro-ph/0204031} \BibitemShut
  {NoStop}%
\bibitem [{\citenamefont {Starobinsky}\ \emph {et~al.}(2001)\citenamefont
  {Starobinsky}, \citenamefont {Tsujikawa},\ and\ \citenamefont
  {Yokoyama}}]{Starobinsky:2001xq}%
  \BibitemOpen
  \bibfield  {author} {\bibinfo {author} {\bibfnamefont {A.~A.}\ \bibnamefont
  {Starobinsky}}, \bibinfo {author} {\bibfnamefont {S.}~\bibnamefont
  {Tsujikawa}}, \ and\ \bibinfo {author} {\bibfnamefont {J.}~\bibnamefont
  {Yokoyama}},\ }\href {\doibase 10.1016/S0550-3213(01)00322-4} {\bibfield
  {journal} {\bibinfo  {journal} {Nucl. Phys.}\ }\textbf {\bibinfo {volume}
  {B610}},\ \bibinfo {pages} {383} (\bibinfo {year} {2001})},\ \Eprint
  {http://arxiv.org/abs/astro-ph/0107555} {arXiv:astro-ph/0107555} \BibitemShut
  {NoStop}%
\bibitem [{\citenamefont {Groot~Nibbelink}\ and\ \citenamefont {van
  Tent}(2002)}]{GrootNibbelink:2001qt}%
  \BibitemOpen
  \bibfield  {author} {\bibinfo {author} {\bibfnamefont {S.}~\bibnamefont
  {Groot~Nibbelink}}\ and\ \bibinfo {author} {\bibfnamefont {B.~J.~W.}\
  \bibnamefont {van Tent}},\ }\href {\doibase 10.1088/0264-9381/19/4/302}
  {\bibfield  {journal} {\bibinfo  {journal} {Class. Quant. Grav.}\ }\textbf
  {\bibinfo {volume} {19}},\ \bibinfo {pages} {613} (\bibinfo {year} {2002})},\
  \Eprint {http://arxiv.org/abs/hep-ph/0107272} {arXiv:hep-ph/0107272}
  \BibitemShut {NoStop}%
\bibitem [{\citenamefont {Bartolo}\ \emph
  {et~al.}(2001{\natexlab{b}})\citenamefont {Bartolo}, \citenamefont
  {Matarrese},\ and\ \citenamefont {Riotto}}]{Bartolo:2001vw}%
  \BibitemOpen
  \bibfield  {author} {\bibinfo {author} {\bibfnamefont {N.}~\bibnamefont
  {Bartolo}}, \bibinfo {author} {\bibfnamefont {S.}~\bibnamefont {Matarrese}},
  \ and\ \bibinfo {author} {\bibfnamefont {A.}~\bibnamefont {Riotto}},\ }\href
  {\doibase 10.1103/PhysRevD.64.083514} {\bibfield  {journal} {\bibinfo
  {journal} {Phys. Rev.}\ }\textbf {\bibinfo {volume} {D64}},\ \bibinfo {pages}
  {083514} (\bibinfo {year} {2001}{\natexlab{b}})},\ \Eprint
  {http://arxiv.org/abs/astro-ph/0106022} {arXiv:astro-ph/0106022} \BibitemShut
  {NoStop}%
\bibitem [{\citenamefont {Hwang}\ and\ \citenamefont
  {Noh}(2002)}]{Hwang:2001fb}%
  \BibitemOpen
  \bibfield  {author} {\bibinfo {author} {\bibfnamefont {J.-c.}\ \bibnamefont
  {Hwang}}\ and\ \bibinfo {author} {\bibfnamefont {H.}~\bibnamefont {Noh}},\
  }\href {\doibase 10.1088/0264-9381/19/3/308} {\bibfield  {journal} {\bibinfo
  {journal} {Class. Quant. Grav.}\ }\textbf {\bibinfo {volume} {19}},\ \bibinfo
  {pages} {527} (\bibinfo {year} {2002})},\ \Eprint
  {http://arxiv.org/abs/astro-ph/0103244} {arXiv:astro-ph/0103244} \BibitemShut
  {NoStop}%
\bibitem [{\citenamefont {Hwang}\ and\ \citenamefont
  {Noh}(2000)}]{Hwang:2000jh}%
  \BibitemOpen
  \bibfield  {author} {\bibinfo {author} {\bibfnamefont {J.-c.}\ \bibnamefont
  {Hwang}}\ and\ \bibinfo {author} {\bibfnamefont {H.}~\bibnamefont {Noh}},\
  }\href {\doibase 10.1016/S0370-2693(00)01253-3} {\bibfield  {journal}
  {\bibinfo  {journal} {Phys. Lett.}\ }\textbf {\bibinfo {volume} {B495}},\
  \bibinfo {pages} {277} (\bibinfo {year} {2000})},\ \Eprint
  {http://arxiv.org/abs/astro-ph/0009268} {arXiv:astro-ph/0009268} \BibitemShut
  {NoStop}%
\bibitem [{\citenamefont {Gordon}\ \emph {et~al.}(2001)\citenamefont {Gordon},
  \citenamefont {Wands}, \citenamefont {Bassett},\ and\ \citenamefont
  {Maartens}}]{Gordon:2000hv}%
  \BibitemOpen
  \bibfield  {author} {\bibinfo {author} {\bibfnamefont {C.}~\bibnamefont
  {Gordon}}, \bibinfo {author} {\bibfnamefont {D.}~\bibnamefont {Wands}},
  \bibinfo {author} {\bibfnamefont {B.~A.}\ \bibnamefont {Bassett}}, \ and\
  \bibinfo {author} {\bibfnamefont {R.}~\bibnamefont {Maartens}},\ }\href
  {\doibase 10.1103/PhysRevD.63.023506} {\bibfield  {journal} {\bibinfo
  {journal} {Phys. Rev.}\ }\textbf {\bibinfo {volume} {D63}},\ \bibinfo {pages}
  {023506} (\bibinfo {year} {2001})},\ \Eprint
  {http://arxiv.org/abs/astro-ph/0009131} {arXiv:astro-ph/0009131} \BibitemShut
  {NoStop}%
\bibitem [{\citenamefont {Taylor}\ and\ \citenamefont
  {Berera}(2000)}]{Taylor:2000ze}%
  \BibitemOpen
  \bibfield  {author} {\bibinfo {author} {\bibfnamefont {A.~N.}\ \bibnamefont
  {Taylor}}\ and\ \bibinfo {author} {\bibfnamefont {A.}~\bibnamefont
  {Berera}},\ }\href {\doibase 10.1103/PhysRevD.62.083517} {\bibfield
  {journal} {\bibinfo  {journal} {Phys. Rev.}\ }\textbf {\bibinfo {volume}
  {D62}},\ \bibinfo {pages} {083517} (\bibinfo {year} {2000})},\ \Eprint
  {http://arxiv.org/abs/astro-ph/0006077} {arXiv:astro-ph/0006077} \BibitemShut
  {NoStop}%
\bibitem [{\citenamefont {Finelli}\ and\ \citenamefont
  {Brandenberger}(2000)}]{Finelli:2000ya}%
  \BibitemOpen
  \bibfield  {author} {\bibinfo {author} {\bibfnamefont {F.}~\bibnamefont
  {Finelli}}\ and\ \bibinfo {author} {\bibfnamefont {R.~H.}\ \bibnamefont
  {Brandenberger}},\ }\href {\doibase 10.1103/PhysRevD.62.083502} {\bibfield
  {journal} {\bibinfo  {journal} {Phys. Rev.}\ }\textbf {\bibinfo {volume}
  {D62}},\ \bibinfo {pages} {083502} (\bibinfo {year} {2000})},\ \Eprint
  {http://arxiv.org/abs/hep-ph/0003172} {arXiv:hep-ph/0003172} \BibitemShut
  {NoStop}%
\bibitem [{\citenamefont {Liddle}\ and\ \citenamefont
  {Mazumdar}(2000)}]{Liddle:1999pr}%
  \BibitemOpen
  \bibfield  {author} {\bibinfo {author} {\bibfnamefont {A.~R.}\ \bibnamefont
  {Liddle}}\ and\ \bibinfo {author} {\bibfnamefont {A.}~\bibnamefont
  {Mazumdar}},\ }\href {\doibase 10.1103/PhysRevD.61.123507} {\bibfield
  {journal} {\bibinfo  {journal} {Phys. Rev.}\ }\textbf {\bibinfo {volume}
  {D61}},\ \bibinfo {pages} {123507} (\bibinfo {year} {2000})},\ \Eprint
  {http://arxiv.org/abs/astro-ph/9912349} {arXiv:astro-ph/9912349} \BibitemShut
  {NoStop}%
\bibitem [{\citenamefont {Bassett}\ \emph {et~al.}(2000)\citenamefont
  {Bassett}, \citenamefont {Gordon}, \citenamefont {Maartens},\ and\
  \citenamefont {Kaiser}}]{Bassett:1999ta}%
  \BibitemOpen
  \bibfield  {author} {\bibinfo {author} {\bibfnamefont {B.~A.}\ \bibnamefont
  {Bassett}}, \bibinfo {author} {\bibfnamefont {C.}~\bibnamefont {Gordon}},
  \bibinfo {author} {\bibfnamefont {R.}~\bibnamefont {Maartens}}, \ and\
  \bibinfo {author} {\bibfnamefont {D.~I.}\ \bibnamefont {Kaiser}},\ }\href
  {\doibase 10.1103/PhysRevD.61.061302} {\bibfield  {journal} {\bibinfo
  {journal} {Phys. Rev.}\ }\textbf {\bibinfo {volume} {D61}},\ \bibinfo {pages}
  {061302} (\bibinfo {year} {2000})},\ \Eprint
  {http://arxiv.org/abs/hep-ph/9909482} {arXiv:hep-ph/9909482} \BibitemShut
  {NoStop}%
\bibitem [{\citenamefont {Pierpaoli}\ \emph {et~al.}(1999)\citenamefont
  {Pierpaoli}, \citenamefont {Garcia-Bellido},\ and\ \citenamefont
  {Borgani}}]{Pierpaoli:1999zj}%
  \BibitemOpen
  \bibfield  {author} {\bibinfo {author} {\bibfnamefont {E.}~\bibnamefont
  {Pierpaoli}}, \bibinfo {author} {\bibfnamefont {J.}~\bibnamefont
  {Garcia-Bellido}}, \ and\ \bibinfo {author} {\bibfnamefont {S.}~\bibnamefont
  {Borgani}},\ }\href@noop {} {\bibfield  {journal} {\bibinfo  {journal}
  {JHEP}\ }\textbf {\bibinfo {volume} {10}},\ \bibinfo {pages} {015} (\bibinfo
  {year} {1999})},\ \Eprint {http://arxiv.org/abs/hep-ph/9909420}
  {arXiv:hep-ph/9909420} \BibitemShut {NoStop}%
\bibitem [{\citenamefont {Langlois}(1999)}]{Langlois:1999dw}%
  \BibitemOpen
  \bibfield  {author} {\bibinfo {author} {\bibfnamefont {D.}~\bibnamefont
  {Langlois}},\ }\href {\doibase 10.1103/PhysRevD.59.123512} {\bibfield
  {journal} {\bibinfo  {journal} {Phys. Rev.}\ }\textbf {\bibinfo {volume}
  {D59}},\ \bibinfo {pages} {123512} (\bibinfo {year} {1999})},\ \Eprint
  {http://arxiv.org/abs/astro-ph/9906080} {arXiv:astro-ph/9906080} \BibitemShut
  {NoStop}%
\bibitem [{\citenamefont {Felder}\ \emph {et~al.}(1999)\citenamefont {Felder},
  \citenamefont {Kofman},\ and\ \citenamefont {Linde}}]{Felder:1999pv}%
  \BibitemOpen
  \bibfield  {author} {\bibinfo {author} {\bibfnamefont {G.~N.}\ \bibnamefont
  {Felder}}, \bibinfo {author} {\bibfnamefont {L.}~\bibnamefont {Kofman}}, \
  and\ \bibinfo {author} {\bibfnamefont {A.~D.}\ \bibnamefont {Linde}},\ }\href
  {\doibase 10.1103/PhysRevD.60.103505} {\bibfield  {journal} {\bibinfo
  {journal} {Phys. Rev.}\ }\textbf {\bibinfo {volume} {D60}},\ \bibinfo {pages}
  {103505} (\bibinfo {year} {1999})},\ \Eprint
  {http://arxiv.org/abs/hep-ph/9903350} {arXiv:hep-ph/9903350} \BibitemShut
  {NoStop}%
\bibitem [{\citenamefont {Perrotta}\ and\ \citenamefont
  {Baccigalupi}(1999)}]{Perrotta:1998vf}%
  \BibitemOpen
  \bibfield  {author} {\bibinfo {author} {\bibfnamefont {F.}~\bibnamefont
  {Perrotta}}\ and\ \bibinfo {author} {\bibfnamefont {C.}~\bibnamefont
  {Baccigalupi}},\ }\href {\doibase 10.1103/PhysRevD.59.123508} {\bibfield
  {journal} {\bibinfo  {journal} {Phys. Rev.}\ }\textbf {\bibinfo {volume}
  {D59}},\ \bibinfo {pages} {123508} (\bibinfo {year} {1999})},\ \Eprint
  {http://arxiv.org/abs/astro-ph/9811156} {arXiv:astro-ph/9811156} \BibitemShut
  {NoStop}%
\bibitem [{\citenamefont {Chiba}\ \emph {et~al.}(1998)\citenamefont {Chiba},
  \citenamefont {Sugiyama},\ and\ \citenamefont {Yokoyama}}]{Chiba:1997ij}%
  \BibitemOpen
  \bibfield  {author} {\bibinfo {author} {\bibfnamefont {T.}~\bibnamefont
  {Chiba}}, \bibinfo {author} {\bibfnamefont {N.}~\bibnamefont {Sugiyama}}, \
  and\ \bibinfo {author} {\bibfnamefont {J.}~\bibnamefont {Yokoyama}},\ }\href
  {\doibase 10.1016/S0550-3213(98)00412-X} {\bibfield  {journal} {\bibinfo
  {journal} {Nucl. Phys.}\ }\textbf {\bibinfo {volume} {B530}},\ \bibinfo
  {pages} {304} (\bibinfo {year} {1998})},\ \Eprint
  {http://arxiv.org/abs/gr-qc/9708030} {arXiv:gr-qc/9708030} \BibitemShut
  {NoStop}%
\bibitem [{\citenamefont {Nakamura}\ and\ \citenamefont
  {Stewart}(1996)}]{Nakamura:1996da}%
  \BibitemOpen
  \bibfield  {author} {\bibinfo {author} {\bibfnamefont {T.~T.}\ \bibnamefont
  {Nakamura}}\ and\ \bibinfo {author} {\bibfnamefont {E.~D.}\ \bibnamefont
  {Stewart}},\ }\href {\doibase 10.1016/0370-2693(96)00594-1} {\bibfield
  {journal} {\bibinfo  {journal} {Phys. Lett.}\ }\textbf {\bibinfo {volume}
  {B381}},\ \bibinfo {pages} {413} (\bibinfo {year} {1996})},\ \Eprint
  {http://arxiv.org/abs/astro-ph/9604103} {arXiv:astro-ph/9604103} \BibitemShut
  {NoStop}%
\bibitem [{\citenamefont {Polarski}\ and\ \citenamefont
  {Starobinsky}(1994)}]{Polarski:1994rz}%
  \BibitemOpen
  \bibfield  {author} {\bibinfo {author} {\bibfnamefont {D.}~\bibnamefont
  {Polarski}}\ and\ \bibinfo {author} {\bibfnamefont {A.~A.}\ \bibnamefont
  {Starobinsky}},\ }\href {\doibase 10.1103/PhysRevD.50.6123} {\bibfield
  {journal} {\bibinfo  {journal} {Phys. Rev.}\ }\textbf {\bibinfo {volume}
  {D50}},\ \bibinfo {pages} {6123} (\bibinfo {year} {1994})},\ \Eprint
  {http://arxiv.org/abs/astro-ph/9404061} {arXiv:astro-ph/9404061} \BibitemShut
  {NoStop}%
\bibitem [{\citenamefont {Lyth}\ \emph {et~al.}(2003)\citenamefont {Lyth},
  \citenamefont {Ungarelli},\ and\ \citenamefont {Wands}}]{Lyth:2002my}%
  \BibitemOpen
  \bibfield  {author} {\bibinfo {author} {\bibfnamefont {D.~H.}\ \bibnamefont
  {Lyth}}, \bibinfo {author} {\bibfnamefont {C.}~\bibnamefont {Ungarelli}}, \
  and\ \bibinfo {author} {\bibfnamefont {D.}~\bibnamefont {Wands}},\ }\href
  {\doibase 10.1103/PhysRevD.67.023503} {\bibfield  {journal} {\bibinfo
  {journal} {Phys. Rev.}\ }\textbf {\bibinfo {volume} {D67}},\ \bibinfo {pages}
  {023503} (\bibinfo {year} {2003})},\ \Eprint
  {http://arxiv.org/abs/astro-ph/0208055} {arXiv:astro-ph/0208055} \BibitemShut
  {NoStop}%
\bibitem [{\citenamefont {Gordon}\ and\ \citenamefont
  {Malik}(2004)}]{Gordon:2003hw}%
  \BibitemOpen
  \bibfield  {author} {\bibinfo {author} {\bibfnamefont {C.}~\bibnamefont
  {Gordon}}\ and\ \bibinfo {author} {\bibfnamefont {K.~A.}\ \bibnamefont
  {Malik}},\ }\href {\doibase 10.1103/PhysRevD.69.063508} {\bibfield  {journal}
  {\bibinfo  {journal} {Phys. Rev.}\ }\textbf {\bibinfo {volume} {D69}},\
  \bibinfo {pages} {063508} (\bibinfo {year} {2004})},\ \Eprint
  {http://arxiv.org/abs/astro-ph/0311102} {arXiv:astro-ph/0311102} \BibitemShut
  {NoStop}%
\bibitem [{\citenamefont {Kawasaki}\ \emph {et~al.}(2012)\citenamefont
  {Kawasaki}, \citenamefont {Miyamoto}, \citenamefont {Nakayama},\ and\
  \citenamefont {Sekiguchi}}]{Kawasaki:2011rc}%
  \BibitemOpen
  \bibfield  {author} {\bibinfo {author} {\bibfnamefont {M.}~\bibnamefont
  {Kawasaki}}, \bibinfo {author} {\bibfnamefont {K.}~\bibnamefont {Miyamoto}},
  \bibinfo {author} {\bibfnamefont {K.}~\bibnamefont {Nakayama}}, \ and\
  \bibinfo {author} {\bibfnamefont {T.}~\bibnamefont {Sekiguchi}},\ }\href
  {\doibase 10.1088/1475-7516/2012/02/022} {\bibfield  {journal} {\bibinfo
  {journal} {JCAP}\ }\textbf {\bibinfo {volume} {1202}},\ \bibinfo {pages}
  {022} (\bibinfo {year} {2012})},\ \Eprint {http://arxiv.org/abs/1107.4962}
  {arXiv:1107.4962 [astro-ph.CO]} \BibitemShut {NoStop}%
\bibitem [{\citenamefont {Di~Valentino}\ \emph {et~al.}(2012)\citenamefont
  {Di~Valentino}, \citenamefont {Lattanzi}, \citenamefont {Mangano},
  \citenamefont {Melchiorri},\ and\ \citenamefont
  {Serpico}}]{DiValentino:2011sv}%
  \BibitemOpen
  \bibfield  {author} {\bibinfo {author} {\bibfnamefont {E.}~\bibnamefont
  {Di~Valentino}}, \bibinfo {author} {\bibfnamefont {M.}~\bibnamefont
  {Lattanzi}}, \bibinfo {author} {\bibfnamefont {G.}~\bibnamefont {Mangano}},
  \bibinfo {author} {\bibfnamefont {A.}~\bibnamefont {Melchiorri}}, \ and\
  \bibinfo {author} {\bibfnamefont {P.}~\bibnamefont {Serpico}},\ }\href
  {\doibase 10.1103/PhysRevD.85.043511} {\bibfield  {journal} {\bibinfo
  {journal} {Phys.Rev.}\ }\textbf {\bibinfo {volume} {D85}},\ \bibinfo {pages}
  {043511} (\bibinfo {year} {2012})},\ \Eprint {http://arxiv.org/abs/1111.3810}
  {arXiv:1111.3810 [astro-ph.CO]} \BibitemShut {NoStop}%
\bibitem [{\citenamefont {Orito}\ \emph {et~al.}(2002)\citenamefont {Orito},
  \citenamefont {Kajino}, \citenamefont {Mathews},\ and\ \citenamefont
  {Wang}}]{Orito:2002hf}%
  \BibitemOpen
  \bibfield  {author} {\bibinfo {author} {\bibfnamefont {M.}~\bibnamefont
  {Orito}}, \bibinfo {author} {\bibfnamefont {T.}~\bibnamefont {Kajino}},
  \bibinfo {author} {\bibfnamefont {G.}~\bibnamefont {Mathews}}, \ and\
  \bibinfo {author} {\bibfnamefont {Y.}~\bibnamefont {Wang}},\ }\href {\doibase
  10.1103/PhysRevD.65.123504} {\bibfield  {journal} {\bibinfo  {journal}
  {Phys.Rev.}\ }\textbf {\bibinfo {volume} {D65}},\ \bibinfo {pages} {123504}
  (\bibinfo {year} {2002})},\ \Eprint {http://arxiv.org/abs/astro-ph/0203352}
  {arXiv:astro-ph/0203352 [astro-ph]} \BibitemShut {NoStop}%
\bibitem [{\citenamefont {Ade}\ \emph {et~al.}(2013{\natexlab{b}})\citenamefont
  {Ade} \emph {et~al.}}]{Ade:2013ydc}%
  \BibitemOpen
  \bibfield  {author} {\bibinfo {author} {\bibfnamefont {P.}~\bibnamefont
  {Ade}} \emph {et~al.} (\bibinfo {collaboration} {Planck Collaboration}),\
  }\href@noop {} {\  (\bibinfo {year} {2013}{\natexlab{b}})},\ \Eprint
  {http://arxiv.org/abs/1303.5084} {arXiv:1303.5084 [astro-ph.CO]} \BibitemShut
  {NoStop}%
\bibitem [{\citenamefont {Feroz}\ \emph {et~al.}(2009)\citenamefont {Feroz},
  \citenamefont {Hobson},\ and\ \citenamefont {Bridges}}]{Feroz:2008xx}%
  \BibitemOpen
  \bibfield  {author} {\bibinfo {author} {\bibfnamefont {F.}~\bibnamefont
  {Feroz}}, \bibinfo {author} {\bibfnamefont {M.}~\bibnamefont {Hobson}}, \
  and\ \bibinfo {author} {\bibfnamefont {M.}~\bibnamefont {Bridges}},\ }\href
  {\doibase 10.1111/j.1365-2966.2009.14548.x} {\bibfield  {journal} {\bibinfo
  {journal} {Mon.Not.Roy.Astron.Soc.}\ }\textbf {\bibinfo {volume} {398}},\
  \bibinfo {pages} {1601} (\bibinfo {year} {2009})},\ \Eprint
  {http://arxiv.org/abs/0809.3437} {arXiv:0809.3437 [astro-ph]} \BibitemShut
  {NoStop}%
\bibitem [{Mul()}]{MultiNest}%
  \BibitemOpen
  \href@noop {} {}\bibinfo {note}
  {{h}ttp://ccpforge.cse.rl.ac.uk/gf/project/multinest/}\BibitemShut {NoStop}%
\bibitem [{\citenamefont {Skilling}(2004)}]{Skilling:2004}%
  \BibitemOpen
  \bibfield  {author} {\bibinfo {author} {\bibfnamefont {J.}~\bibnamefont
  {Skilling}},\ }\href@noop {} {\bibfield  {journal} {\bibinfo  {journal} {AIP
  conference proceedings of the 24th international workshop on Bayesian
  inference and maximum entropy methods in science and engineering}\ }
  (\bibinfo {year} {2004})}\BibitemShut {NoStop}%
\bibitem [{\citenamefont {Mukherjee}\ \emph {et~al.}(2006)\citenamefont
  {Mukherjee}, \citenamefont {Parkinson},\ and\ \citenamefont
  {Liddle}}]{Mukherjee:2005wg}%
  \BibitemOpen
  \bibfield  {author} {\bibinfo {author} {\bibfnamefont {P.}~\bibnamefont
  {Mukherjee}}, \bibinfo {author} {\bibfnamefont {D.}~\bibnamefont
  {Parkinson}}, \ and\ \bibinfo {author} {\bibfnamefont {A.~R.}\ \bibnamefont
  {Liddle}},\ }\href@noop {} {\bibfield  {journal} {\bibinfo  {journal}
  {Astrophys. J.}\ }\textbf {\bibinfo {volume} {638}},\ \bibinfo {pages} {L51}
  (\bibinfo {year} {2006})},\ \Eprint {http://arxiv.org/abs/astro-ph/0508461}
  {arXiv:astro-ph/0508461} \BibitemShut {NoStop}%
\bibitem [{\citenamefont {Shaw}\ \emph {et~al.}(2007)\citenamefont {Shaw},
  \citenamefont {Bridges},\ and\ \citenamefont {Hobson}}]{Shaw:2007jj}%
  \BibitemOpen
  \bibfield  {author} {\bibinfo {author} {\bibfnamefont {R.}~\bibnamefont
  {Shaw}}, \bibinfo {author} {\bibfnamefont {M.}~\bibnamefont {Bridges}}, \
  and\ \bibinfo {author} {\bibfnamefont {M.~P.}\ \bibnamefont {Hobson}},\
  }\href {\doibase 10.1111/j.1365-2966.2007.11871.x} {\bibfield  {journal}
  {\bibinfo  {journal} {Mon. Not. Roy. Astron. Soc.}\ }\textbf {\bibinfo
  {volume} {378}},\ \bibinfo {pages} {1365} (\bibinfo {year} {2007})},\ \Eprint
  {http://arxiv.org/abs/astro-ph/0701867} {arXiv:astro-ph/0701867} \BibitemShut
  {NoStop}%
\bibitem [{\citenamefont {Feroz}\ and\ \citenamefont
  {Hobson}(2008)}]{Feroz:2007kg}%
  \BibitemOpen
  \bibfield  {author} {\bibinfo {author} {\bibfnamefont {F.}~\bibnamefont
  {Feroz}}\ and\ \bibinfo {author} {\bibfnamefont {M.}~\bibnamefont {Hobson}},\
  }\href {\doibase 10.1111/j.1365-2966.2007.12353.x} {\bibfield  {journal}
  {\bibinfo  {journal} {Mon.Not.Roy.Astron.Soc.}\ }\textbf {\bibinfo {volume}
  {384}},\ \bibinfo {pages} {449} (\bibinfo {year} {2008})},\ \Eprint
  {http://arxiv.org/abs/0704.3704} {arXiv:0704.3704 [astro-ph]} \BibitemShut
  {NoStop}%
\bibitem [{\citenamefont {Lewis}\ \emph {et~al.}(2000)\citenamefont {Lewis},
  \citenamefont {Challinor},\ and\ \citenamefont {Lasenby}}]{CAMB}%
  \BibitemOpen
  \bibfield  {author} {\bibinfo {author} {\bibfnamefont {A.}~\bibnamefont
  {Lewis}}, \bibinfo {author} {\bibfnamefont {A.}~\bibnamefont {Challinor}}, \
  and\ \bibinfo {author} {\bibfnamefont {A.}~\bibnamefont {Lasenby}},\
  }\href@noop {} {\bibfield  {journal} {\bibinfo  {journal} {Astrophys. J.}\
  }\textbf {\bibinfo {volume} {538}},\ \bibinfo {pages} {473} (\bibinfo {year}
  {2000})},\ \Eprint {http://arxiv.org/abs/astro-ph/9911177} {astro-ph/9911177}
  \BibitemShut {NoStop}%
\bibitem [{\citenamefont {Lewis}\ and\ \citenamefont {Bridle}(2002)}]{COSMOMC}%
  \BibitemOpen
  \bibfield  {author} {\bibinfo {author} {\bibfnamefont {A.}~\bibnamefont
  {Lewis}}\ and\ \bibinfo {author} {\bibfnamefont {S.}~\bibnamefont {Bridle}},\
  }\href@noop {} {\bibfield  {journal} {\bibinfo  {journal} {Phys. Rev.}\
  }\textbf {\bibinfo {volume} {D66}},\ \bibinfo {pages} {103511} (\bibinfo
  {year} {2002})},\ \Eprint {http://arxiv.org/abs/astro-ph/0205436}
  {astro-ph/0205436} \BibitemShut {NoStop}%
\bibitem [{\citenamefont {Bucher}\ \emph {et~al.}(2002)\citenamefont {Bucher},
  \citenamefont {Moodley},\ and\ \citenamefont {Turok}}]{Bucher:2000kb}%
  \BibitemOpen
  \bibfield  {author} {\bibinfo {author} {\bibfnamefont {M.}~\bibnamefont
  {Bucher}}, \bibinfo {author} {\bibfnamefont {K.}~\bibnamefont {Moodley}}, \
  and\ \bibinfo {author} {\bibfnamefont {N.}~\bibnamefont {Turok}},\ }\href
  {\doibase 10.1103/PhysRevD.66.023528} {\bibfield  {journal} {\bibinfo
  {journal} {Phys. Rev.}\ }\textbf {\bibinfo {volume} {D66}},\ \bibinfo {pages}
  {023528} (\bibinfo {year} {2002})},\ \Eprint
  {http://arxiv.org/abs/astro-ph/0007360} {arXiv:astro-ph/0007360} \BibitemShut
  {NoStop}%
\bibitem [{\citenamefont {Bucher}\ \emph {et~al.}(2001)\citenamefont {Bucher},
  \citenamefont {Moodley},\ and\ \citenamefont {Turok}}]{Bucher:2000hy}%
  \BibitemOpen
  \bibfield  {author} {\bibinfo {author} {\bibfnamefont {M.}~\bibnamefont
  {Bucher}}, \bibinfo {author} {\bibfnamefont {K.}~\bibnamefont {Moodley}}, \
  and\ \bibinfo {author} {\bibfnamefont {N.}~\bibnamefont {Turok}},\ }\href
  {\doibase 10.1103/PhysRevLett.87.191301} {\bibfield  {journal} {\bibinfo
  {journal} {Phys. Rev. Lett.}\ }\textbf {\bibinfo {volume} {87}},\ \bibinfo
  {pages} {191301} (\bibinfo {year} {2001})},\ \Eprint
  {http://arxiv.org/abs/astro-ph/0012141} {arXiv:astro-ph/0012141} \BibitemShut
  {NoStop}%
\bibitem [{\citenamefont {Kasanda}\ \emph {et~al.}(2012)\citenamefont
  {Kasanda}, \citenamefont {Zunckel}, \citenamefont {Moodley}, \citenamefont
  {Bassett},\ and\ \citenamefont {Okouma}}]{Kasanda:2011np}%
  \BibitemOpen
  \bibfield  {author} {\bibinfo {author} {\bibfnamefont {S.~M.}\ \bibnamefont
  {Kasanda}}, \bibinfo {author} {\bibfnamefont {C.}~\bibnamefont {Zunckel}},
  \bibinfo {author} {\bibfnamefont {K.}~\bibnamefont {Moodley}}, \bibinfo
  {author} {\bibfnamefont {B.}~\bibnamefont {Bassett}}, \ and\ \bibinfo
  {author} {\bibfnamefont {P.}~\bibnamefont {Okouma}},\ }\href {\doibase
  10.1088/1475-7516/2012/07/021} {\bibfield  {journal} {\bibinfo  {journal}
  {JCAP}\ }\textbf {\bibinfo {volume} {1207}},\ \bibinfo {pages} {021}
  (\bibinfo {year} {2012})},\ \Eprint {http://arxiv.org/abs/1111.2572}
  {arXiv:1111.2572 [astro-ph.CO]} \BibitemShut {NoStop}%
\bibitem [{\citenamefont {Zunckel}\ \emph {et~al.}(2011)\citenamefont
  {Zunckel}, \citenamefont {Okouma}, \citenamefont {Kasanda}, \citenamefont
  {Moodley},\ and\ \citenamefont {Bassett}}]{Zunckel:2010mm}%
  \BibitemOpen
  \bibfield  {author} {\bibinfo {author} {\bibfnamefont {C.}~\bibnamefont
  {Zunckel}}, \bibinfo {author} {\bibfnamefont {P.}~\bibnamefont {Okouma}},
  \bibinfo {author} {\bibfnamefont {S.}~\bibnamefont {Kasanda}}, \bibinfo
  {author} {\bibfnamefont {K.}~\bibnamefont {Moodley}}, \ and\ \bibinfo
  {author} {\bibfnamefont {B.}~\bibnamefont {Bassett}},\ }\href {\doibase
  10.1016/j.physletb.2011.01.012} {\bibfield  {journal} {\bibinfo  {journal}
  {Phys.Lett.}\ }\textbf {\bibinfo {volume} {B696}},\ \bibinfo {pages} {433}
  (\bibinfo {year} {2011})},\ \Eprint {http://arxiv.org/abs/1006.4687}
  {arXiv:1006.4687 [astro-ph.CO]} \BibitemShut {NoStop}%
\end{thebibliography}%

\onecolumngrid

\appendix

\section{Tables}
\label{app:Tables1}

\begin{table*}[b]
\scriptsize
\centering
\begin{tabular}{|c|c|c|c|c|c|c|}
\hline 
      Parameter 	&	          C.L 	& Mixed NDI, general corr. 		&	 Mixed NDI,  $\gamma=0$ 	&	  Mixed NDI,  $\gamma=1$ & 	Mixed  NDI, $\gamma=-1$	  &	   Adiabatic    \\ \hline
    
   $   \omega_b $ &         68\% & 0.0239 (0.0233, 0.0246)&  0.0239 (0.0232, 0.0247) & 0.0233 (0.0227, 0.0238) & 0.0224 (0.0219, 0.0229) & 0.0231 (0.0225, 0.0237)     \\ \hline 
$ \omega_{c} $ &         68\% & 0.1054 (0.1004, 0.1104)&  0.1049 (0.0995, 0.1102) & 0.1094 (0.1048, 0.1140) & 0.1145 (0.1102, 0.1191) & 0.1093 (0.1040, 0.1146)     \\ \hline 
$   100\theta$ &         68\% & 1.0488 (1.0447, 1.0526)&  1.0452 (1.0425, 1.0482) & 1.0457 (1.0428, 1.0486) & 1.0389 (1.0364, 1.0411) & 1.0419 (1.0398, 1.0440)     \\ \hline 
$        \tau$ &         68\% & 0.0896 (0.0753, 0.1035)&  0.0911 (0.0776, 0.1065) & 0.0872 (0.0746, 0.1011) & 0.0884 (0.0756, 0.1023) & 0.0899 (0.0757, 0.1045)     \\ \hline 
$\Omega_\Lambda$ &         68\% & 0.7852 (0.7586, 0.8090)&  0.7800 (0.7525, 0.8070) & 0.7602 (0.7346, 0.7835) & 0.7147 (0.6865, 0.7388) & 0.7508 (0.7221, 0.7769)     \\ \hline 
$         H_0$ &         68\% & 77.56 (74.31, 81.01)&  76.49 (73.35, 80.16) & 74.35 (71.69, 77.11) & 69.28 (67.02, 71.47) & 72.95 (70.27, 75.69)     \\ \hline 
$\ln[10^{10} A_0^2]$ &         68\% & 3.0713 (3.0260, 3.1323)&  3.2032 (3.1360, 3.2840) & 3.0717 (3.0342, 3.1120) & 3.1649 (3.1300, 3.2067) & 3.1038 (3.0641, 3.1422)     \\ \hline 
$n_\textrm{ad}^\textrm{eff}$ &         68\% & 0.9835 (0.9640, 1.0042)&  0.9997 (0.9811, 1.0204) & 0.9850 (0.9718, 0.9992) & 0.9638 (0.9509, 0.9769) & 0.9859 (0.9685, 1.0041)     \\ \hline 
$    \gamma_1$ &         95\% & $>$ -0.1137&  & & &     \\ \hline 
$    \gamma_2$ &         95\% & $>$ -0.0894&  & & &     \\ \hline 
$    \alpha_1$ &         95\% & $<$ 0.0980&  $<$ 0.2421	 & $<$ 0.0303	 & $<$ 0.0093&     \\ \hline 
$    \alpha_2$ &         95\% & $<$ 0.2713&  $<$ 0.4003	 & $<$ 0.0303	 & $<$ 0.0093	 &     \\ \hline 
$ \tilde{r}_0$ &         95\% & $<$ 0.4751&  $<$ 0.2866	 & & & $<$ 0.3334	      \\ \hline 
$    \gamma_0$ &         95\% & $>$ -0.1004&  & & &     \\ \hline 
$    \alpha_0$ &         95\% & $<$ 0.1414&  $<$ 0.2816	 & $<$ 0.0303	 & $<$ 0.0093	 &     \\ \hline 
$n_\textrm{ar}$ &         95\% & 0.9846 (0.8025, 1.1361)&  0.9997 (0.9651, 1.0427) & & & 0.9859 (0.9545, 1.0236)      \\ \hline 
$n_\textrm{as}$ &         95\% & 0.9837 (0.6240, 1.2896)&  & 0.9850 (0.9593, 1.0116) & 0.9638 (0.9377, 0.9889) &     \\ \hline 
$n_\textrm{iso}$ &         95\% & 1.4508 (0.5678, 2.1718)&  1.2465 (0.6096, 1.8885) & 0.9850 (0.9593, 1.0116) & 0.9638 (0.9377, 0.9889) &     \\ \hline 
$         r_0$ &         95\% & $<$ 0.2414&  $<$ 0.2866	 & & & $<$ 0.3334	      \\ \hline 
$\alpha_\textrm{cor0}$ &         95\% & 0.0961 (-0.0824, 0.1808)&  & 0.0846 (0.0184, 0.1713) & -0.0389 (-0.0960, -0.0067) &     \\ \hline 
$    \alpha_T$ &         95\% & 0.0580 (-0.0486, 0.1028)&  $<$ 0.0531 & 0.0462 (0.0104, 0.0916) & -0.0230 (-0.0593, -0.0039) &     \\ \hline

$-\ln{\mathcal{Z}}$& &3905.78 & 3902.38 & 3902.73 & 3905.35 & 3901.17 \\ \hline
$\ln(\mathcal{Z}_{\mtr{adiab.}}/\mathcal{Z})$& &4.6 &   1.2 &   1.6  &  4.2 & 0 \\ \hline
$\mathcal{Z}_{\mtr{adiab.}}/\mathcal{Z}$& &100 &   3.4 &    4.8 &   65 &   1  \\ \hline
\end{tabular}
\caption{\textbf{Amplitude parametrization, neutrino density isocurvature (NDI)}. The median values and 68\% or 95\% confidence level (C.L.) intervals (in parenthesis) are given for a selection of parameters. For the fully (anti)correlated models, $\gamma=\pm 1$, we have $r_0=0$, so these models are without tensor contribution. The last line,  $\mathcal{Z}_{\mtr{adiab.}}/\mathcal{Z}$, shows the ratio of the probability of the pure adiabatic model compared to the model of each column.
\label{tab:NDIamp}}
\end{table*}

\begin{table*}
\scriptsize
\centering
\begin{tabular}{|c|c|c|c|c|}
\hline 
       Parameter 	&	          C.L 	& Mixed NVI, general corr. 	&	  NVI, $\gamma=0$   & Adiabatic   \\ \hline
  $   \omega_b $ &         68\% & 0.0228 (0.0222, 0.0235)&  0.0238 (0.0231, 0.0245) & 0.0231 (0.0225, 0.0237)     \\ \hline 
$ \omega_{c} $ &         68\% & 0.1071 (0.1020, 0.1122)&  0.1075 (0.1024, 0.1126) & 0.1093 (0.1040, 0.1146)     \\ \hline 
$   100\theta$ &         68\% & 1.0484 (1.0451, 1.0521)&  1.0412 (1.0390, 1.0433) & 1.0419 (1.0398, 1.0440)     \\ \hline 
$        \tau$ &         68\% & 0.0898 (0.0760, 0.1040)&  0.0893 (0.0765, 0.1041) & 0.0899 (0.0757, 0.1045)     \\ \hline 
$\Omega_\Lambda$ &         68\% & 0.7744 (0.7482, 0.7984)&  0.7596 (0.7326, 0.7840) & 0.7508 (0.7221, 0.7769)     \\ \hline 
$         H_0$ &         68\% & 75.89 (73.05, 78.85)&  73.87 (71.33, 76.70) & 72.95 (70.27, 75.69)     \\ \hline 
$\ln[10^{10} A_0^2]$ &         68\% & 3.1895 (3.1399, 3.2402)&  3.1462 (3.1005, 3.1930) & 3.1038 (3.0641, 3.1422)     \\ \hline 
$n_\textrm{ad}^\textrm{eff}$ &         68\% & 0.9948 (0.9680, 1.0265)&  0.9958 (0.9777, 1.0171) & 0.9859 (0.9685, 1.0041)     \\ \hline 
$    \gamma_1$ &         95\% & $<$ -0.0588&  &     \\ \hline 
$    \gamma_2$ &         95\% & $<$ -0.0604&  &     \\ \hline 
$    \alpha_1$ &         95\% & $<$ 0.1245&  $<$ 0.1961	 &     \\ \hline 
$    \alpha_2$ &         95\% & $<$ 0.1349&  $<$ 0.1976	 &     \\ \hline 
$   \tilde{r}$ &         95\% & $<$ 0.6213&  $<$ 0.3088	 & $<$ 0.3334	      \\ \hline 
$    \gamma_0$ &         95\% & $<$ -0.0637&  &     \\ \hline 
$    \alpha_0$ &         95\% & $<$ 0.0997&  $<$ 0.1617	 &     \\ \hline 
$n_\textrm{ar}$ &         95\% & 0.9935 (0.7785, 1.1985)&  0.9958 (0.9607, 1.0390) & 0.9859 (0.9545, 1.0236)      \\ \hline 
$n_\textrm{as}$ &         95\% & 1.0009 (0.7213, 1.3083)&  &     \\ \hline 
$n_\textrm{iso}$ &         95\% & 1.1454 (0.0329, 2.4865)&  1.0528 (0.4302, 1.7779) &     \\ \hline 
$         r_0$ &         95\% & $<$ 0.2489&  $<$ 0.3088	 & $<$ 0.3334	      \\ \hline 
$\alpha_\textrm{cor0}$ &         95\% & -0.0907 (-0.1636, -0.0333)&  &     \\ \hline 
$    \alpha_T$ &         95\% & -0.0635 (-0.1267, -0.0091)&  $<$ 0.0576 &     \\ \hline 

$-\ln \mathcal{Z}$& & 3905.57 & 3903.96& 3901.17\\ \hline
$\ln(\mathcal{Z}_{\mtr{adiab.}}/\mathcal{Z})$& & 4.4 & 2.8 & 0 \\ \hline
$\mathcal{Z}_{\mtr{adiab.}}/\mathcal{Z}$& & 81 & 16 & 1  \\ \hline
\end{tabular}
\caption{\textbf{Amplitude parametrization, neutrino velocity isocurvature (NVI)}. The median values and 68\% or 95\% confidence level (C.L.) intervals (in parenthesis) are given for a selection of parameters. 
 \label{tab:NVIamp}}
 
\end{table*}

\begin{table*}
\scriptsize
\centering
\begin{tabular}{|c|c|c|c|c|c|c|}
 \hline Parameter &	 C.L 	& Mixed CDI, general corr. 	 	&	 Mixed CDI,  $\gamma=0$ 	&	Mixed CDI,  $\gamma=1$ & 	Mixed CDI, $\gamma=-1$	   &	   Adiabatic  \\ \hline
    
  $   \omega_b $ &         68\% & 0.0229 (0.0223, 0.0236)&  0.0237 (0.0230, 0.0244) & 0.0230 (0.0225, 0.0235) & 0.0225 (0.0221, 0.0230) & 0.0231 (0.0225, 0.0237)     \\ \hline 
$ \omega_{c} $ &         68\% & 0.1052 (0.1002, 0.1105)&  0.1062 (0.1007, 0.1113) & 0.1079 (0.1030, 0.1126) & 0.1162 (0.1113, 0.1212) & 0.1093 (0.1040, 0.1146)     \\ \hline 
$   100\theta$ &         68\% & 1.0472 (1.0435, 1.0505)&  1.0436 (1.0415, 1.0459) & 1.0438 (1.0415, 1.0462) & 1.0391 (1.0370, 1.0412) & 1.0419 (1.0398, 1.0440)     \\ \hline 
$        \tau$ &         68\% & 0.0906 (0.0772, 0.1046)&  0.0922 (0.0790, 0.1067) & 0.0866 (0.0741, 0.1007) & 0.0903 (0.0764, 0.1046) & 0.0899 (0.0757, 0.1045)     \\ \hline 
$\Omega_\Lambda$ &         68\% & 0.7798 (0.7526, 0.8029)&  0.7700 (0.7430, 0.7976) & 0.7611 (0.7366, 0.7849) & 0.7076 (0.6768, 0.7342) & 0.7508 (0.7221, 0.7769)     \\ \hline 
$         H_0$ &         68\% & 76.24 (73.26, 79.26)&  75.16 (72.40, 78.60) & 74.02 (71.59, 76.67) & 68.85 (66.59, 71.17) & 72.95 (70.27, 75.69)     \\ \hline 
$\ln[10^{10} A_0^2]$ &         68\% & 3.0931 (3.0540, 3.1404)&  3.1725 (3.1211, 3.2281) & 3.0859 (3.0498, 3.1223) & 3.1689 (3.1316, 3.2074) & 3.1038 (3.0641, 3.1422)     \\ \hline 
$n_\textrm{ad}^\textrm{eff}$ &         68\% & 0.9961 (0.9788, 1.0149)&  1.0030 (0.9839, 1.0261) & 0.9902 (0.9751, 1.0064) & 0.9575 (0.9430, 0.9712) & 0.9859 (0.9685, 1.0041)     \\ \hline 
$    \gamma_1$ &         95\% & $>$ -0.1275&  & & &     \\ \hline 
$    \gamma_2$ &         95\% & $>$ -0.1042&  & & &     \\ \hline 
$    \alpha_1$ &         95\% & $<$ 0.0454&  $<$ 0.1114	 & $<$ 0.0148	 & $<$ 0.0073	 &     \\ \hline 
$    \alpha_2$ &         95\% & $<$ 0.3815&  $<$ 0.5012	 & $<$ 0.0148	 & $<$ 0.0073	 &     \\ \hline 
$ \tilde{r}_0$ &         95\% & $<$ 0.5238&  $<$ 0.2925	 & & & $<$ 0.3334	      \\ \hline 
$    \gamma_0$ &         95\% & $>$ -0.1126&  & & &     \\ \hline 
$    \alpha_0$ &         95\% & $<$ 0.0955&  $<$ 0.2110	 & $<$ 0.0148	 & $<$ 0.0073	 &     \\ \hline 
$n_\textrm{ar}$ &         95\% & 0.9964 (0.8057, 1.1627)&  1.0030 (0.9673, 1.0506) & & & 0.9859 (0.9545, 1.0236)      \\ \hline 
$n_\textrm{as}$ &         95\% & 0.9995 (0.6086, 1.3376)&  & 0.9902 (0.9611, 1.0218) & 0.9575 (0.9284, 0.9845) &     \\ \hline 
$n_\textrm{iso}$ &         95\% & 2.0523 (0.8075, 3.3592)&  1.7338 (0.9795, 2.7873) & 0.9902 (0.9611, 1.0218) & 0.9575 (0.9284, 0.9845) &     \\ \hline 
$         r_0$ &         95\% & $<$ 0.2571&  $<$ 0.2925	 & & & $<$ 0.3334	      \\ \hline 
$\alpha_\textrm{cor0}$ &         95\% & 0.0747 (-0.0770, 0.1499)&  & 0.0531 (0.0100, 0.1206) & -0.0368 (-0.0852, -0.0063) &     \\ \hline 
$    \alpha_T$ &         95\% & 0.0209 (-0.0295, 0.0467)& $ <$ 0.0358 & 0.0272 (0.0055, 0.0582) & -0.0214 (-0.0532, -0.0035) &     \\ \hline 

$-\ln{\mathcal{Z}}$ & & 3905.93 & 3902.82 & 3904.71 & 3905.49& 3901.17 \\ \hline
$\ln(\mathcal{Z}_{\mtr{adiab.}}/\mathcal{Z})$& &4.8 &  1.7 &   3.5 &    4.3 &   0 \\ \hline
$\mathcal{Z}_{\mtr{adiab.}}/\mathcal{Z}$& & 117 &    5.2 &   34 &   75 &   1  \\ \hline 
\end{tabular}
\caption{{\bf Amplitude parametrization, cold dark matter isocurvature (CDI)}.  The median values and 68\% or 95\% confidence level (C.L.) intervals (in parenthesis) are given for a selection of parameters. For the fully (anti)correlated models, $\gamma=\pm 1$, we have $r_0=0$, so these models are without tensor contribution. 
\label{tab:CDIamp}}
\end{table*}

\begin{table*}
\scriptsize
\centering  
\begin{tabular}{|c|c|c|c|c|c|c|}
\hline 
      Parameter 	&	C.L. & Mixed NDI, gen. corr. 	 	&	  Mixed NDI, $  \gamma=0$ 	&	  Mixed NDI,     $\gamma=1$ & 	 Mixed NDI,    $ \gamma=-1$	 &	   Adiabatic    \\ \hline
   $   \omega_b $ &         68\% & 0.0234 (0.0229, 0.0240)&  0.0236 (0.0229, 0.0243) 	 & 0.0233 (0.0228, 0.0239) 	 & 0.0224 (0.0219, 0.0229) 	 &  0.0231 (0.0226, 0.0236)      \\ \hline 
$    \omega_c$ &         68\% & 0.1077 (0.1030, 0.1122)&  0.1064 (0.1010, 0.1114) 	 & 0.1092 (0.1046, 0.1136) 	 & 0.1145 (0.1102, 0.1190) 	 &  0.1095 (0.1045, 0.1142)      \\ \hline 
$      \theta$ &         68\% & 1.0454 (1.0428, 1.0484)&  1.0437 (1.0413, 1.0464) 	 & 1.0457 (1.0431, 1.0486) 	 & 1.0390 (1.0365, 1.0411) 	 &  1.0418 (1.0400, 1.0437)      \\ \hline 
$        \tau$ &         68\% & 0.0876 (0.0752, 0.1014)&  0.0901 (0.0768, 0.1041) 	 & 0.0884 (0.0749, 0.1014) 	 & 0.0891 (0.0755, 0.1025) 	 &  0.0899 (0.0763, 0.1034)      \\ \hline 
$\ln[10^{10} A_0^2]$ &         68\% & 3.0686 (3.0312, 3.1081)&  3.1739 (3.1167, 3.2499) 	 & 3.0710 (3.0355, 3.1098) 	 & 3.1663 (3.1307, 3.2057) 	 &  3.1059 (3.0690, 3.1409)      \\ \hline 
$\Omega_\Lambda$ &         68\% & 0.7669 (0.7427, 0.7905)&  0.7699 (0.7427, 0.7966) 	 & 0.7611 (0.7364, 0.7847) 	 & 0.7151 (0.6865, 0.7397) 	 &  0.7495 (0.7238, 0.7750)      \\ \hline 
$         H_0$ &         68\% & 75.02 (72.46, 77.89)&  75.13 (72.19, 78.48) 	 & 74.47 (71.84, 77.26) 	 & 69.33 (67.06, 71.53) 	 & 72.75 (70.40, 75.36)      \\ \hline 
$n_\textrm{ad}^\textrm{eff}$ &         68\% & 0.9879 (0.9742, 1.0026)&  0.9982 (0.9805, 1.0178) 	 & 0.9853 (0.9720, 0.9990) 	 & 0.9639 (0.9509, 0.9766) 	 &  0.9843 (0.9697, 1.0021)      \\ \hline 
$    \gamma_0$ &         95\% & $>$ -0.3178&  & & &     \\ \hline 
$    \alpha_0$ &         95\% & $<$ 0.0568&  $<$ 0.2482	 & $<$ 0.0280	 & $<$ 0.0104	 &     \\ \hline 
$\eta_{\sigma\sigma}$ &         95\% & p.r.&  0.0146 (-0.0129, 0.0641) 	 & & & 0.0102 (-0.0175, 0.0637)      \\ \hline 
$\eta_{\sigma s}$ &         95\% & p.r.&  & & &     \\ \hline 
$   \eta_{ss}$ &         95\% & p.r.&  p.r.& &  &     \\ \hline 
$    \varepsilon$ &         95\% & $<$ 0.0570&  $<$ 0.0152	 & & & $<$ 0.0193	      \\ \hline 
$\eta_{ss}-\varepsilon$ &         95\% & &  & -0.0073 (-0.0202, 0.0061) 	 & -0.0180 (-0.0311, -0.0056) 	 &     \\ \hline 
$n_\textrm{ar}$ &         95\% & 0.9486 (0.6564, 1.0805)&  0.9982 (0.9649, 1.0400) 	 & & & 0.9843 (0.9561, 1.0183)      \\ \hline 
$n_\textrm{as}$ &         95\% & 0.9981 (0.8459, 1.1719)&  & 0.9853 (0.9595, 1.0122) 	 & 0.9639 (0.9378, 0.9888) 	 &     \\ \hline 
$n_\textrm{iso}$ &         95\% & 0.9957 (0.8548, 1.1206)&  1.0129 (0.8475, 1.1325) 	 & 0.9853 (0.9595, 1.0122) 	 & 0.9639 (0.9378, 0.9888) 	 &     \\ \hline 
$         r_0$ &         95\% & $<$ 0.1909&  $<$ 0.2439	 & & & $<$ 0.3083	      \\ \hline 
$\alpha_\textrm{cor0}$ &         95\% & 0.0729 (-0.0424, 0.1535)&  & 0.0848 (0.0194, 0.1649) 	 & -0.0378 (-0.1015, -0.0061) 	 &     \\ \hline 
$    \alpha_T$ &         95\% & 0.0405 (-0.0222, 0.0840)&  $<$ 0.0503 	 & 0.0463 (0.0110, 0.0881) 	 & -0.0222 (-0.0623, -0.0035) 	 &     \\ \hline 

$-\ln{\mathcal{Z}}$ & & 3901.52  & 3900.19 & 3901.09 & 3903.60 & 3898.94 \\ \hline
$\ln(\mathcal{Z}_{\mtr{adiab.}}/\mathcal{Z})$& &2.6  &  1.3  &2.2 & 4.7 &  0\\ \hline
$\mathcal{Z}_{\mtr{adiab.}}/\mathcal{Z}$& & 13 &    3.5 &    8.6 &  106 &    1\\ \hline 
\end{tabular}
\caption{\textbf{Slow-roll parametrization, neutrino density isocurvature (NDI)}. The median values and 68\% or 95\% confidence level (C.L.) intervals (in parenthesis) are given for a selection of parameters. For some parameters the whole prior range (p.r.) is allowed by the data. Note: $\gamma=\pm 1$ models do not have a tensor contribution (since $r_0=0$ automatically), and the only ``slow-roll parameter'' is then the combination $\eta_{ss}-\varepsilon$ for which we assume a uniform prior $(-0.075,\,0.075)$.
 \label{tab:NDIslow}}
\end{table*}

\begin{table*}
\scriptsize
\centering  
\begin{tabular}{|c|c|c|c|c|c|c|}
 \hline  Parameter &	C.L.& Mixed CDI, gen. corr. 	  	&	 Mixed CDI,  $\gamma=0$ 	&	Mixed CDI,   $\gamma=1$ & 	 Mixed CDI, $\gamma=-1$	 &	   Adiabatic   \\ \hline
   $   \omega_b $ &         68\% & 0.0231 (0.0226, 0.0236)&  0.0234 (0.0228, 0.0241) 	 & 0.0230 (0.0225, 0.0235) 	 & 0.0225 (0.0220, 0.0230) 	 &  0.0231 (0.0226, 0.0236)      \\ \hline 
$    \omega_c$ &         68\% & 0.1088 (0.1029, 0.1150)&  0.1068 (0.1018, 0.1119) 	 & 0.1077 (0.1029, 0.1129) 	 & 0.1164 (0.1113, 0.1215) 	 &  0.1095 (0.1045, 0.1142)      \\ \hline 
$      \theta$ &         68\% & 1.0421 (1.0390, 1.0452)&  1.0426 (1.0405, 1.0447) 	 & 1.0440 (1.0417, 1.0464) 	 & 1.0390 (1.0368, 1.0413) 	 &  1.0418 (1.0400, 1.0437)      \\ \hline 
$        \tau$ &         68\% & 0.0894 (0.0763, 0.1030)&  0.0894 (0.0757, 0.1037) 	 & 0.0862 (0.0733, 0.1003) 	 & 0.0904 (0.0757, 0.1059) 	 &  0.0899 (0.0763, 0.1034)      \\ \hline 
$\ln[10^{10} A_0^2]$ &         68\% & 3.1042 (3.0558, 3.1660)&  3.1207 (3.0812, 3.1595) 	 & 3.0836 (3.0469, 3.1227) 	 & 3.1706 (3.1333, 3.2105) 	 &  3.1059 (3.0690, 3.1409)      \\ \hline 
$\Omega_\Lambda$ &         68\% & 0.7539 (0.7162, 0.7851)&  0.7650 (0.7379, 0.7896) 	 & 0.7620 (0.7360, 0.7858) 	 & 0.7062 (0.6725, 0.7356) 	 &  0.7495 (0.7238, 0.7750)      \\ \hline 
$         H_0$ &         68\% & 73.15 (69.62, 76.75)&  74.43 (71.72, 77.31) 	 & 74.09 (71.62, 76.91) 	 & 68.78 (66.23, 71.27) 	 & 72.75 (70.40, 75.36)      \\ \hline 
$n_\textrm{ad}^\textrm{eff}$ &         68\% & 0.9848 (0.9623, 1.0042)&  0.9973 (0.9796, 1.0168) 	 & 0.9905 (0.9757, 1.0071) 	 & 0.9565 (0.9412, 0.9716) 	 &  0.9843 (0.9697, 1.0021)      \\ \hline 
$    \gamma_0$ &         95\% & p.r.&  & & &     \\ \hline 
$    \alpha_0$ &         95\% & $<$ 0.0298&  $<$ 0.1171	 & $<$ 0.0125	 & $<$ 0.0064	 &     \\ \hline 
$\eta_{\sigma\sigma}$ &         95\% &p.r.&  0.0154 (-0.0134, 0.0628) 	 & & & 0.0102 (-0.0175, 0.0637)      \\ \hline 
$\eta_{\sigma s}$ &         95\% &p.r.&  & & &     \\ \hline 
$   \eta_{ss}$ &         95\% & p.r.&  p.r. 	 &  &  &     \\ \hline 
$    \varepsilon$ &         95\% & $<$ 0.0378&  $<$ 0.0157	 & & & $<$ 0.0193	      \\ \hline 
$ \eta_{ss} - \varepsilon $ &         95\% & &  & -0.0048 (-0.0193,  0.0113 ) 	 & -0.2176 (-0.0369, -0.0068) 	 &     \\ \hline 
$n_\textrm{ar}$ &         95\% & 0.9657 (0.7220, 1.0712)&  0.9973 (0.9636, 1.0370) 	 & & & 0.9843 (0.9561, 1.0183)      \\ \hline 
$n_\textrm{as}$ &         95\% & 0.9938 (0.6802, 1.1844)&   & 0.9905 (0.9613, 1.0226) 	 & 0.9565 (0.9263, 0.9863) 	 &     \\ \hline 
$n_\textrm{iso}$ &         95\% & 0.9889 (0.8459, 1.1219)&  1.0199 (0.8481, 1.1355) 	 & 0.9905 (0.9613, 1.0226) 	 & 0.9565 (0.9263, 0.9863) 	 &     \\ \hline 

$         r_0$ &         95\% & $<$ 0.2033&  $<$ 0.2516	 & & & $<$ 0.3083	      \\ \hline 
$\alpha_\textrm{cor0}$ &         95\% & 0.0167 (-0.0836, 0.1004)&  & 0.0542 (0.0113, 0.1249) 	 & -0.0379 (-0.0871, -0.0076) 	 &     \\ \hline 

$    \alpha_T$ &         95\% & 0.0093 (-0.0448, 0.0497)&  $<$ 0.0404 	 & 0.3607 (0.0083, 0.0541) 	 & -0.0219 (-0.0550, -0.0043) 	 &     \\ \hline 

$-\ln\mathcal{Z}$ & & 3902.87  & 3901.74 & 3902.40 & 3903.86 & 3898.94 \\ \hline
$\mathcal{Z}_{\mtr{adiab.}}/\mathcal{Z}$& &51 & 16 & 33 & 137 & 1\\ \hline
\end{tabular}
\caption{\textbf{Slow-roll parametrization, cold dark matter density isocurvature (CDI)}. The median values and 68\% or 95\% confidence level (C.L.) intervals (in parenthesis) are given for a selection of parameters.  For some parameters the whole prior range (p.r.) is allowed by the data. Note: $\gamma=\pm 1$ models do not have a tensor contribution (since $r_0=0$ automatically), and the only ``slow-roll parameter'' is then the combination $\eta_{ss}-\varepsilon$ for which we assume a uniform prior $(-0.075,\,0.075)$.
 \label{tab:CDIslow}}

\end{table*}

\end{document}